\documentclass[aps,prb,twocolumn]{revtex4}
\usepackage{amsmath,amsthm}             
\usepackage[pdftex]{graphicx}           

\let\abs=\envert

\newcommand{\beq}{\begin{equation}}
\newcommand{\eeq}{\end{equation}}
\newcommand{\eel}[1]{\label{#1}\end{equation}}
\newcommand{\bea}{\begin{eqnarray}}
\newcommand{\eea}{\end{eqnarray}}
\newcommand{\eeal}[1]{\label{#1}\end{eqnarray}}
\newcommand{\beac}{\begin{equation}\begin{array}{rcl}}
\newcommand{\eeacn}[1]{\end{array}\label{#1}\end{equation}}
\newcommand{\beqs}[1]{\begin{equation}\label{#1}\begin{split}}
\newcommand{\eeqs}{\end{split}\begin{equation}}

\newcommand\ket[1]{\big|\,#1\,\big\rangle}
\newcommand\bra[1]{\big\langle\,#1\,\big|}
\newcommand\braket[2]{\big\langle\,#1\,\big|\,#2\,\big\rangle}

\newcommand\pe{{\bf {\cdot}}}    
\newcommand\op[1]{{{\bf #1}}}  
\newcommand\vect[1]{\overrightarrow{#1}}    
\newcommand\vers[1]{\op{\hat{#1}}}    
\newcommand\vop[1]{\overrightarrow{\bf{#1}}}    
\newcommand\opc[1]{{\cal #1}}

\newcommand\eq[1]{Eq.~(\ref{#1})}
\newcommand\numeq[1]{(\ref{#1})}
\newcommand\eqs[1]{Eqs.~(\ref{#1})}
\newcommand\fig[1]{Figure~\ref{#1}}
\newcommand\numfig[1]{\ref{#1}}
\newcommand\figs[1]{Figures~\ref{#1}}
\newcommand\sect[1]{Section~\ref{#1}}
\newcommand\apen[1]{Appendix~\ref{#1}}

\newcommand\insertfig[5]
{\begin{figure}[#1]
    \begin{center}
        \includegraphics[bb = 0 0 #3cm #4cm]{#2.jpg}
    \end{center}
    \caption{#5}
    \label{#2}
\end{figure}}

\begin{document}

    \title{Quantum irreversible decoherence behaviour in open quantum systems with few degrees of freedom. Application to $^1$H NMR reversion experiments in nematic liquid crystals.}
    \author{H. H. Segnorile and  R. C. Zamar}
    \affiliation{Instituto de F\'isica Enrique Gaviola - CONICET,
    Facultad de Matem\'atica, Astronom\'{\i}a y F\'{\i}sica,
    Universidad Nacional de C\'ordoba \\ M.Allende y H. de la Torre - Ciudad Universitaria,
    X5016LAE - C\'ordoba, Argentina}
    \date{\today}

    \begin{abstract}
    {\small
        An experimental study of NMR spin decoherence in nematic liquid crystals is presented. Decoherence dynamics can be put in evidence by means of refocusing experiments of the dipolar interactions. The experimental technique used in this work is based on the MREV8 pulse sequence.  The aim of the work is to detect  the main features of the irreversible quantum decoherence in liquid crystals,  on the basis of the theory presented by the authors  recently. The focus is laid on experimentally probing the eigen-selection process in the intermediate time scale, between quantum interference of a closed system and thermalization, as a signature of the quantum spin decoherence of the open quantum system,  as well as on quantifying the effects of non-idealities as possible sources of signal decays which could mask  the intrinsic decoherence.  In order to contrast experiment and theory, the theory was adapted to obtain the decoherence function corresponding to the MREV8 reversion experiments.  Non-idealities of the experimental setting, like external field inhomogeneity, pulse misadjustments and the presence of non-reverted spin interaction terms are analysed in detail within this framework, and their effects on the observed signal decay are numerically estimated.
        It is found that, though all these non-idealities could in principle affect the evolution of the spin dynamics, their influence can be mitigated and they do not present the characteristic behaviour of the irreversible spin decoherence. As unique characteristic of decoherence, the experimental results clearly show the occurrence of eigen-selectivity in the intermediate timescale, in complete agreement with the theoretical predictions. We conclude that the eigen-selection effect is the fingerprint of decoherence associated with a quantum open spin system in liquid crystals. Besides, these features of the results account for the quasi-equilibrium states of the spin system, which were observed previously in these mesophases, and lead to conclude that the quasi-equilibrium is a definite stage of the spin dynamics during its evolution towards equilibrium.
    }
    \end{abstract}
    \maketitle


\section{Introduction}\label{se:intro}

    Derivation of a thermodynamic-like stationary state, of equilibrium or quasi-equilibrium,  from a microscopic quantum mechanical starting point, is a major problem faced today by the physics of irreversible processes in nonequilibrium systems, with impact in a variety of fields, from fundamental research to applications in areas of high current interest like quantum computing and quantum information theory \cite{popescu06,polkonikov11,yukareview,rieman,rigol08,privman98,palma96,reina02,vanBeek11,charpentier07}.

    Particularly, the occurrence of quasi-equilibrium spin states in nematic liquid crystals (LC's) poses the problem of the irreversible evolution of a finite open quantum system of interacting particles, coupled with a large quantum environment, towards a quasi-stationary state \cite{SegZam2011}.
    In LC's, due to the rapid molecular motions that average the intermolecular spin interactions to zero, the  effective spin system comprises a small number of magnetic degrees of freedom, namely the intramolecular dipolar interactions, which remain because of the typical orientational order of these mesophases \cite{degene93}. However, the proton Nuclear Magnetic Resonance (NMR) response in LC's is consistent with a true quasi-equilibrium in spite of the small number of the degrees of freedom of the observed system \cite{gonza11}. This topic has attracted the interest of many researchers in the NMR field \cite{pintar74,ernst97,waugh98,waugh04,walls06,halse2012} as well as other areas of physics \cite{yukalov11}.

    In our recent publication on NMR quantum decoherence in LC's, which we shall refer to as QD-I \cite{SegZam2011}, a theory was presented which describes the irreversible quantum decoherence processes undergone by an observed and controlled system of quantum interacting particles because of its coupling with an unobserved lattice or environment.
    It provided a comprehensive explanation, compatible with  previous experimental evidence \cite{buljuba09,gonza11}, on the mechanisms which turn the proton spin system density matrix of nematic LC's into a quasi-equilibrium form after an arbitrary initial coherent state .

    Accordingly, the decoherence process transforms the initial state into a block diagonal matrix in the eigenbasis of the spin-environment interaction Hamiltonian. This evolution occurs over a time scale intermediate between that of the Liouvillian evolution of an isolated observed system and the long time scale where evolution is governed by relaxation and thermalization processes driven by thermal fluctuations.

    The fingerprint of quantum decoherence theory is the eigen-selection process, which causes  a selective decay of the off-diagonal components of the density matrix, preserving the block diagonal part of the initial state.
    Such explanation relies on general requirements on the eigenvalue distribution functions of the  relevant quantum operators that represent the interaction of the system with the environment.
    Hence, experimental observation of eigen-selectivity would provide a direct evidence of the correlated dynamics between the spin system and the quantum environment that preludes the reach to quasi-equilibrium.

    The aim of this work is the experimental study of decoherence spin dynamics, within the theoretical framework of  QD-I. Particularly, we focus on the most relevant aspects of this phenomenon, namely eigen-selectivity and the occurrence of the intermediate time scale.
    Irreversible decoherence is studied by means of refocusing experiments, designed to counteract the effect of  the spin dynamics generated by the dipolar spin interactions.
    Besides of the quantum interference corresponding to a closed system, the intrinsic decoherence coexists with other sources of the signal decay, like the line broadening due to a distribution of the order parameter \cite{Limmer}, non-idealities of the experimental setting like inhomogeneity of the static and rf mangetic fields or pulse imperfections, and the effects  of non-secular terms of the dipolar Hamiltonian that cannot be experimentally reversed  \cite{RhimPV71}. Accordingly, to isolate the distinctive features of decoherence associated only with the microscopic dynamics, we explore the intermediate time scale by combining dipolar refocusing with a meticulous theoretical and numerical analysis of the experiments based on hypotheses of general character. This strategy enables us to visualize the physical processes involved and to quantify the influence of the non-idealities.

    Confirmation of the irreversibility of the spin dynamics within the intermediate time scale has  relevant practical and basic consequences.
    It implies that the state attained by the spin system over the intermediate scale has a representation in the form of a block diagonal density operator, containing the information of the initial preparation which does not undergone subsequent evolution, except for the slow dynamics imposed by the  spin-relaxation process, no matter the details of the coherences present in the initial condition.
    The traditional introduction of the spin-temperature hypothesis amounts to assuming that the off-diagonal density matrix elements can be forgotten after quantum interference cancels their contribution to the observable expectation value \cite{goldman70}.
    In this work we provide a quantum explanation for the damping of the off-diagonal elements and for the time scale where it occurs. Clarifying this aspect enables to replace the phenomenological assumption by  a specific condition satisfied by the spin system state.
    Also, understanding the nature of the decoherence mechanism  can contribute to the set up of a quantum spin-lattice relaxation theory beyond the Markovian limit, providing insight into the interplay of the quantum correlations developed in the microscopic irreversible dynamics and the dissipative macroscopic evolution.
    Finally, the analysis carried out in this work may contribute to the current discussion about the role of system-environment entanglement as basic mechanism of quantum decoherence of interacting particle systems \cite{helm09,pernice12,popescu06,Makri13}.

    The article is organized as follows. In \sect{se:Exp_Eigen}, by means of experiments of refocusing of the dipolar interactions, we show how the eigen-selection process is evidenced during the partial reversion of the spin dynamics, which occurs in an intermediate time scale, between the time scales of the Liouvillian free evolution or adiabatic by one hand, and the thermalization and relaxation processes by the other hand.
    \sect{se:Theory_Back} is devoted to a detailed summary of the concepts developed and results obtained in QD-I, which are
    exhaustively used in this work.
    In \sect{subse:MedRevEigen_DefExp} a theoretical description of the decoherence dynamics driven by reversion experiments is presented, while a series of experimental measurement are shown in \sect{subse:MedRevEigen_Med} and the results compared with the theoretical prediction.
    \sect{subse:MedRevEigen_ErrMed} is concerned with the analysis of possible sources of eigen-selectivity like effects coming from a closed-spin-system dynamics under experimental misadjustments or non-idealities in the theoretical approach used for its description.
    A supplementary material\cite{suppmatDQII} is attached to this work, where the details about the origin of the theoretical expression for the reversion evolution operator and the signals obtained in \sect{subse:MedRevEigen_ErrMed} are presented together with the results of numerical calculations for such signals under different experimental settings.
    Besides, an experimental analysis of the effects of the inhomogeneity of the static magnetic field is presented there.\\
    Finally, \sect{se:conclu} is devoted to discuss and conclude about the concepts developed and results obtained along this work.

\section{Eigen-selectivity effects on coherence evolution}\label{se:Exp_Eigen}

    This section is dedicated to the experimental detection of the eigen-selection process which characterizes the irreversible time evolution of the spin coherence in LC's, due to the coupling of the spin system to a quantum lattice or environment. This will be done within the framework of the theoretical proposal presented in QD-I.

\subsection{Theoretical background}\label{se:Theory_Back}

    For the convenience of the reader, hereinafter we  include a brief summary of the steps followed in QD-I to work out the nonequilibrium irreversible dynamics of an observed and controlled system coupled to an external unobserved environment. This problem was approached using the following Hamiltonian:
    \begin{equation}\label{HamGen}
        \opc{H} = \opc{H}_S+\opc{H}_{SL}+\opc{H}_L,\\
    \end{equation}
    where $\opc{H}_S = \opc{H}_S^{(s)}\otimes\op{1}^{(f)}$ and $\opc{H}_L = \op{1}^{(s)}\otimes\opc{H}_L^{(f)}$ are respectively the Hamiltonians of the observed system (i.e. the spin system) and the lattice or unobserved environment. Symbols of the form  $\op{O}^{(s)}$ and $\op{O}^{(f)}$ indicate operators acting exclusively on the Hilbert space of the system and the lattice, respectively. The interaction Hamiltonian  acts on both Hilbert spaces, and is represented by   $\opc{H}_{SL} = \sum_q \op{F}_q \op{A}_q$, with $\op{F}_q = \op{1}^{(s)}\otimes\op{F}_q^{(f)}$,
    and $\op{A}_q = \op{A}_q^{(s)}\otimes\op{1}^{(f)}$. Index $q$ can label different characteristics,
    like a spin pair or a tensor component of the interaction Hamiltonian.

    The existence of distinct time scales associated with different physical processes, which become important as the spin dynamics evolves, was postulated. The occurrence of such time scales is subjected to certain general conditions that the relevant Hamiltonians must fulfill. Such requirements are associated with commutation relations between $\opc{H}_S$, $\opc{H}_L$ and $\opc{H}_{SL}$ (it is worth to note that $[\opc{H}_S,\opc{H}_L]=0$).
    In this way, the dynamics of the observed system in the earlier time scale is obtained by assuming that
    $[\opc{H}_S,\opc{H}_{SL}]=0$ and $[\opc{H}_L,\opc{H}_{SL}]=0$, which implies a system model called {\it essentially isolated system}, {\it where the observed system evolves reversibly and the expectation values of the observables decay due to quantum interference.}
    A later time scale is obtained by assuming $[\opc{H}_S,\opc{H}_{SL}]=0$ and $[\opc{H}_L,\opc{H}_{SL}]\neq0$, where the system model was referred to as {\it essentially adiabatic system}, {\it and the corresponding dynamics of its observables is irreversible.}
    Finally, the latest time scale is obtained by assuming $[\opc{H}_S,\opc{H}_{SL}]\neq0$ and $[\opc{H}_L,\opc{H}_{SL}]\neq0$, and the system model was named {\it system in thermal contact}.

    The focus in QD-I was laid on the study of pure decoherence processes of the observables due to the coupling of the system to a quantum environment, without thermalization and relaxation effects. Therefore, the dynamics of the observed system was studied  assuming $[\opc{H}_S,\opc{H}_{SL}]=0$, namely, the system dynamics was described under {\it essentially isolated} or {\it essentially adiabatic} models, where the observed system conserves its energy.

    Within this framework, in the following we will present the Hamiltonians which describe the NMR experiments on a wide variety of nematic LC's.
    Also, we will explain the procedure for averaging over the lattice variables involved in the definitions.
    Hereinafter, all the Hamiltonians are expressed in units of $\hbar$.
    First, we write the full-quantum dipolar Hamiltonian of nematic LC's as
    \begin{equation}\label{sum_HamDip_CL}
        \opc{H}_d = \sum_i \opc{H}_{di},
    \end{equation}
    where
    \begin{equation}\label{HamDip_CL}
    \begin{split}
        \opc{H}_{di} &= \op{H}_{\op{di}}^{(s)}
        \otimes\op{S}_{\op{zzi}}^{(f)},
    \end{split}
    \end{equation}
    with
    \[\op{H}_{\op{di}}^{(s)} =  \op{1}^{(s_1)}\otimes\cdots\otimes\op{H}_{\op{d}}^{(s_i)}
    \otimes\cdots\otimes\op{1}^{(s_N)}.\]
    In the former equations, index $i$ labels the $i$-th molecule and the sum runs over the molecules of the whole sample.
    The superscript $(s_i)$ indicates an operator acting on the spin Hilbert space of the $i$-th molecule.
    The operator $\op{H}_{\op{d}}^{(s_i)}$ is the spin part of the contribution to the dipolar Hamiltonian of the $i$-th molecule, which is written as
    \begin{equation}\label{dipolar_sol}
    \begin{split}
        \op{H}_{\op{d}}^{(s_i)} &= -\frac{3}{2}\gamma^{2}\hbar\sum_{j \neq k}\frac{1}{r_{jk}^{3}}
        \left(\frac{3}{2}\cos^{2}{\beta_{jk}}-\frac{1}{2}\right)\\
        &\times\left(\op{I_{zj}I_{zk}}-\frac{1}{3}\vop{I_j}\pe\vop{I_k}\right)^{(s_i)}.
    \end{split}
    \end{equation}
    The expression \numeq{dipolar_sol} is the secular part of the dipolar Hamiltonian, which is adequate for describing the spin dynamics in the high intensity external static magnetic field approximation.
    The indices $j,k$ run over all the proton sites within the $i$-th molecule, $r_{jk}$ is the internuclear distance between spins $j$ and $k$, $\beta_{jk}$ stand for the polar angle of the vector $\vect{r_{jk}}$ with respect to the system fixed to the molecule.
    The spin angular momentum is $\vop{I_j} = \op{I_{xj}}\,\vers{x} + \op{I_{yj}}\,\vers{y} + \op{I_{zj}}\,\vers{z}$.
    The introduction of the quantum character of the environment variables in \eq{HamDip_CL} is carried out by the operators $\op{S}_{\op{zzi}}^{(f)}$.
    That is a novel contribution of QD-I, which allows a general description of the dipolar Hamiltonian in liquid crystal NMR with the important consequence that the quantum correlation between the observed system and the environment can be included in the dynamics.
    Precisely, as shown in QD-I, the irreversibility of the spin dynamics naturally emerges in the decoherence time scale when the full-quantum character of the spin-environment interaction energy is assumed.
    The meaning of $\op{S}_{\op{zzi}}^{(f)}$ is that of a molecular orientational operator whose eigenvalues $S_i$ are related with the angle $\theta_i$ between the long molecular axis of the $i$-th molecule and the external static magnetic field $\vect{B_0}$, where $S_i = \left(\frac{3}{2}\cos^{2}{\theta_i}-\frac{1}{2}\right)$.

    In the definition of \eq{sum_HamDip_CL}, we used a kind of motionally averaged dipolar Hamiltonian approach\cite{SegZam2011}.
    This means, in a quantum language, that the eigenvalues of the intermolecular lattice operators of the dipolar Hamiltonian are negligible in comparison with the intramolecular ones and the time scale of the dynamics originated from the intermolecular dipolar terms is much longer than that of the intramolecular one.
    Accordingly, such slow dynamics can be neglected by eliminating the intermolecular terms from the dipolar Hamiltonian.

    The average dipolar energy is obtained through the equilibrium lattice density operator as
    \begin{equation}\label{HamDip_CL_promf}
    \langle\opc{H}_{di}\rangle_f = tr_f\left\{\opc{H}_{di}\,\rho_{L(eq)}\right\},
    \end{equation}
    where the trace is taken over the lattice variables, $\langle\cdot\rangle_f \equiv tr_f\left\{\cdot\,\rho_{L(eq)}\right\}$
    denotes the expectation value, and $\rho_{L(eq)}$ is the lattice density operator at thermal equilibrium,
    \begin{equation}\label{rho_f_equil}
    \rho_{L(eq)} = \op{1}^{(s)}\otimes\rho_{L(eq)}^{(f)} = \op{1}^{(s)}\otimes e^{-\beta_T\,\opc{H}_{L}^{(f)}}/\opc{N}_f,
    \end{equation}
    where $\beta_T \equiv \frac{1}{k_B T}$,  $k_B$ the Boltzmann constant, $T$ the absolute temperature, and $\opc{N}_f~\equiv~tr_f\left\{e^{-\beta_T\,\opc{H}_{L}^{(f)}}\right\}$.
    The molecular dipolar Hamiltonian corresponding to the closed spin system is obtained by tracing over the lattice variables in \eq{HamDip_CL_promf}:
    \begin{equation}\label{HamDipi_CL_promf}
    \begin{split}
    \langle\opc{H}_{di}\rangle_f &= S_{zz}\,\op{H}_{\op{di}}^{(s)}
    \otimes\op{1}^{(f)},
    \end{split}
    \end{equation}
    with the definition of the nematic order parameter $S_{zz}$ \cite{degene93} as
    \begin{equation}\label{order_param_def}
    \begin{split}
    S_{zz} \equiv \langle \op{S}_{\op{zzi}}^{(f)} \rangle_f &= tr_f\left\{\op{S}_{\op{zzi}}^{(f)}\,\rho_{L(eq)}^{(f)}\right\}\\
    &= \sum_f\, S_i(f)\, \bra{f}\rho_{L(eq)}^{(f)}\ket{f},
    \end{split}
    \end{equation}
    where $\left\{\ket{f}\right\}$ is an eigenbasis of the operator $\op{S}_{\op{zzi}}^{(f)}$ with
    $\op{S}_{\op{zzi}}^{(f)}\ket{f} = S_i(f)\ket{f}$.
    If we consider an homogeneous environment for each molecule (i.e. absence of border effects) and also assume that the environment states form a continuous and dense space, following the results in \apen{app:Def_Order_Param}, the order parameter \numeq{order_param_def} has the same value for different molecules and it can be expressed as

    \begin{equation}\label{order_param_homogenv}
    \begin{split}
    S_{zz} = \int dS_1\,S_1\,p^{\{S\}}_1\left(S_1\right),
    \end{split}
    \end{equation}
    where $p^{\{S\}}_1\left(S_1\right)$ is the distribution function of the eigenvalues $S_1$ of one molecule, which is the same distribution for each molecule in the sample, and satisfies $\int dS_1\,p^{\{S\}}_1\left(S_1\right) = 1$.
    It is worth to note that $p^{\{S\}}_1\left(S_1\right)$ is the orientational molecular distribution function (OMDF) studied in QD-I (in this reference the OMDF is written as $p\left(S_1\right)$).
    Besides, the expression \numeq{order_param_homogenv} has the form of the usual definition of the order parameter in the literature\cite{degene93}.

    To complete the description of NMR experiments in nematic LC's under a full-quantum approach with a Hamiltonian form like \eq{HamGen}, together with the dipolar Hamiltonian \numeq{sum_HamDip_CL} we have to include the Zeeman and the lattice Hamiltonians.
    The Zeeman Hamiltonian is written as
    \begin{equation}\label{HamZeeman_tot}
    \opc{H}_Z = -\omega_0\,\op{I_z} = -\omega_0\,\sum_i\op{I_{zi}},
    \end{equation}
    where $\omega_0 = \gamma B_0$ is the Larmor frequency, $\gamma$ is the proton gyromagnetic ratio and $B_0$ is the strength of the static magnetic field, which is applied along the laboratory $\vers{z}$ axis, and $\op{I_{zi}}$ is the $\vers{z}$ projection of the total proton spin angular momentum of the $i$-th molecule.\\
    The environment or lattice Hamiltonian $\opc{H}_L$ is associated with the potential energy of the mechanical interaction between the molecules of the whole sample. This Hamiltonian takes into account the molecules as an ensemble of correlated quantum objets. The details of the mechanical intermolecular interaction are not needed to extract some general conclusions about the influence of the environment on the spin dynamics.
    In fact, knowing whether $\opc{H}_L$ commutes or not with the interaction Hamiltonian $\opc{H}_{SL}$ is enough for concluding about the reversible or irreversible character of decoherence in some time scale, as was shown in QD-I.
    However, the influence of $\opc{H}_L$ is indirectly taken into account in the molecular averaged values of lattice operators, as is seen in \eq{HamDip_CL_promf}, and also in the distribution probability function of the eigenvalues of such operators, as we will show in \sect{subse:MedRevEigen_DefExp}.

    Finally, using the average dipolar Hamiltonian of the spin as a closed system
    \begin{equation}\label{HamDipSpin_tot}
    \langle\opc{H}_d\rangle_f = \sum_i \langle\opc{H}_{di}\rangle_f,
    \end{equation}
    we define the Hamiltonians in \eq{HamGen} as
    \begin{equation}\label{HamInte_S_tot}
        \opc{H}_{S} = \opc{H}_Z + \langle\opc{H}_d\rangle_f = \sum_i \opc{H}_{Si},
    \end{equation}
    with
    \begin{equation}\label{HamMol_S}
        \opc{H}_{Si} = -\omega_0\,\op{I_{zi}} + \langle\opc{H}_{di}\rangle_f,
    \end{equation}
    and
    \begin{equation}\label{HamInte_SL_tot}
        \opc{H}_{SL} = \opc{H}_d - \langle\opc{H}_d\rangle_f = \sum_i \opc{H}_{SLi},
    \end{equation}
    with
    \begin{equation}\label{HamInte_SL}
    \begin{split}
        &\opc{H}_{SLi} = \op{H}_{\op{di}}^{(s)}
        \otimes\left(\op{S}_{\op{zzi}}^{(f)}-S_{zz}\op{1}^{(f)}\right).
    \end{split}
    \end{equation}

    Below, we discuss how the environment induced decoherence emerges when the dynamics of the system is described through the Hamiltonian of \eq{HamGen}.
    Our strategy is the calculation of the reduced spin density matrix from the time-evolved density operator of the whole system, under the total Hamiltonian \numeq{HamGen}.

    It is convenient for comparison with NMR experiments to use the rotating-frame representation, namely a frame whose $z$-axis is parallel to the external magnetic field and rotates about it with an angular frequency equal to the Larmor frequency. Thus, in this frame the time evolution of the density matrix is
    \begin{equation}\label{rho_CL_t_rot}
    \widehat{\rho}(t) = \op{\widehat{U}}(t)\,\widehat{\rho}_S(0)\rho_{L(eq)}\op{\widehat{U}}^\dag(t),
    \end{equation}
    where $\widehat{\op{O}}\equiv e^{-i\omega_0\op{I_z}t}\,\op{O}\,e^{i\omega_0\op{I_z}t}$ is the representation of an operator $\op{O}$
    in the rotating-frame, and
    \begin{equation}\label{U_rot}
    \op{\widehat{U}}(t) = e^{-i\,\opc{\widehat{H}}_S\,t}\,
    e^{-i\,\left(\sum_i \opc{H}_{SLi} + \opc{H}_{L}\right)\,t},
    \end{equation}
    where $\opc{\widehat{H}}_S = \sum_i \langle\opc{H}_{di}\rangle_f$ is the transformed spin Hamiltonian.

    Here, we introduce the eigenbasis of the operators $\op{H}_{\op{di}}^{(s)}$,
    $\{\ket{\zeta s} \equiv \ket{\zeta_1 s_1}\otimes\cdots\otimes\ket{\zeta_i s_i}\otimes\cdots\otimes\ket{\zeta_N s_N}\}$,
    which span the spin Hilbert space of the $N$ molecules of the sample. The symbols $\zeta_i$'s indicate the different eigenvalues and $s_i$ label their degeneration, thus $\op{H}_{\op{di}}^{(s)}\ket{\zeta s} = \zeta_i \ket{\zeta s}$.
    Hereinafter, the symbol $\zeta$ will represent dependence on the set of eigenvalues $\{\zeta_i\}$ belonging to the molecules of the whole sample.
    Consequently, in the rotating-frame, $\{\ket{\zeta s}\}$ is an eigenbasis for both, the spin part of the interaction Hamiltonian $\opc{H}_{SL}$ (i.e. the dipolar Hamiltonian) and the spin (or observed system) Hamiltonian $\opc{\widehat{H}}_S$.
    The initial state of the spin system $\widehat{\rho}_S(0)$ is obtained, for example, after applying a radiofrequency pulse sequence.
    Therefore, the matrix elements of the reduced density operator, $\widehat{\sigma}(t)$, in the basis $\{\ket{\zeta s}\}$ are
    \begin{equation}\label{sigma_CL_t}
    \begin{split}
    &\bra{\zeta s}\widehat{\sigma}(t)\ket{\zeta' s'} = tr_f\left\{\bra{\zeta s}\widehat{\rho}(t)\ket{\zeta' s'}\right\}\\
    &\quad = e^{-i \sum_i (\zeta_i^{} - \zeta_i') S_{zz} t}\,\bra{\zeta s}\widehat{\rho}^{(s)}_S(0)\ket{\zeta' s'}\,G_{\{\zeta,\zeta'\}}(t),
    \end{split}
    \end{equation}
    where
    \begin{equation}\label{G_sigma_CL_t}
    G_{\{\zeta,\zeta'\}}(t) = tr_f\left\{\op{U}^{\dag(f)}(\zeta', t)\op{U}^{(f)}(\zeta, t)
    \rho_{L(eq)}^{(f)}\right\},
    \end{equation}
    is the decoherence function associated with the time evolution operator
    \begin{equation}\label{U_sigma_CL_t}
    \op{U}^{(f)}(\zeta, t) = e^{-i\,\left(\sum_i \zeta_i\opc{H}^{(f)}_{SLi} +
    \opc{H}^{(f)}_{L}\right)\,t}.
    \end{equation}
    It is worth to note that the evolution operator \numeq{U_sigma_CL_t} and thus the decoherence function \numeq{G_sigma_CL_t} depend on the spin eigenvalues of the molecules of the whole sample, this dependence is represented by the symbols $\zeta$ and $\zeta'$.\\
    In order to explicit the spin dynamics of a representative molecule, say the $i$-th one, we use the operator expansion technique due to Zassenhaus  to factorize the
    evolution operator \numeq{U_sigma_CL_t} into a product of exponential operators, as follows \cite{kumar65,SegZam2011} :
    \begin{equation}\label{U_sigma_CL_t_des}
    \begin{split}
    \op{U}^{(f)}(\zeta, t) &= e^{-i\opc{H}_{Ri}^{(f)}(\zeta)t}\,e^{-i\zeta_i\opc{H}^{(f)}_{SLi}t}
    \,\op{U}_C^{(f)(SLi)}(\zeta,t),
    \end{split}
    \end{equation}
    where  we defined the  operator
    \begin{equation}\label{HamRi}
    \opc{H}_{Ri}^{(f)}(\zeta) \equiv  \sum_{j\neq i} \zeta_j\opc{H}^{(f)}_{SLj}+\opc{H}^{(f)}_{L}.
    \end{equation}
    Again, the dependence of operator \numeq{HamRi} with the spin eigenvalues $\{\zeta_j\}$ of all the molecules excepting the $i$-th is taken into account by the general symbol $\zeta$.
    At the same time, the third factor in \numeq{U_sigma_CL_t_des} can be expanded as
    \begin{equation}\label{USLi_Zass}
    \op{U}_C^{(f)(SLi)}(\zeta,t) = e^{t^2\op{C}^{(f)(SLi)}_{2}(\zeta)}
    e^{t^3\op{C}^{(f)(SLi)}_{3}(\zeta)}\cdots,
    \end{equation}
    where $\op{C}^{(f)(SLi)}_{q}(\zeta)$ represents some ($q$-1)-order nested commutator between Hamiltonians
    $\opc{H}_{Ri}^{(f)}(\zeta)$ and $\zeta_i\opc{H}^{(f)}_{SLi}$.
    It is worth to anticipate that these nested commutators, which emerge from the quantum character of the environmental variables, will produce an irreversible decoherence spin dynamics in an intermediate time scale between the coherence and the thermalization processes, along which the quasi-equilibrium develops\cite{SegZam2011}.

    In keeping with the same spirit of separating the molecular spin dynamics, we  consider the cases where  the initial condition
    $\widehat{\rho}_S(0)$ can be expressed as
    \begin{equation}\label{rhoS_CL_0_aprox}
    \widehat{\rho}_S(0) = \frac{1}{\opc{N}_{S_1}^{N-1}}\sum_i \widehat{\rho}^{(s)}_{Si}(0)\otimes\op{1}^{(f)},
    \end{equation}
    where $\opc{N}_{S_1} \equiv tr_{s_1}\left\{\op{1}^{(s_1)}\right\}$
    is the trace of the identity operator in the Hilbert space
    of the spins belonging to a molecule,
    $\widehat{\rho}^{(s)}_{Si}(0) = \op{1}^{(s_1)}\otimes\cdots\otimes\widehat{\rho}^{(s_i)}(0)\otimes\cdots\otimes\op{1}^{(s_N)}$,
    where $\widehat{\rho}^{(s_i)}(0)$ is the initial spin density matrix of the $i$-th molecule, which is assumed identical for
    all molecules.
    Then, the matrix elements of \numeq{rhoS_CL_0_aprox} in the spin space are
    \begin{equation}\label{rhoS_CL_0_aprox_elem}
    \begin{split}
    &\bra{\zeta s}\widehat{\rho}^{(s)}_S(0)\ket{\zeta' s'} =
    \frac{1}{\opc{N}_{S_1}^{N-1}}\sum_i \bra{\zeta s}\rho^{(s)}_{Si}(0)\ket{\zeta' s'}\\
    &\,=\frac{1}{\opc{N}_{S_1}^{N-1}}\sum_i \bra{\zeta_i^{}s_i^{}}\widehat{\rho}^{(s_i)}(0)\ket{\zeta_i's_i'}
    \prod_{j\neq i}\delta_{\zeta_j'^{}s_j'^{}}^{\zeta_js_j},
    \end{split}
    \end{equation}
    where we defined
    \[\prod_{j\neq i}\delta_{\zeta_j'^{}s_j'^{}}^{\zeta_js_j}\equiv \delta_{\zeta_1^{}s_1^{},\zeta_1's_1'}
    \cdots\delta_{\zeta_N^{}s_N^{},\zeta_N's_N'},\]
    as the product of the $N-1$ Kr\"{o}necker deltas associated with all the molecules different from the $i$-th one. Then, by using \eqs{U_sigma_CL_t_des} and \numeq{rhoS_CL_0_aprox_elem}, \eq{sigma_CL_t} can be expressed as
    \begin{equation}\label{sigma_CL_t_summol_rot}
    \begin{split}
    \bra{\zeta s}\widehat{\sigma}(t)\ket{\zeta's'} &= \sum_i e^{-i (\zeta_i^{} - \zeta_i') S_{zz} t}\,
    \bra{\zeta_i^{}s_i^{}}\widehat{\rho}^{\;(s_i)}(0)\ket{\zeta_i's_i'}\\
    &\times\prod_{j\neq i}\delta_{\zeta_j'^{}s_j'^{}}^{\zeta_j^{}s_j^{}}\;
    G_{\{\zeta,\zeta_i'\}}(t)/\opc{N}_{S_1}^{N-1},
    \end{split}
    \end{equation}
    where $G_{\{\zeta,\zeta_i'\}}(t)$ is the decoherence function \numeq{G_sigma_CL_t} for the cases where the Kr\"{o}necker deltas present in \eq{rhoS_CL_0_aprox_elem} do not cancel.
    Because of the delta functions, \numeq{G_sigma_CL_t}
    is evaluated when $\zeta_j=\zeta_j',\,\forall j\,/j\neq i$, for
    the expression shown in \eq{U_sigma_CL_t_des}, hence
    $\opc{H}_{Ri}^{(f)}(\zeta) = \opc{H}_{Ri}^{(f)}(\zeta')$ and in
    $\op{U}_C^{(f)(SLi)}(\zeta,t)$ only $\zeta_i$ is unaffected by such delta functions (i.e. it can be $\zeta_i \neq \zeta_i'$). This reasoning leads to the following expression for the decoherence function
    \begin{equation}\label{G_sigma_CL_t_mol}
    \begin{split}
    G_{\{\zeta,\zeta_i'\}}(t) &= tr_f\bigg\{
    \op{U}_{C,\zeta_i'}^{\dag(f)(SLi)}(\zeta,t)\,e^{-i (\zeta_i^{} - \zeta_i')\opc{H}^{(f)}_{SLi}\,t}\\
    &\qquad\qquad\times\op{U}_C^{(f)(SLi)}(\zeta,t)\,\rho_{L(eq)}^{(f)}\bigg\},
    \end{split}
    \end{equation}
    where we introduced the operator $\op{U}_{C,\zeta_i'}^{(f)(SLi)}(\zeta,t)$
    which can differ from $\op{U}_C^{(f)(SLi)}(\zeta,t)$ only by the replacement of $\zeta_i$ by $\zeta_i'$.
    It is worth to remark that if $\zeta_i=\zeta'_i$ in \numeq{G_sigma_CL_t_mol}, then the decoherence function satisfies
    \[ G_{\{\zeta,\zeta_i\}}(t) = tr_f\left\{\rho_{L(eq)}^{(f)}\right\} = 1, \]
    which indicates that decoherence does not affect the subspace associated with a given degenerate eigenvalue $\zeta_i$.
    An important consequence of this is that, if in the matrix representation of the spin density operator $\widehat{\rho}^{\;(s_i)}(0)$ in  the eigenbasis of $\opc{H}^{(s_i)}_{SL}$ the eigenstates are ordered, in such way that blocks associated with a given eigenvalue $\zeta_i$ are formed in the diagonal, these blocks remain invariant under decoherence.
    This property, which was already reflected in \eq{G_sigma_CL_t} for $\zeta = \zeta'$,
    is of great importance in the dynamics that brings the spin system into a quasi-equilibrium state.
    In particular, this means that the quasi-invariant spin operators have to commute with the spin part of the interaction Hamiltonian\cite{SegZam2011} (i.e. the dipolar Hamiltonian in our case), but they need not commute with the Zeeman Hamiltonian. Therefore, several quasi-invariants that do not present evolution in the intermediate time scale could present a time dependence under the Zeeman evolution operator and they will not be quasi-invariants out of the rotating-frame anymore. Accordingly, the matrix form of the spin density operator \numeq{sigma_CL_t_summol_rot} (or \numeq{sigma_CL_t}) in the eigenbasis $\{\ket{\zeta s}\}$ does not depend on time when the quasi-equilibrium state is reached. It is worth to note that in the common eigenbasis $\{\ket{E s}\}$ that diagonalize both the dipolar and the Zeeman Hamiltonians, which was presented in QD-I, the matrix form of such quasi-invariants could present time-dependent complex exponential factors with frequencies given by the Zeeman eigenvalues.

    Finally, a zero-trace spin observable acting on the spin space of individual molecules, and which is time independent in the rotating-frame, has the general form:
    \begin{equation}\label{Obs_trn}
        \op{\widehat{O}} = \sum_i \op{\widehat{O}}^{(s)}_i\otimes\op{1}^{(f)},
    \end{equation}
    with
    $\op{\widehat{O}}^{(s)}_i = \op{1}^{(s_1)}\otimes\cdots\otimes\op{\widehat{O}}^{(s_i)}\otimes\cdots\otimes\op{1}^{(s_N)}$.
    By using \eq{sigma_CL_t_summol_rot}, the expectation value of $\op{\widehat{O}}$ can be written as
    \begin{equation}\label{Val_Obs_trn_fin_rot}
    \begin{split}
        &\left<\widehat{\op{O}}\right>(t) = \sum_i \sum_{\zeta_i^{}s_i^{},\zeta_i's_i'}
        \bra{\zeta_i^{}s_i^{}}\widehat{\rho}^{\;(s_i)}(0)\ket{\zeta_i's_i'}\\
        &\qquad\times\bra{\zeta_i's_i'}\op{\widehat{O}}^{(s_i)}\ket{\zeta_i^{}s_i^{}}
         e^{-i (\zeta_i^{} - \zeta_i') S_{zz} t}\,G_{\{\zeta_i^{},\zeta_i'\}}(t),
    \end{split}
    \end{equation}
    with
    \begin{equation}\label{G_sigma_CL_t_summol}
        G_{\{\zeta_i^{},\zeta_i'\}}(t) \equiv \sum_{\zeta_1^{}s_1^{},\dots,\zeta_N^{}s_N^{}}^{\zeta_k^{}\neq \zeta_i^{},s_k^{}\neq s_i^{}} G_{\{\zeta,\zeta_i'\}}(t)/\opc{N}_{S_1}^{N-1},
    \end{equation}
    where the sum in \eq{G_sigma_CL_t_summol} runs over all the values of $\zeta_k^{}$ and $s_k^{}$ with $k\neq i$, thus this sum only conserves the dependence with the spin eigenvalues $\zeta_i^{}$ and $\zeta_i'$ of the $i$-th molecule.
    It is worth to note that the decoherence function \numeq{G_sigma_CL_t_summol} characterizes the decoherence process of the $i$-th molecule but it preserves correlations with the remaining molecules of the sample through the functions $G_{\{\zeta,\zeta_i'\}}(t)$; such correlations are produced by the environment operators in the Hamiltonians through the nested commutators in the evolution operator \numeq{USLi_Zass}, thus they have a quantum character.

    The irreversibility of the spin dynamics is introduced through the environment induced quantum decoherence associated with an open quantum system characterized as an {\it essentially adiabatic system}. According to the theory, the eigen-selection process, prominently involved in the decoherence dynamics, leaves invariant the block diagonal space of the density operator drawing it to a quasi-equilibrium representation.
    As will be shown below, the experimental measurements confirm the eigen-selection effects during the coherence evolution under a process of partial reversion of the spin dynamics. Besides, two different time scales associated to different sources of decoherence will be distinguished, one of them during free evolution and the other under refocusing of the coherences.
    The experiments designed with the propose of observing such effects are shown in \sect{subse:MedRevEigen_DefExp}, which are based on the FID single quantum coherence signal. However, the procedure and the results can be extended to any kind of coherence.

\subsection{Definition of the experiments}\label{subse:MedRevEigen_DefExp}

    In order to highlight the contribution to the coherence signal decay coming from sources other than the Liouvillian  dynamics corresponding to an isolated spin system (i.e. quantum interference), we will measure the single coherence signal evolution of the FID, when the spins are subjected to a rf pulse sequence configured to compensate the Liouvillian evolution.
    An experiment of reversion of the spin dynamics under the high field secular dipolar Hamiltonian could be performed, for instance, by means of the single sequence shown in \fig{FIG1} (a) or the sequence known as MREV8\cite{Mans71,REV73a,REV73b,Abrag_Gold_cap2}, which is shown in  \fig{FIG1} (b).
    An explanation of the effect on the spin dynamics due to such kind of reversion sequences can be found in the supplementary material (see section II in supplementary material\cite{suppmatDQII} for details on the reversion spin dynamics).
    These kind of sequences were selected because the efficiency of the reversion process relies on the relationship between the total time of the FID evolution and the setting time between pulses.
    In these reversion pulse sequences, the smaller the setting times $\tau_1$ and $\tau_2$ are, the better the reversion will be. Therefore, to access to long reversion periods while keeping the time parameters as small as possible, we apply a pulse sequence consisting of a chain of blocks of the same reversion unit, as is shown in \fig{FIG1} (c).
    On the other hand, in other techniques like the  `magic-sandwich' (MS) \cite{RhimPV70,RhimPV71,RhimK71,RhimK72,nielsen97} the effectiveness of the reversion depends on the use of high intensity rf pulses with increasing duration, which could become inadequate for cases of long time evolutions, as is the case of the nematics PAA$_{d6}$ and PAA studied in this work.

    Using the results obtained throughout section III.D.2 of QD-I, we can calculate the evolution operator that represents the spin dynamics due to the pulse sequence of \fig{FIG1} (c) in the rotating-frame representation, that is
        \begin{equation}\label{U_rot_evol_RT_evol}
        \op{\widehat{U}}(t, t_2, t_1) = \op{\widehat{U}}(t)\op{\widehat{U}_{rt}}(t_2)\op{\widehat{U}}(t_1),
        \end{equation}
    where $\op{\widehat{U}}$ is the free evolution operator \numeq{U_rot} and $\op{\widehat{U}_{rt}}$ is the reversion dynamics evolution operator, which is defined as
        \begin{equation}\label{U_rot_RT}
        \op{\widehat{U}_{rt}}(t) = e^{i\,\opc{\widehat{H}}_S\,t/\kappa}\,
        e^{i\,\left(\sum_i \opc{H}_{SLi}/\kappa\, - \opc{H}_{L}\right)\,t},
        \end{equation}
    with $\kappa$ as a positive constant whose value depends on the particular refocusing technique used.
    For the pulse sequences (a) and (b) of \fig{FIG1}, $\kappa =2$.
    In the case of a single-block reversion sequence, we have $t_1 \equiv \tau_1$ (or $t_1 \equiv 4\tau_1$) and $t_2 \equiv \tau_2$ (or $t_2 \equiv 4\tau_2$), with $\tau_1$ and $\tau_2$ defined in \fig{FIG1} (a) (or (b)). On the other hand, in the case of the chaining block sequence shown in \fig{FIG1} (c), we have $t_1 \equiv n\tau_1$ (or $t_1 \equiv 4n\tau_1$) and $t_2 \equiv n\tau_2$ (or $t_2 \equiv 4n\tau_2$), being $n$ the number of blocks of the pulse sequence of \fig{FIG1} (a) (or (b)).
    \eq{U_rot_RT} represents the dynamics under an MREV8-like sequence when the experimental setting has  $\tau_1$ and $\tau_2$ small enough to neglect the non-secular terms of the dipolar Hamiltonian which arises from applying the $\pi/2$ pulse pairs \cite{Abrag_Gold_cap2} (see also section II in supplementary material\cite{suppmatDQII} for details about the emergence of the operator factor $e^{i\,\opc{\widehat{H}}_S\,t/\kappa}$).
    This constraint is imposed over each block in the sequence of \fig{FIG1} (c) but the total times $n\tau_1$ (or $4n\tau_1$) and $n\tau_2$ (or $4n\tau_2$) do not need to be so small.

    By applying the evolution operator \numeq{U_rot_evol_RT_evol} to the state $\ket{\zeta s}$, it is obtained

        \begin{equation}\label{U_rot_evol_RT_evol_ket}
        \begin{split}
        \op{\widehat{U}}(t, t_2, t_1)\ket{\zeta s} &=
        e^{-i \sum_i\zeta_i^{} S_{zz} \left(t + t_1 - t_2/\kappa\right)}\ket{\zeta s}\\&\quad\otimes
        \op{U}^{(f)}(\zeta, t, t_2, t_1),
        \end{split}
        \end{equation}
    where we defined
        \begin{equation}\label{U_evol_RT_evol_f}
        \op{U}^{(f)}(\zeta, t, t_2, t_1) \equiv \op{U}^{(f)}(\zeta, t) \op{U}^{(f)}(\zeta, t_2, t_1).
        \end{equation}
    Truncating the Zassenhaus expansion of \eq{U_evol_RT_evol_f} to the first order gives

       \begin{equation}\label{U_evol_RT_evol_f_C}
        \begin{split}
        &\op{U}^{(f)}(\zeta, t, t_2, t_1) \cong e^{-i\,\opc{H}_{Ri}^{(f)}(\zeta)\,t} e^{-i\,\opc{H}_{Ri}^{(f)}(-\zeta/\kappa)\,t_2}\\
        &\quad\times e^{-i\,\opc{H}_{Ri}^{(f)}(\zeta)\,t_1}\,
        e^{-i\,\zeta_i\opc{H}^{(f)}_{SLi}\,\left(t + t_1 - t_2/\kappa\right)}\\
        &\quad\times e^{\zeta_i\op{C}^{(f)}_{i,SL}(\zeta)\,\left(t^2 - 2 t t_2/\kappa + 2 t t_1 + t_1^2 + t_2^2/\kappa^2 - 2 t_1 t_2/\kappa\right)/2}\\
        &\quad\times e^{\zeta_i\op{C}^{(f)}_{i,L}\,\left(t^2 + 2 t t_2 + 2 t t_1 + t_1^2 - t_2^2/\kappa - 2 t_1 t_2/\kappa\right)/2},
        \end{split}
       \end{equation}
    where the operators
        \begin{eqnarray}
        \op{C}^{(f)}_{i,SL}(\zeta) &\equiv& [\sum_{j\neq i}\zeta_j\opc{H}^{(f)}_{SLj},\opc{H}^{(f)}_{SLi}]\label{Ci2_Zass_SL}\\
        \op{C}^{(f)}_{i,L} &\equiv& [\opc{H}^{(f)}_{L},\opc{H}^{(f)}_{SLi}],\label{Ci2_Zass_L}
        \end{eqnarray}
    are anti-Hermitian and thus they have pure imaginary eigenvalues.
    The expression in \eq{U_evol_RT_evol_f_C} is valid for the intermediate time scale, under the postulate of the existence of different time scales described in \sect{se:Theory_Back}. Similar to \eq{HamRi}, the symbol $\zeta$ in the commutator \numeq{Ci2_Zass_SL} represents dependence with the spin eigenvalues of all the molecules except for the $i$-th one.

    Now, using \eq{U_evol_RT_evol_f_C} we can obtain, with the same spirit as in QD-I, the decoherence function for the spin dynamics valid for a first time scale longer than the Liouville characteristic time scale, that is

        \begin{equation}\label{Gi_evol_RT_evol_C}
        \begin{split}
        &G_{\{\zeta,\zeta_i'\}}(t, t_2, t_1) = tr_f\bigg\{\op{U}^{\dag(f)}_{\zeta_i'}(\zeta, t, t_2, t_1)\\
        &\qquad\qquad\qquad\qquad\qquad\times \op{U}^{(f)}(\zeta, t, t_2, t_1)\rho_{L(eq)}^{(f)}\bigg\}\\
        &\; = tr_f\bigg\{e^{-i\,(\zeta_i^{}-\zeta_i')\opc{H}^{(f)}_{SLi}\,\left(t + t_1 - t_2/\kappa\right)}\\
        &\;\times e^{-i\,(\zeta_i^{}-\zeta_i')\,i\op{C}^{(f)}_{i,SL}(\zeta)\,
        \left(t^2 - 2 t t_2/\kappa + 2 t t_1 + t_1^2 + t_2^2/\kappa^2 - 2 t_1 t_2/\kappa\right)/2}\\
        &\;\times e^{-i\,(\zeta_i^{}-\zeta_i')\,i\op{C}^{(f)}_{i,L}\,\left(t^2 + 2 t t_2 + 2 t t_1 + t_1^2 - t_2^2/\kappa - 2 t_1 t_2/\kappa\right)/2}\,
        \rho_{L(eq)}^{(f)}\bigg\},
        \end{split}
        \end{equation}
    where we introduced the operator $\op{U}^{(f)}_{\zeta_i'}(\zeta, t, t_2, t_1)$
    which can differ from $\op{U}^{(f)}(\zeta, t, t_2, t_1)$ only by the replacement of $\zeta_i$ by $\zeta_i'$.
    The operators $i\op{C}^{(f)}_{i,SL}$ and $i\op{C}^{(f)}_{i,L}$ appearing in \eq{Gi_evol_RT_evol_C},  are Hermitian  due to the anti-Hermitian character of the commutators, as we have seen in \numeq{Ci2_Zass_SL} and \numeq{Ci2_Zass_L}, thus their eigenvalues are pure real numbers.
    By making $t_2 = \kappa t_1$ (i.e. $\tau_2 = \kappa \tau_1$), we eliminate the Liouville dynamics during the reversion and \eq{Gi_evol_RT_evol_C} becomes in this case

        \begin{equation}\label{Gi_evol_RT_evol_C_eval}
        \begin{split}
        &G_{\{\zeta,\zeta_i'\}}(t, t_1) = tr_f\bigg\{e^{-i\,(\zeta_i^{}-\zeta_i')\opc{H}^{(f)}_{SLi}\,t}\\
        &\quad\times e^{-i\,(\zeta_i^{}-\zeta_i')\,i\op{C}^{(f)}_{i,L}\,\left\{\left[t + (\kappa+1)\,t_1\right]^2 - \left(\kappa^2 + 3\kappa + 2\right) \, t^2_1 \right\}/2}\\
        &\quad\times e^{-i\,(\zeta_i^{}-\zeta_i')\,i\op{C}^{(f)}_{i,SL}(\zeta)\,t^2/2}\,\rho_{L(eq)}^{(f)}\bigg\},
        \end{split}
        \end{equation}
    with the definition $G_{\{\zeta,\zeta_i'\}}(t, t_1) \equiv G_{\{\zeta,\zeta_i'\}}(t, \kappa t_1, t_1)$, which represents that the above condition gives a decoherence function which depends just on the reversion time $t_1$.

    To extract a final expression for the decoherence function, we use that the environment states form a continuous and dense space, so that we can replace in \eq{Gi_evol_RT_evol_C_eval} the sum in the trace by an integral over the lattice space.
    This procedure is detailed in \apen{app:Deco_Fun_CDSpace}, obtaining

        \begin{equation}\label{Gi_evol_RT_evol_C_eval_aprox_int_eig}
        \begin{split}
        &G_{\{\zeta,\zeta_i'\}}(t, \tau) = \int d\Delta S_i \; e^{-i\,(\zeta_i^{}-\zeta_i')\Delta S_i\,t}\\
        &\quad\times \int dC^L_i \; e^{-i\,(\zeta_i^{}-\zeta_i')C^L_i\,\left\{\left(t + \tau\right)^2 - \left[1+(\kappa+1)^{-1}\right] \, \tau^2 \right\}/2}\\
        &\quad\times \int dC^{SL}_{i\,\zeta} \; e^{-i\,(\zeta_i^{}-\zeta_i')C^{SL}_{i\,\zeta}\,t^2/2} \; p_i(\Delta S_i, C^L_i, C^{SL}_{i\,\zeta}),
        \end{split}
        \end{equation}
    where $\tau = (\kappa+1)\,t_1$ is the total time under reversion, $\Delta S_i$, $C^L_i$ and $C^{SL}_{i\,\zeta}$ are respectively the eigenvalues of the operators $\opc{H}^{(f)}_{SLi}$, $i\op{C}^{(f)}_{i,L}$ and $i\op{C}^{(f)}_{i,SL}(\zeta)$, which are real numbers.
    The index $\zeta$ in $C^{SL}_{i\,\zeta}$ stands for dependence with the spin eigenvalues in the same sense as in \eq{Ci2_Zass_SL}.
    Accordingly, $d\Delta S_i$, $dC^L_i$ and $dC^{SL}_{i\,\zeta}$ are the differentials of those eigenvalues.
    In \eq{Gi_evol_RT_evol_C_eval_aprox_int_eig}, the function $p_i(\Delta S_i, C^L_i, C^{SL}_{i\,\zeta})$ satisfies
    \[\int d\Delta S_i \int dC^L_i \int dC^{SL}_{i\,\zeta} \; p_i(\Delta S_i, C^L_i, C^{SL}_{i\,\zeta}) = 1,\]
    thus it can be interpreted as a probability distribution function of the eigenvalues given by $\Delta S_i$, $C^L_i$ and $C^{SL}_{i\,\zeta}$.

    We can see from \eq{Gi_evol_RT_evol_C_eval_aprox_int_eig} that the decoherence function is the result of a superposition of complex exponential functions weighted by a distribution of their frequencies.
    We are concerned with studying the decoherence function produced by the coupling of the spin system with an environment whose states belong to a continuous and dense Hilbert space, where a distribution of the eigenvalues of each complex exponential in \eq{Gi_evol_RT_evol_C_eval_aprox_int_eig} can be defined. In that physic system, such distribution is supposed to have a general bell-shape form around the mean value of the eigenvalues, converging to zero fast enough so that integrations in \eq{Gi_evol_RT_evol_C_eval_aprox_int_eig} can be extended to $\pm\infty$. Presumably, these conditions on the eigenvalue distribution functions are similar to those imposed in the work of reference \cite{yukalov11} to derive the equilibration of `quasi-isolated quantum systems' in the strict sense.
    Due to the general bell-shape form assumed for the distribution functions $p_i(\Delta S_i, C^L_i, C^{SL}_{i\,\zeta})$ characterizing the environment of our physical system, the superposition of complex exponentials functions which constitutes \numeq{Gi_evol_RT_evol_C_eval_aprox_int_eig} will have a form of a decay function in the time $t$ and/or $\tau$.
    It is easy to see that the bigger the value of $\Delta \zeta_i$ is, the faster the decay of such function will be. In the case of $\Delta \zeta_i = 0$ (i.e. $\zeta_i^{} = \zeta_i'$) we have $G_{\{\zeta,\zeta_i'\}}(t, \tau) = 1$ and we do not have a decay. All these characteristics are described as eigen-selectivity or eigen-selection effect over the dynamics of the observed system, as we have seen in QD-I and at the end of \sect{se:Theory_Back} (see \eq{G_sigma_CL_t_mol}).

    This feature of the decoherence function, produced by the eigen-selectivity, is still obtained from a non-truncated time evolution operator, having a more complex form than  \numeq{U_evol_RT_evol_f_C}.
    In such case, the form of the decoherence function will be like \numeq{Gi_evol_RT_evol_C_eval_aprox_int_eig} but with a more extensive development of integrals due to a bigger amount (maybe infinite) of eingenvalues coming from nested commutators of increasing order, and with a more complex distribution function $p_i$. Therefore, a complete form (i.e. without truncating) of the decoherence function \numeq{Gi_evol_RT_evol_C_eval_aprox_int_eig} can be expressed as

        \begin{equation}\label{Gi_evol_RT_evol_C_eig_complete}
        \begin{split}
        &G_{\{\zeta,\zeta_i'\}}(t, \tau) = \iiint \prod_{D,n}\left\{dC^{D,n}_{i\,\zeta}\;e^{-i\,\Delta \zeta_i\,\opc{Y}_{D,n}(t,\tau)\,C^{D,n}_{i\,\zeta}}\right\}\\
        &\quad\times \int d\Delta S_i \; e^{-i\,\Delta \zeta_i\Delta S_i\,t}\;p_i\left(\Delta S_i, \left\{C^{D,n}_{i\,\zeta}\right\}\right),
        \end{split}
        \end{equation}
    with
    \[\iiint \prod_{D,n}\left\{dC^{D,n}_{i\,\zeta}\right\} \int d\Delta S_i \; p_i\left(\Delta S_i, \left\{C^{D,n}_{i\,\zeta}\right\}\right)= 1,\]
    where  $\iiint \prod_{D,n}\left\{\cdot\right\}$ represents multiple integrals over the eigenvalues of different class of nested commutators labeled by $D$ and with a nesting order $n$. Such eigenvalues are generically termed $C^{D,n}_{i\,\zeta}$. The polynomials $\opc{Y}_{D,n}(t,\tau)$ are functions of $t$ and/or $\tau$ which multiply the corresponding $C^{D,n}_{i\,\zeta}$ eingenvalue in the complex exponential functions. The symbol $\left\{C^{D,n}_{i\,\zeta}\right\}$ represents the set of such eingenvalues.
    It is worth to note that the general expression \numeq{Gi_evol_RT_evol_C_eig_complete} is valid for any kind of experimental setting, the only difference between experiments can be the forms of the polynomials $\opc{Y}_{D,n}$ with a possibly different time dependence.

    In the following, in order to extract a more handleable version for the decoherence function, which in turn is useful for taking account of some features of the experimental results, we will introduce the hypothesis of  the existence of different time scales in the dynamics, as  was assumed previously in \sect{se:Theory_Back}.
    Under such hypothesis, the dynamics produced by the exponential operator with $\opc{H}^{(f)}_{SLi}$, which was called \emph{reversible adiabatic quantum decoherence} in QD-I, is faster than the dynamics produced by the remaining exponential operators containing nested commutators, which was called \emph{irreversible adiabatic quantum decoherence} in QD-I. Hence, the decay time produced by the complex exponentials with $\Delta S_i$ is smaller than the ones associated to the remainder exponentials with eigenvalues $C^{D,n}_{i\,\zeta}$.
    The experimental evidences\cite{gonza11,SegZam2011} show that these two time scales are very well separated. Therefore, in \eq{Gi_evol_RT_evol_C_eig_complete}, for a time $t$ over which the decay function produced by the complex exponential with $\Delta S_i$ is close to vanish, the values of the exponentials with eigenvalues $C^{D,n}_{i\,\zeta}$ do not significantly deviate from their values for $t=0$, thus the decoherence function can be very well approximated replacing the multiple integrals over such eigenvalues by their values in $t=0$, namely, it can be approximated using
    $\iiint \prod_{D,n}\left\{\cdot\right\}_{t=0}$.\\
    Finally, applying this approximation to the particular case of \eq{Gi_evol_RT_evol_C_eval_aprox_int_eig} gives

        \begin{equation}\label{Gi_evol_RT_evol_C_eval_aprox}
        \begin{split}
        &G_{\{\zeta,\zeta_i'\}}(t, \tau) \simeq \int d\Delta S_i \; e^{-i\,(\zeta_i^{}-\zeta_i')\Delta S_i\,t}\\
        &\quad\times \int dC^L_i \; e^{i\,(\zeta_i^{}-\zeta_i')C^L_i\,\tau^2/[2(\kappa+1)]} \; p_i(\Delta S_i, C^L_i),
        \end{split}
        \end{equation}
    where it is defined the function
    \[p_i(\Delta S_i, C^L_i) \equiv \int dC^{SL}_{i\,\zeta} \; p_i(\Delta S_i, C^L_i, C^{SL}_{i\,\zeta}).\]
    From \eq{Gi_evol_RT_evol_C_eval_aprox}, we see that under such approximation the dependence of the decoherence function on $C^{SL}_{i\,\zeta}$ is eliminated in \eq{Gi_evol_RT_evol_C_eval_aprox_int_eig}, thus the decoherence function only depends on the difference of the values
    $\zeta_i - \zeta_i'$ (i.e. the differences of the eigenvalues of the interaction Hamiltonian spin part belonging to the $i$-th molecule).
    Therefore, under the mentioned approximation, we will have from \eq{G_sigma_CL_t_summol} that
    \[G_{\{\zeta_i^{},\zeta_i'\}}(t, \tau) \equiv G_{\{\zeta,\zeta_i'\}}(t, \tau),\]
    where in the decoherence function $G_{\{\zeta,\zeta_i'\}}$ the dependence of the spin eigenvalues of the molecules other than the $i$-th (taken into account with the symbol $\zeta$) is removed.
    However, notice that the environmental quantum correlations between the molecules of the sample is preserved.

    Two additional hypotheses of general character will be assumed on the decoherence function, as follows:
    \newcounter{chypoG}

    \begin{list}{\textbf{HypoG--\Roman{chypoG}:}}{\usecounter{chypoG}}
        \item \label{hypoG_SI}\textbf{Statistical independence between the eigenvalues $\Delta S_i$ and $C^L_i$, $\forall i$.}
        This hypothesis is reasonable due to the different nature of the spectral properties of their operators\cite{SegZam2011}. This is written as
        $p_i(\Delta S_i, C^L_i) \cong p^{\{\Delta S\}}_i\left(\Delta S_i\right)\,p^{\{C^L\}}_i\left(C^L_i\right)$ and it brings us the possibility of writing
        \eq{Gi_evol_RT_evol_C_eval_aprox} as
        \[G_{\{\zeta_i^{},\zeta_i'\}}(t, \tau) \equiv G_{\{\zeta_i^{},\zeta_i'\}}(t)\,G_{\{\zeta_i^{},\zeta_i'\}}^{\,(rt)}(\tau),\]
        with
        \begin{equation}\label{AQD_func_DeltaS}
        G_{\{\zeta_i^{},\zeta_i'\}}(t) \equiv \int d\Delta S_i \;e^{-i\,(\zeta_i^{}-\zeta_i')\,\Delta S_i\,t}\; p^{\{\Delta S\}}_i\left(\Delta S_i\right),
        \end{equation}
        and
        \begin{equation}\label{irrAQD_func_CL}
        G_{\{\zeta_i^{},\zeta_i'\}}^{\,(rt)}(\tau) \equiv \int dC^L_i \;e^{i\,(\zeta_i^{}-\zeta_i') \, C^L_i \,\tau^2/[2(\kappa+1)]}\; p^{\{C^L\}}_i\left(C^L_i\right).
        \end{equation}
        \item \label{hypoG_HE}\textbf{Homogeneous environment for each molecule, i.e. absence of border effects.}
        This is written as $p_i(\Delta S_i, C^L_i) \equiv p_j(\Delta S_j, C^L_j)$, $\forall i,j$, and it brings us the relationship
        \[G_{\{\zeta_i^{},\zeta_i'\}}(t, \tau) \equiv G_{\{\zeta_j^{},\zeta_j'\}}(t, \tau),\; \forall i,j.\]
    \end{list}

    At this point,  some comments about the MREV8 pulse sequence used in the experiments are pertinent.
    This sequence is composed of two sequences WHH-4 \cite{WHH68}, which correspond to each half in \fig{FIG1} (b).
    Due to the symmetric disposition of the pulses in the MREV8 sequence, the expression of the evolution operator \numeq{U_rot_evol_RT_evol} in this case actually has the form:
    \[\op{\widehat{U}}(t, t_2, t_1) = \op{\widehat{U}}(t)\op{\widehat{U}}(t_1/2)\op{\widehat{U}_{rt}}(t_2)\op{\widehat{U}}(t_1/2),\]
    which is symmetric in the reversion times (i.e. $t_1$ and $t_2$).
    Accordingly, it can be seen that the dynamics under reversion does not depend on the first-order nested commutator $\op{C}^{(f)(SLi)}_{2}(\zeta)$ in \eq{USLi_Zass}, which means that the dynamics under reversion produced by $\op{C}^{(f)}_{i,L}$ in the decoherence function \numeq{Gi_evol_RT_evol_C_eval} is completely reverted when $t_2 = \kappa t_1$.
    Therefore, in such reversion experiment, the decoherence function with the evolution operator truncated up to the first-order commutator will be equal to the decoherence function \numeq{Gi_evol_RT_evol_C_eval} putting $t_1 = 0$, and it will be equal to the function \numeq{Gi_evol_RT_evol_C_eval_aprox_int_eig} putting $\tau = 0$, so they will be independent of the reversion time.
    On the other hand, such independence of the dynamics with $\op{C}^{(f)(SLi)}_{2}(\zeta)$ does not occur for the asymmetric single reversion sequence of \fig{FIG1} (a).
    Obtaining the decoherence function under MREV8 using a truncated expression for \eq{USLi_Zass} up to the second-order conmutator $\op{C}^{(f)(SLi)}_{3}(\zeta)$ would involve a tough calculation.
    However,  the approximations and conclusions extracted from the general decoherence function form \numeq{Gi_evol_RT_evol_C_eig_complete} will still be valid and the result obtained in \eq{Gi_evol_RT_evol_C_eval_aprox} will differ in the complex exponential function involving $C^L_i$.
    This function will have a dependence with other additional eigenvalues $C^{D,n}_{i\,\zeta}$ and their corresponding polynomials $\opc{Y}_{D,n}(t,\tau)$.
    Since that aim is out of the scope of our work, we will conserve the decoherence function \numeq{Gi_evol_RT_evol_C_eval_aprox} for our succeeding analysis. Given that hypotheses \textbf{HypoG-I} and \textbf{HypoG-II} will be valid in the general case of \eq{Gi_evol_RT_evol_C_eig_complete} and the final form of decoherence depends on a proposed distribution function of some eigenvalues (as we will see the in the following text), the conclusion will be unchanged if we use the asymmetric single reversion sequence of pulses shown in \fig{FIG1} (a) and the evolution operator in \eq{U_rot_evol_RT_evol}.

    In addition, it is worth to mention that the approximation of small $\tau_1$ can be improved by adjusting the time between pulses and the pulse widths as small as possible.
    As was mentioned, the factor $\kappa$ in \eq{U_rot_RT} is equal two, thus ideally the sequence MREV8 of \fig{FIG1} (b) will revert the spin dynamics when the condition $\tau_2 = 2\tau_1$ is satisfied.
    The reversion of the FID with MREV8 is carried out with the pulse sequence in \fig{FIG1} (c). There, each block $RS(i)$ is a sequence like that presented in \fig{FIG1} (b). This configuration allows to set the time between pulses as small as allowed by the experimental apparatus, in order to minimize the effects of non-idealities of the sequence.
    In this case, the time period of evolution under reversion is $\tau = n\tau_c$, where $n$ is the number of MREV8-blocks, and the FID is acquired during  time $t$.

    The observed FID corresponds to the expectation value of the observable $\op{I}_{\op{y}}$, ensuing the reversion sequence. From the theory presented in QD-I, which was summarized in \sect{se:Theory_Back}, using \eq{Val_Obs_trn_fin_rot} with
    $\widehat{\rho}^{\;(s_1)}(0) = \frac{\beta_T\,\omega_0}{\opc{N}_{S_1}}\,\op{I}_{\op{y}}^{(s_1)}$
    and $\op{\widehat{O}}^{(s_1)} = \op{I}_{\op{y}}^{(s_1)}$,
    and the decoherence function \numeq{Gi_evol_RT_evol_C_eval_aprox} under \textbf{HypoG-I} and \textbf{HypoG-II}, in the `on-resonance' condition, we finally obtain:

        \begin{equation}\label{Val_Obs_MREV_FID}
        \begin{split}
        \left<\widehat{\op{I}}_{\op{y}}(t, \tau)\right> &=  \frac{\beta_T\,\omega_0\,N}{\opc{N}_{S_1}} \sum_{\zeta_1^{}s_1^{},\zeta_1's_1'}
        \abs{\bra{\zeta_1^{}s_1^{}}\op{I}_{\op{y}}^{(s_1)}\ket{\zeta_1's_1'}}^2\\
        &\qquad\times e^{-i\,(\zeta_1^{}-\zeta_1')\, S_{zz}\,t}\,
        G_{\zeta_1^{},\zeta_1'}(t) \,G_{\zeta_1^{},\zeta_1'}^{\,(rt)}(\tau),
        \end{split}
        \end{equation}
    where the values of $\tau$ are multiples of $\tau_c$.\\
    The result of \eq{Val_Obs_MREV_FID} is similar to the FID obtained in Section III.E in QD-I, with the addition that  the signal  is attenuated by a decoherent factor given by $G_{\zeta_1^{},\zeta_1'}^{\,(rt)}(\tau)$, which depends on eigenvalues  $\zeta_1^{}$ and $\zeta_1'$ of the dipolar Hamiltonian. Ultimately, the agreement of the experimental results with the predictions formulated using \eq{Val_Obs_MREV_FID} will confirm the validity of the introduced hypotheses.

    In \eq{Val_Obs_MREV_FID}, the function $G_{\zeta_1^{},\zeta_1'}(t)$ represent the main decoherence process under a free evolution of the system. This decoherence was called adiabatic quantum decoherence (AQD) in QD-I, with a time scale shorter than the rest of the decoherence processes and a reversible character of its dynamics.
    As was analyzed in QD-I, in nematic liquid crystals the AQD is associated with the OMDF. That distribution is expressed as a distribution of the values of the order parameter $S_i$.
    The distribution function associated here with such decoherence is $p^{\{\Delta S\}}_i\left(\Delta S_i\right)$, being $\Delta S_i = S_i - S_{zz}$ and $S_{zz} = \int dS_i\,S_i\,p^{\{S\}}_i\left(S_i\right)$.
    Using the random variable transformation (RVT) theorem\cite{Gillespie83}, we have that
    $p^{\{S\}}_i\left(S_i\right) \equiv p^{\{\Delta S\}}_i\left(S_i - S_{zz}\right)$ (i.e. the distribution function $p^{\{S\}}_i$ is the same that the one given by $p^{\{\Delta S\}}_i$ but with its statistical variable shifted in the mean value $S_{zz}$),
    thus the AQD function in \numeq{Val_Obs_MREV_FID} can be expressed as

        \begin{equation}\label{AQD_func_S}
        \begin{split}
        &G_{\{\zeta_1^{},\zeta_1'\}}(t) = \int^{\infty}_{-\infty} d\Delta S_1 \; e^{-i\,(\zeta_1^{}-\zeta_1')\,\Delta S_1\,t} \; p^{\{\Delta S\}}_1\left(\Delta S_1\right)\\
        &\quad = e^{i\,(\zeta_1^{}-\zeta_1')\,S_{zz}\,t}\,\int^{\infty}_{-\infty} dS_1 \; e^{-i\,(\zeta_1^{}-\zeta_1')\,S_1\,t}\; p^{\{S\}}_1\left(S_1\right).
        \end{split}
        \end{equation}
    From \eq{AQD_func_S}, we can see that $G_{\{\zeta_1^{},\zeta_1'\}}(t)$ is the same as the one obtained in Section III.C in QD-I, which exposes the relation between the AQD function and the OMDF given by $p^{\{S\}}_1\left(S_1\right)$.

    In QD-I, we have seen that a gaussian distribution is a suitable approximation for the OMDF in nematics, namely
       \begin{equation}\label{probDeltaS_Gauss}
        p^{\{\Delta S\}}_1\left(\Delta S_1\right) = \frac{1}{\sqrt{2\pi\,\sigma_{S_1}^2}}\;
        e^{-\frac{(\Delta S_1)^2}{2\,\sigma_{S_1}^2}},
        \end{equation}
    with $\sigma_{S_1}$ the standard deviation of $S_1$ which is the same for $\Delta S_1$.

    Finally, we can see that the distribution \numeq{probDeltaS_Gauss} yields a gaussian form for the AQD function \numeq{AQD_func_S} as
        \begin{equation}\label{G_E1E1'_DCA_gauss}
        G_{\{\zeta_1^{},\zeta_1'\}}(t) = e^{-\frac{1}{2}\,(\zeta_1^{} - \zeta_1')^2\, \sigma_{S_1}^2\, t^2}.
        \end{equation}

    By the other hand, in \eq{Val_Obs_MREV_FID} the function $G_{\zeta_1^{},\zeta_1'}^{\,(rt)}(\tau)$ represents the decoherence function produced by the first-order nested commutator belonging to a set of nested commutators which conform the complete evolution operator. The dynamics produced just by this decoherence function could be eventually reverted, by designing a particular pulse setting, but it is impossible to revert simultaneously the dynamics of the whole set of commutators, as was demonstrated in Section III.D.2 of QD-I.
    Therefore, such decoherence function produced by the complete set of nested commutator is irreversible; it was called
    \emph{essentially adiabatic quantum decoherence} in QD-I and it has a slower dynamics than the reversible AQD, introducing an intermediate time scale between the AQD or Liouvillian process and the process of thermalization.

    We can obtain different decay functions for $G_{\zeta_1^{},\zeta_1'}^{\,(rt)}(\tau)$ depending on the form of $p^{\{C^L\}}_1\left(C^L_1\right)$.
    For instance, for a gaussian form similar to \numeq{probDeltaS_Gauss}, replacing $\Delta S_1$ and $\sigma_{S_1}$  by $C^L_1$ and $\sigma_{C^L_1}$, respectively, the decoherence function adopts the following gaussian behavior
        \begin{equation}\label{Gi_evol_RT_C_Gauss}
        G_{\zeta_1^{},\zeta_1'}^{\,(rt)}(\tau) = e^{-(\zeta_1^{}-\zeta_1')^2\,\sigma^2_{C^L_1}\,\tau^4/\left[8\,\left(\kappa+1\right)^2\right]}.
        \end{equation}
    By other hand, for a Lorentzian distribution
       \begin{equation}\label{probC_Lorenz}
        p^{\{C^L\}}_1\left(C^L_1\right) = \frac{1}{\pi}\;
        \frac{\delta C^L_1}{(\delta C^L_1)^2+(C^L_1)^2},
       \end{equation}
    it is obtained an exponential decay behavior
        \begin{equation}\label{Gi_evol_RT_C_Lorentz}
        G_{\zeta_1^{},\zeta_1'}^{\,(rt)}(\tau) = e^{-\abs{\zeta_1^{}-\zeta_1'}\,\delta C^L_1\,\tau^2/\left[2\,\left(\kappa+1\right)\right]}.
        \end{equation}

    Finally, with the aim of analyzing the spectral properties of the FID function under reversion, using \eq{FT_AQD_int_result} (see \apen{app:FT_FIDrev}) in \numeq{Val_Obs_MREV_FID}, we obtain the Fourier transform on the time $t$ of the signal produced by the reversion experiment:

        \begin{equation}\label{Val_Obs_MREV_FID_FT}
        \begin{split}
        &\opc{F}_t\left\{\left<\widehat{\op{I}}_{\op{y}}(t, \tau)\right>\right\}(\omega)\\
        &= \frac{\beta_T\,\omega_0\,N}{\opc{N}_{S_1}} \sum_{\zeta_1^{}s_1^{},\zeta_1's_1'}\abs{\bra{\zeta_1^{}s_1^{}}\op{I}_{\op{y}}^{(s_1)}\ket{\zeta_1's_1'}}^2\\
        &\;\times \frac{2\pi}{\abs{\zeta_1' - \zeta_1^{}}}\,p^{\{\Delta S\}}_1\left(\frac{\omega-(\zeta_1'-\zeta_1^{})S_{zz}}{\zeta_1' - \zeta_1^{}}\right) \, G_{\zeta_1^{},\zeta_1'}^{\,(rt)}(\tau).
        \end{split}
        \end{equation}

    At this point, we analyze the meaning of \eq{Val_Obs_MREV_FID_FT}.
    First, we note that if in \numeq{Val_Obs_MREV_FID} we make $G_{\{\zeta_1^{},\zeta_1'\}}(t) = 1$ and $G_{\zeta_1^{},\zeta_1'}^{\,(rt)}(\tau) = 1$, $\forall (t, \tau)$, the resulting reverted FID is that of a closed system.
    Therefore, the Fourier transform of such signal will be a superposition of Dirac deltas shifted at the spectrum characteristic frequencies $(\zeta_1'-\zeta_1^{})S_{zz}$, like \numeq{FT_expoc}.
    By other hand, if in \numeq{Val_Obs_MREV_FID} and \numeq{Val_Obs_MREV_FID_FT} we just assumed  $G_{\zeta_1^{},\zeta_1'}^{\,(rt)}(\tau) = 1$, $\forall \tau$, we would obtain the resulting signal of a FID under the AQD dynamics, which is the same result obtained in Section III.E of QD-I.

    The AQD produces the line-shape of the spectrum. Such spectrum is obtained as a superposition of copies of the OMDF shifted to the frequencies $(\zeta_1'-\zeta_1^{})S_{zz}$ and scaled by the factor $\abs{\zeta_1' - \zeta_1^{}}$, as can be seen in \numeq{Val_Obs_MREV_FID_FT}.
    This scaling effect of $\abs{\zeta_1' - \zeta_1^{}}$ produces the eigen-selection effect in the time domain, due to which the bigger the value of $\abs{\zeta_1' - \zeta_1^{}}$ is, the faster the decay of the decoherence function will be. Such effect is reflected in the spectrum making that for higher frequencies $(\zeta_1'-\zeta_1^{})S_{zz}$  the corresponding copy of the OMDF is less intense and wider.

    Now we gained more insight to understand the resulting signal under reversion in \eqs{Val_Obs_MREV_FID} and \numeq{Val_Obs_MREV_FID_FT}.
    We can see that the spectrum \numeq{Val_Obs_MREV_FID_FT} resembles the one obtained for the AQD in QD-I, but in the present case there is a modulation produced by decoherence during the reversion dynamics. Such decoherence under reversion produces faster decays for higher vales of $\abs{\zeta_1' - \zeta_1^{}}$, as can be seen from \eqs{Gi_evol_RT_C_Gauss} and \numeq{Gi_evol_RT_C_Lorentz}.
    Therefore, if we have different spectra for different values of $\tau$ we should see that the higher the frequency $(\zeta_1'-\zeta_1^{})S_{zz}$ of the spectrum line is, the faster the decay in $\tau$ will be, provoking a kind of compression of the spectrum.
    This feature of the dynamics under reversion will provide us a clear method to detect the effects of the eigen-selectivity due to the coupling of the system with the environment.

    As a final comment, from \eq{irrAQD_func_CL} and \apen{app:Rel_Deco_Dist_Funs}, we can observe that the decoherence function under reversion is the Fourier transform in the variable $C^L_i$ of the distribution function $p^{\{C^L\}}_i$, valuated in
    \[-(\zeta_i^{}-\zeta_i')\,\tau^2/[2(\kappa+1),\]
    this is

        \begin{equation}\label{irrAQD_func_CL_FT}
        \begin{split}
        &G_{\{\zeta_i^{},\zeta_i'\}}^{\,(rt)}(\tau) = \left[\opc{F}_{C^L_i}\left\{p^{\{C^L\}}_i\left(C^L_i\right)\right\}(\alpha)\right]_{\alpha = -\frac{(\zeta_i^{}-\zeta_i')\,\tau^2}{2(\kappa+1)}}\\
        &= \left[\int^{\infty}_{-\infty} dC^L_i \;e^{-i\,\alpha\,C^L_i}\; p^{\{C^L\}}_i\left(C^L_i\right)\right]_{\alpha = -\frac{(\zeta_i^{}-\zeta_i')\,\tau^2}{2(\kappa+1)}}.
        \end{split}
        \end{equation}
    The expressions \numeq{AQD_func_S} and \numeq{irrAQD_func_CL_FT} show the close relation existing between the decoherence process and the distribution function of the variables associated with the environment.

    The numerical calculations of \eqs{Val_Obs_MREV_FID} and \numeq{Val_Obs_MREV_FID_FT} are shown in \figs{FIG2} and \numfig{FIG4} for 5CB and PAA$_{d6}$, respectively, where a frequency selective gaussian decay was proposed for decoherence along $t$ and $\tau$ time scales.
    In the case of 5CB, we used the ten-spin model (core protons plus the $\alpha-CH_2$ proton pair) from reference \cite{ssnmr09}
    because of the high effort that would involve the inclusion of more spins, which is beyond of the current computational facilities.
    For PAA$_{d6}$, the eight-spin model presented in QD-I was used.
    In \fig{FIG2} (a) it can be seen the calculated FID signals for different values of $\tau$ in 5CB, while \fig{FIG2} (b) shows the corresponding amplitude spectra. The frontal view of the spectra in \fig{FIG2} (c) shows that the faster decays correspond to the components of higher frequency. This feature is more evident for the normalized spectra of \fig{FIG2} (d), where it can be appreciated the compression of the spectrum for increasing values of $\tau$, in correspondence with the smoothing of the signals as seen in \fig{FIG2} (a).

    The details of the calculated variation of the normalized amplitude of the spectrum lines of frequencies 5.65kHZ and 8.50kHz are shown in \fig{FIG3}, where it can be appreciated the higher decay rate of the high frequency component.
    The calculations presented in \fig{FIG2} show the effects on the dynamics introduced by the eigen-selectivity of the decoherence process.
    The choice of the 5CB sample for demonstrating the effects of the eigen-selectivity was motivated by the clear separation existing between the spectral peaks of the two groups of spectral lines around 5.65kHZ and 8.50kHz found in the numerical calculation, corresponding to the strong dipolar couplings of the molecular core and the $\alpha$-pair respectively.
    On the contrary, the spectrum of PAA$_{d6}$ does not exhibit a clear distinction among the frequency lines, since the strong dipolar couplings have similar values. However, a compression of the spectra as a function of the reversal time $\tau$ similar to that found in 5CB is still visible in this compound, as can be seen in \figs{FIG4} (a) and (b),  because of which it is used to show that this effect is part of the evidence of the eigen-selectivity of the decoherence process.
    Notice that if decoherence did not present eigen-selectivity during the reversion period, all the spectral lines would decay with the same rate and therefore the spectral compression would not occur.

    Finally, it is worth to remark that the occurrence of decoherent factors is a consequence of incorporating the `mechanical' variables into the description of the dynamics within a full-quantum view, and that the eigen-selectivity is a direct consequence of this fact. As will be shown in \sect{subse:MedRevEigen_ErrMed}, the relevant experimental errors do not entail eigen-selectivity.
    The experimental measurements corresponding to the experiments proposed are presented in \sect{subse:MedRevEigen_Med} .

\subsection{Measurements}\label{subse:MedRevEigen_Med}

    In this section we present the experimental results obtained by application of the refocusing experiment of \fig{FIG1} (c) in the FID dynamics, which was discussed in detail in \sect{subse:MedRevEigen_DefExp}.

    The experiments were carried out in a home-built spectrometer, based on a magnet of a Varian EM360, of 60MHz for protons, with the probe adapted for application of pulsed radiofrequency. The electronic setup allows the complete control of the phase of the pulses, with a precision of $0.022^o$, a time step of 40ns, and a minimum configurable time of 240ns. The rf power used permits $\pi/2$ pulses of about 6.5$\mu$s. The homogeneity of the magnetic field is controlled with shimming coils, with a minimum half width of 470Hz approximately, for the spectral lines (using a model of gaussian line-form for each spectral line). The maximum dead time in the signal acquisition is about 18$\mu$s. The temperature can be set between 25$^o$C and 150$^o$C, with a medium accuracy of $\pm$1$^o$C and a stability of $\pm$0.1$^o$C.

    We made experiments in samples of the nematic liquid crystals 5CB (4'-pentyl-4-biphenyl-carbonitrile), PAA$_{d6}$ (methyl deuterated para-azoxyanisole) and PAA (para-azoxyanisole), and the solid adamantane. The solid sample was included for comparison, since this system contains an unlimited number of interacting spins, in contrast with the finite `clusters' comprising the protons of LC molecules.
    Degradation of spin coherence could also occur in a solid induced by non-spin degrees of freedom, for instance due to lattice phonons. However, this point is out of scope of the present work.

    With the purpose of reference for subsequent discussion, in \fig{FIG5} we show the measured FID's with the corresponding spectra, for all the samples studied. These signals are the result of eight acquisitions, except in PAA$_{d6}$, where the signal was acquired 208 times.
    PAA$_{d6}$ and PAA molecules differ in that in the first the methyl groups are replaced by $CD_3$ groups.
    In adamantane, while the  molecules as a whole are fixed on the solid network, they undergo rapid motions, due to which the crystalline spin system can be effectively represented by a lattice of spins $1/2$ \cite{schnell2001}.

    It should be noted that the experimental spectrum of 5CB in \fig{FIG5} (a2) is different from the spectrum numerically calculated with the ten-spin model used in \sect{subse:MedRevEigen_DefExp}.
    It can be seen that in the measured spectrum the higher amplitude peaks are shifted to higher frequencies, while in the calculated spectrum the situation is the opposite.
    This is so because the ten-spin calculation does not include the remaining $CH_2$ proton pairs of the alkyl chain, which contribute to the high frequency peak.
    However, these discrepancies do not invalid the usefulness of the ten-spin model for appreciating the effects on the spin dynamics of the many-body quantum character of the spin interactions, like the eigen-selectivity, as was shown in \sect{subse:MedRevEigen_DefExp}.

    The measurements in  PAA$_{d6}$ and PAA were made at temperatures T = 115$^o$C and T = 110$^o$C, respectively. In the remaining cases, the experiments were carried out at room temperature, namely T = 27$^o$C.
    The MREV8 sequence shown in \fig{FIG1} (b) was set up to mitigate the effects of the finite width of the pulses.
    The total time of each sequence is $\tau_c = 2\,(2 \tau_2 + 2 \tau_1 + 4 t_w)$\cite{Abrag_Gold_cap2}, then  $\tau_1 = \tau_c / 12 - t_w$ and $\tau_2 = \tau_c / 6 - t_w$.
    For a pulse of $\pi/2$ with a width of $t_w = 6.56\mu s$ and a minimum setting time of
    $\tau_1 = 1.6\mu s$, we have $\tau_c = 12\,(\tau_1 + t_w) = 97.92\mu s$ and
    $\tau_2 = 9.76\mu s$.
    By comparing the time scale of the FID's shown in \fig{FIG5}, it can be anticipated that the evolution under the dipolar Hamiltonian during the pulse duration will not have relevance in the experimental results.

    In \fig{FIG6} we present experimental results corresponding to the refocusing experiments discussed in \sect{subse:MedRevEigen_DefExp} in 5CB under the sequence shown in \fig{FIG1} (c).
    From the spectral evolution of the `time-reversed' FID's of \fig{FIG6} (a), given in \fig{FIG6} (b) or in more detail in \fig{FIG6} (c) and  \fig{FIG6} (d), it can be seen how the spin dynamics is affected by the eigen-selection process during the reversion (in the last figure the spectra as a function of $\tau$ are normalized to facilitate the comparison).
    This becomes more evident when comparing the amplitude variation of two groups of spectral lines, those around the frequencies 5.55kHz and 10.60kHz, as shown in \fig{FIG8} (a). There it can be seen that the peak at 10.60kHz presents a higher decay rate for all $\tau$.
    It can be appreciated in \fig{FIG6} (c) how the two peaks match their amplitudes at $\tau \approx 361.4 \mu s$, reverting subsequently their initial amplitude relation.
    It is worth to note that in the spectra the frequencies lines close to 0kHz are affected by instrumental artifacts like noise in the baseline of the signals. Therefore, the variations of such lines should not be given any physical meaning.

    The results of the same experiment performed in nematics PAA$_{d6}$ and PAA, and solid adamantane, are shown in \fig{FIG7}.
    In PAA$_{d6}$, in agreement with the results of the numerical calculation shown in \sect{subse:MedRevEigen_DefExp}, it is observed a narrowing of the spectrum, in consistency with a more pronounced decay of the higher frequencies lines.
    By other hand, in the PAA spectrum  low and high frequency groups of lines are resolved. \fig{FIG8} (b) shows the detail of the normalized decay of two lines at 1.60kHz and 7.35kHz.
    It should be noticed that the decay rate depends on the line position in the frequency spectrum, and the `bell' form associated with the homogeneous broadening of every line (the lineshape produced by AQD) is equally affected by decoherence or non-ideal effects like static field inhomogeinity.
    Due to this, for general cases the time evolution of the different parts of the spectrum  will be masked  by the superposition of broadened lines, preventing the experimental resolution of the decay rates of different parts of the spectrum.
    Therefore, it can be expected that eigen-selectivity would be more easily observe in samples with spectrum having groups of lines with certain degree of resolution, like nematics 5CB and PAA.

    The experiment sketched in \fig{FIG1} (c) can be modified by changing the MREV8-train by a single block, where the time parameter $\tau_1$ is varied independently, and the value of $\tau_2$ is defined as to obtain the maximum signal, being  $\tau_2 > \tau_1$. For these values of $\tau_2$ as a function of $\tau_1$ the slope of a straight line is fitted, whose ideal value is two. By using this dependence, the reversion experiment can be done by using a single block, in such a way to minimize non-idealities present in the pulse setting of the experiments. In this experiment, the reversion time $\tau$ is a linear function of $\tau_1$. The advantage of this method is that it permits a continuous variation of $\tau_1$ yielding a smooth profile for the signal amplitude in the reversion experiments. On the other hand, long-time settings in such experiment can introduce contributions from the non-secular dipolar Hamiltonian on the dynamics (see section II in supplementary material\cite{suppmatDQII} for details about the influence of the non-secular dipolar part on the MREV8 reversion dynamics).

    The results of the reversion experiment using the continuous MREV8 method in 5CB are shown in  \fig{FIG9}. In \fig{FIG10} is shown the normalized variation with the reversion time of the amplitude of two peaks at distinct frequencies, corresponding to the measurement of \fig{FIG9}. There it can be appreciated the frequency selective decay.
    Due to the finite pulse widths, the initial measured amplitudes correspond to a time $\tau$ significantly shifted from zero. For these reversion times the frequency selective decay is  noticeably, accordingly the first spectra in \fig{FIG9} already presents a larger decay of the higher frequency. This explains the difference with the spectra shown in \fig{FIG6}.
    Except for this last detail, the obtained results show the same features than the reversion experiments using blocks of the MREV8 sequence. Similar results were obtained for the other compounds studied in this work.

    By means of the continuous MREV8 experiment it is possible to obtain a detail of the maximum  amplitude of the reversed FID's, as a function of the reversion time $\tau$.
    In \fig{FIG11}, the results obtained by averaging data near the maximum of each reversed FID are shown, where the signal to noise ratio is larger. Besides, the plots were normalized with respect to the first FID obtained in each experiment.
    We added three gaussian profiles with different values of  standard deviation ($\sigma = 350\mu s,\,580\mu s,\,800\mu s$), used as a guide to the eye.
    Since the first FID signal already present a significant attenuation, the gaussian curves have a value greater than one for $\tau = 0$.
    It can be seen that the decay time is longer than the characteristic decay time of the FID for every compound (see \fig{FIG5}).
    This is consistent with the theoretical approach proposed in QD-I, where it is considered that the spin dynamics during the reversion period is governed by a different mechanism than the one which control the spin dynamics during the FID evolution. This quantum process is characterized by a longer time scale, since it is associated with higher order terms of a perturbative treatment.
    Besides, we can see that for times shorter than 780$\mu s$ the PAA$_{d6}$ (with 8 spins per molecule) presents a similar decay behavior that the PAA (with 14 spins per molecule) and the 5CB (with 19 spins per molecule) presents a similar decay behaviour that the adamantane (a solid array of spins) for all $\tau$ time. These results reflect that decoherence is not associated with the nature of the spins as a closed system.

    It is also observed in \fig{FIG11} that the decay in PAA$_{d6}$  is more similar to that of  PAA than to the one of 5CB.
    The fact of observing similar responses to decoherence for these two samples is consistent with decoherence being controlled by the coupling of the spin system to external degrees of freedom. For values of $\tau$ greater than 780$\mu s$, decoherence in PAA turns stronger, which could be indicating the occurrence of an additional decoherent mechanism associated with the protons of the methyl groups. However, because of the low signal to noise ratio of PAA$_{d6}$  and other sources of error, which will be discussed in \sect{subse:MedRevEigen_ErrMed}, more experiments  are necessary to confirm this assertion.

    It is worth to note that, using a single MREV8 pulse sequence with a continuous variation of time settings, for long $\tau$ values the decay can be affected by the influence of the non-secular dipolar Hamiltonian part in the dynamics under reversion (see section II in supplementary material\cite{suppmatDQII}). However, such experiment is able to show the intermediate time scale of decoherence as it is the aim in this work. For a more exactly measurement of the decoherence time other techniques can be used, like the MS\cite{gonza11,Abrag_Gold_cap2,RhimPV70,RhimPV71,RhimK71,RhimK72,nielsen97}, when they are adequate due to constraints in the experimental setting like the duration of the rf pulses. In particular, using a combined technique of MS with z-rotational decoupling, namely MSHOT-3\cite{nielsen97}, would improve the performance of the reversion sequence, which could be particularly useful for extending the study to solids where the intensity of the dipolar coupling is higher than in LC. With this technique, it would be possible to eliminate higher order terms (up to fifth order) in the dipolar decoupling than that using MREV8 (up to the third).\\

    Summarizing, from the  results presented in this section, we have the following conclusions:
    \begin{enumerate}
    \item the time scale of the attenuation of the time reversed FID signal in MREV8 experiments occurs in a longer time scale than the one of the FID evolution,
    \item the high-frequency spectral components decay faster, showing an eigen-selection process,
    \end{enumerate}
    in total agreement with the theory presented in QD-I.
    Besides, such intermediate time scale for decoherence, between the FID evolution and the thermalization process time scales, is consistent with the hypothesis about the existence of dynamics with different time scales used in the theoretical approach of this work and QD-I.\\

\subsection{Non-ideal behaviors and approximations in the experiments}\label{subse:MedRevEigen_ErrMed}

    The MREV8 sequence shown in  \fig{FIG1} (b) has been optimized by adjusting the time parameters $\tau_1$ and $\tau_2$, to mitigate the effects of the finite width $t_w$ of the pulses, as was commented in \sect{subse:MedRevEigen_Med}. The shortest interval $\tau_1$ between pulses  was fixed in 1.6$\mu$s. It is worth to note that in a time of 6.56$\mu$s, corresponding with a $\pi/2$ pulse in our experiment, the free evolution dynamics produced by the dipolar interaction is not significant as can be seen from the FID's shown in \fig{FIG5}.

    The non-ideality arising in the control of the reversion parameter $\kappa$ would come from deviations from the exact value $\kappa$=2 due to misadjustments of the $\pi/2$ pulses (see \sect{subse:MedRevEigen_DefExp} for details about this reversion parameter).
    In a similar fashion, a correct setting of the $\pi/2$ pulses with $\kappa$=2 could be affected by an incorrect adjustment of the time $\tau_2$ (i.e. with $\tau_2 \neq \kappa\tau_1$), causing an imperfect reversion of the spin dynamics.
    These errors could introduce an additional dynamics of the {\it essentially isolated system}\cite{SegZam2011}, generated by the molecular dipolar Hamiltonian, producing perturbations of the signals and their spectra as a function of the reversion parameter $\tau$.

    With the aim of estimating the effects of experimental misadjustments, in the following we will obtain several analytical expressions for different kind of errors in the reversion sequence shown in \fig{FIG1} (c) using MREV8 blocks of pulses.
    We will consider the spin system as a closed system to study the possibility of obtaining some eigen-selection effect in such case, which could overlap with the effects of the coupling with the environment.
    We relegate to the section II in supplementary material\cite{suppmatDQII} the complete analytical demonstration, as well as the numerical calculation of the signal expressions which will be used in this section.

    We introduce the error $\epsilon$ in the setting of the time under the reversion dynamics by writting $\tau_2 = (\kappa + \epsilon)\,\tau_1$.
    It is worth to note that such error takes account of misadjustments in the time $\tau_2$ as well as deviations from a perfect setting of the $\pi/2$ pulses (see section II in supplementary material\cite{suppmatDQII} for details about the equivalence between $\tau_2$ time misadjustments and $\pi/2$ pulses misadjustments).
    Therefore, we have $t_2 = (\kappa + \epsilon)\,t_1$ and the total reversion time $\tau = (\kappa + 1 + \epsilon)\,t_1$.
    Using such erroneous setting of the $t_2$ time, we obtain for the expectation value of $\widehat{\op{I}}_{\op{y}}$ corresponding to the closed spin system, under `on-resonance' condition, the following result

    \begin{equation}\label{Val_Obs_MREV_FID_errHd}
    \begin{split}
    &\left<\widehat{\op{I}}_{\op{y}}(t, \tau)^{\epsilon}\right> = \frac{\beta_T\,\omega_0\,N}{\opc{N}_{S_1}}\, tr_{s_1}\bigg\{\op{I}_{\op{y}}^{(s_1)}\,e^{-i\,\opc{\widehat{H}}_{S1}^{(s_1)}\,t}\\
    &\qquad\times e^{i\,\opc{\widehat{H}}_{S1}^{(s_1)}\,\frac{\epsilon/\kappa}{\kappa+1+\epsilon}\,\tau}\,
    \op{I}_{\op{y}}^{(s_1)}\,e^{-i\,\opc{\widehat{H}}_{S1}^{(s_1)}\,\frac{\epsilon/\kappa}{\kappa+1+\epsilon}\,\tau}\,e^{i\,\opc{\widehat{H}}_{S1}^{(s_1)}\,t}\bigg\}\\
    &\qquad\simeq \frac{\beta_T\,\omega_0\,N}{\opc{N}_{S_1}} \sum_{\zeta_1^{}s_1^{},\zeta_1's_1'}
    \abs{\bra{\zeta_1^{}s_1^{}}\op{I}_{\op{y}}^{(s_1)}\ket{\zeta_1's_1'}}^2\\
    &\qquad\qquad\times e^{-i\,(\zeta_1^{}-\zeta_1')\, S_{zz}\,\left(t-\frac{\epsilon/\kappa}{\kappa+1}\,\tau\right)},
    \end{split}
    \end{equation}
    where we used the approximation $\epsilon/\kappa \ll 1$ which implies $t_1 \simeq \tau/(\kappa + 1)$.
    The Fourier transform of \numeq{Val_Obs_MREV_FID_errHd} is

    \begin{equation}\label{Val_Obs_MREV_FID_FT_errHd}
    \begin{split}
    &\opc{F}_t\left\{\left<\widehat{\op{I}}_{\op{y}}(t, \tau)^{\epsilon}\right>\right\}(\omega)\\
    &= \frac{2\pi\,\beta_T\,\omega_0\,N}{\opc{N}_{S_1}} \sum_{\zeta_1^{}s_1^{},\zeta_1's_1'}\abs{\bra{\zeta_1^{}s_1^{}}\op{I}_{\op{y}}^{(s_1)}\ket{\zeta_1's_1'}}^2\\
    &\quad\times \delta\left[\omega - (\zeta_1'-\zeta_1^{})\,S_{zz}\,\right]\,
    e^{i\,(\zeta_1^{}-\zeta_1')\, S_{zz}\,\frac{\epsilon/\kappa}{\kappa+1}\,\tau}.
    \end{split}
    \end{equation}
    The expression in \eq{Val_Obs_MREV_FID_FT_errHd}, for the Fourier transform of the FID signal for a closed spin system, does not present a decay with some eigen-selection effect, but it shows an oscillatory behaviour of each spectral line proportional to the product of the eigenvalue difference $\zeta_1^{}-\zeta_1'$ with the error factor $\epsilon/\kappa$.\\

    Other interesting case to consider, is when the FID signal is thought as produced by the different signals coming from a distributed molecular orientation of the main molecular axis of the LC system\cite{schmie82}.
    In such case, with the same misadjustment as in \eq{Val_Obs_MREV_FID_errHd}, we have for the FID

    \begin{equation}\label{Val_Obs_MREV_FID_errHddist}
    \begin{split}
    &\overline{\left<\widehat{\op{I}}_{\op{y}}(t, \tau)^{\epsilon}\right>}
    = \frac{\beta_T\,\omega_0\,N}{\opc{N}_{S_1}} \sum_{\zeta_1^{}s_1^{},\zeta_1's_1'}
    \abs{\bra{\zeta_1^{}s_1^{}}\op{I}_{\op{y}}^{(s_1)}\ket{\zeta_1's_1'}}^2\\
    &\qquad\times \int^{\infty}_{-\infty} dS_1 \; e^{-i\,(\zeta_1^{}-\zeta_1')\,S_1\,\left(t-\frac{\epsilon/\kappa}{\kappa+1}\,\tau\right)}\; p^{\{S\}}_1\left(S_1\right)\\
    &\quad= \frac{\beta_T\,\omega_0\,N}{\opc{N}_{S_1}} \sum_{\zeta_1^{}s_1^{},\zeta_1's_1'}
    \abs{\bra{\zeta_1^{}s_1^{}}\op{I}_{\op{y}}^{(s_1)}\ket{\zeta_1's_1'}}^2\\
    &\qquad\times e^{-i\,(\zeta_1^{}-\zeta_1')\, S_{zz}\,\left(t-\frac{\epsilon/\kappa}{\kappa+1}\,\tau\right)} \, e^{-\frac{1}{2}\,(\zeta_1^{} - \zeta_1')^2\, \sigma_{S_1}^2\,\left(t-\frac{\epsilon/\kappa}{\kappa+1}\,\tau\right)^2},
    \end{split}
    \end{equation}
    where $p^{\{S\}}_1\left(S_1\right)$ is the OMDF (see \sect{subse:MedRevEigen_DefExp}). The expression \numeq{Val_Obs_MREV_FID_errHddist} is the classical representation of the AQD seen in \sect{subse:MedRevEigen_DefExp}. Since the dynamics produced by the error factor $\epsilon/\kappa$ under the reversion time $\tau$ is very slow in comparison with the one produced under the free-evolution time $t$, we can write

    \begin{equation}\label{Val_Obs_MREV_FID_errHddist_approx}
    \begin{split}
    &\overline{\left<\widehat{\op{I}}_{\op{y}}(t, \tau)^{\epsilon}\right>}
    \simeq \frac{\beta_T\,\omega_0\,N}{\opc{N}_{S_1}} \sum_{\zeta_1^{}s_1^{},\zeta_1's_1'}
    \abs{\bra{\zeta_1^{}s_1^{}}\op{I}_{\op{y}}^{(s_1)}\ket{\zeta_1's_1'}}^2\\
    &\quad\times e^{-i\,(\zeta_1^{}-\zeta_1')\, S_{zz}\,\left(t-\frac{\epsilon/\kappa}{\kappa+1}\,\tau\right)} \, e^{-\frac{1}{2}\,(\zeta_1^{} - \zeta_1')^2\, \sigma_{S_1}^2\, \left[t^2+\left(\frac{\epsilon/\kappa}{\kappa+1}\,\tau\right)^2\right]}.
    \end{split}
    \end{equation}
    The Fourier transform of \numeq{Val_Obs_MREV_FID_errHddist_approx} is

    \begin{equation}\label{Val_Obs_MREV_FID_FT_errHddist_approx}
    \begin{split}
    &\opc{F}_t\left\{\overline{\left<\widehat{\op{I}}_{\op{y}}(t, \tau)^{\epsilon}\right>}\right\}(\omega)\\
    &\simeq \frac{\beta_T\,\omega_0\,N}{\opc{N}_{S_1}} \sum_{\zeta_1^{}s_1^{},\zeta_1's_1'}\abs{\bra{\zeta_1^{}s_1^{}}\op{I}_{\op{y}}^{(s_1)}\ket{\zeta_1's_1'}}^2\\
    &\quad\times \frac{2\pi}{\abs{\zeta_1' - \zeta_1^{}}}\,\frac{1}{\sqrt{2\pi\,\sigma_{S_1}^2}}\,
    e^{-\frac{1}{2}\,\left[\frac{\omega-(\zeta_1'-\zeta_1^{})S_{zz}}{(\zeta_1' - \zeta_1^{})\,\sigma_{S_1}}\right]^2}\\
    &\quad\times e^{i\,(\zeta_1^{}-\zeta_1')\, S_{zz}\,\frac{\epsilon/\kappa}{\kappa+1}\,\tau} \, e^{-\frac{1}{2}\,(\zeta_1^{} - \zeta_1')^2\, \sigma_{S_1}^2\,\left(\frac{\epsilon/\kappa}{\kappa+1}\,\tau\right)^2}.
    \end{split}
    \end{equation}
    We can see from \eq{Val_Obs_MREV_FID_FT_errHddist_approx} that there is a decay factor with eigen-selection effect for each spectral line, but it is masked by an oscillatory behaviour of each frequency line, instead of the monotonous decay that is experimentally observed (see for instance \figs{FIG6} and \numfig{FIG8}).
    Besides, such predicted decay is reversible and it would, in principle, be possible to increase the decay time by correcting the misadjustment represented by the factor $\epsilon/\kappa$, since the dynamics is very susceptible to changes of $\epsilon$.
    The factor $\frac{\epsilon/\kappa}{\kappa+1}$ is acting like a scale factor in the reversion time $\tau$ for the decoherence function as well as for the oscillation function. Therefore, changing this factor simultaneously modifies the time behaviours of both functions in such way that if a decay of the signal is observed, the oscillations have to be observed as well.
    Indeed, this effect is not observed in the experimental measurements, where the correction of the misadjustments produces the extinction of the oscillatory behaviour under the reversion dynamics presenting an observable decay in the signals in the intermediate time scale.\\
    If we see the signal under the reversion time $\tau$ obtained for the time $t=0$ in \eq{Val_Obs_MREV_FID_errHddist_approx}, we have

    \begin{equation}\label{Val_Obs_MREV_FID_errHddist_tzero}
    \begin{split}
    &\overline{\left<\widehat{\op{I}}_{\op{y}}(0, \tau)^{\epsilon}\right>}
    = \frac{\beta_T\,\omega_0\,N}{\opc{N}_{S_1}} \sum_{\zeta_1^{}s_1^{},\zeta_1's_1'}
    \abs{\bra{\zeta_1^{}s_1^{}}\op{I}_{\op{y}}^{(s_1)}\ket{\zeta_1's_1'}}^2\\
    &\quad\times e^{i\,(\zeta_1^{}-\zeta_1')\, S_{zz}\,\frac{\epsilon/\kappa}{\kappa+1}\,\tau} \, e^{-\frac{1}{2}\,(\zeta_1^{} - \zeta_1')^2\, \sigma_{S_1}^2\,\left(\frac{\epsilon/\kappa}{\kappa+1}\,\tau\right)^2}.
    \end{split}
    \end{equation}
    The expression \numeq{Val_Obs_MREV_FID_errHddist_tzero} has the form of a FID but with a slower time dependance. Such behaviour under reversion depends on the time $\frac{\epsilon/\kappa}{\kappa+1}\,\tau$ instead of the time $t$. The dynamics represented by \eq{Val_Obs_MREV_FID_errHddist_tzero} is reversible and the oscillations in the signal are unavoidable. Clearly, this is not a feature of the experiments, as was mentioned.\\

    If the setting of the $\pi/2$ pulses were very erroneous or if the dynamics under reversion were strongly influenced by the non-secular dipolar Hamiltonian (for instance, when the time $\tau_1$ is not small enough to neglect the dynamics produced by the non-secular dipolar Hamiltonian), the signal under reversion could be written as

    \begin{equation}\label{Val_Obs_MREV_FID_Hdnonsec}
    \begin{split}
    &\left<\widehat{\op{I}}_{\op{y}}(t, \tau)^\ddag\right> = \frac{\beta_T\,\omega_0\,N}{\opc{N}_{S_1}}\, tr_{s_1}\bigg\{\op{I}_{\op{y}}^{(s_1)}\,e^{-i\,\opc{\widehat{H}}_{S1}^{(s_1)}\,t}\\
    &\qquad\qquad\times e^{-i\,\opc{\widehat{H}}^{\ddag(s_1)}_{S1}\,\tau}\, \op{I}_{\op{y}}^{(s_1)}\,e^{i\,\opc{\widehat{H}}^{\ddag(s_1)}_{S1}\,\tau}\,e^{i\,\opc{\widehat{H}}_{S1}^{(s_1)}\,t}\bigg\}\\
    &= \frac{\beta_T\,\omega_0\,N}{\opc{N}_{S_1}} \sum_{\zeta_1^{}s_1^{},\zeta_1's_1'}
    \bra{\zeta_1's_1'}\op{I}_{\op{y}}^{(s_1)}\ket{\zeta_1^{}s_1^{}}\,
    e^{-i\,(\zeta_1^{}-\zeta_1')\, S_{zz}\,t}\\
    &\times \sum_{\alpha_1,\alpha_1'} \braket{\zeta_1s_1}{\alpha_1}\bra{\alpha_1}\op{I}_{\op{y}}^{(s_1)}\ket{\alpha_1'}
    \braket{\alpha_1'}{\zeta_1's_1'}\\
    &\qquad\qquad\times e^{-i\,(\alpha_1-\alpha_1')\,\tau},
    \end{split}
    \end{equation}
    where $\opc{\widehat{H}}^{\ddag(s_1)}_{S1}$ is the resulting Hamiltonian under reversion, which is different of $\opc{\widehat{H}}_{S1}^{(s_1)}$
    (see section II in supplementary material\cite{suppmatDQII} for details of the definition of $\opc{\widehat{H}}^{\ddag(s_1)}_{S1}$).
    In \eq{Val_Obs_MREV_FID_Hdnonsec} the eigenbase $\left\{\ket{\alpha_1}\right\}$ of the Hamiltonian $\opc{\widehat{H}}^{\ddag(s_1)}_{S1}$ is defined, where $\opc{\widehat{H}}^{\ddag(s_1)}_{S1}\ket{\alpha_1} = \alpha_1\ket{\alpha_1}$.
    The Fourier transform of \numeq{Val_Obs_MREV_FID_Hdnonsec} is

    \begin{equation}\label{Val_Obs_MREV_FID_FT_Hdnonsec}
    \begin{split}
    &\opc{F}_t\left\{\left<\widehat{\op{I}}_{\op{y}}(t, \tau)^\ddag\right>\right\}(\omega) =
    \frac{2\pi\,\beta_T\,\omega_0\,N}{\opc{N}_{S_1}}\\
    &\times \sum_{\zeta_1^{}s_1^{},\zeta_1's_1'}
    \bra{\zeta_1's_1'}\op{I}_{\op{y}}^{(s_1)}\ket{\zeta_1^{}s_1^{}}\;
    \delta\left[\omega - (\zeta_1'-\zeta_1^{})\,S_{zz}\,\right]\\
    &\times \sum_{\alpha_1,\alpha_1'} \braket{\zeta_1^{}s_1^{}}{\alpha_1}\bra{\alpha_1}\op{I}_{\op{y}}^{(s_1)}\ket{\alpha_1'}
    \braket{\alpha_1'}{\zeta_1's_1'}\\
    &\qquad\qquad\times e^{-i\,(\alpha_1-\alpha_1')\,\tau},
    \end{split}
    \end{equation}
    where we can see that the decay corresponding to the spectral lines does not present any eigen-selection effect.
    These effects of misadjustments in the reversion time, with the spin system considered as a closed system, were numerically simulated and the results are shown in the section II in supplementary material\cite{suppmatDQII}, where we obtain numerical conclusions which coincide with the analytical analysis.\\

    Finally, to contrast with the commented analytical results about misadjustments in the experimental setting in the closed spin system, we derived the analytical signal produced by the same reversion experiment affected by an erroneous setting of the reversion time, for the open quantum system conformed by the protons coupled to the environment.
    Using the time setting $t_2 = (\kappa + \epsilon)\,t_1$ in the evolution operator \numeq{U_evol_RT_evol_f_C}, we obtain the decoherence function

    \begin{equation}\label{Gi_evol_RT_evol_C_eval_aprox_int_eig_errt2}
    \begin{split}
    &G_{\{\zeta,\zeta_i'\}}(t, \tau)^{\epsilon} = \int d\Delta S_i \; e^{-i\,(\zeta_i^{}-\zeta_i')\,\Delta S_i\,\opc{Y}_{\Delta S}(t,\tau)^{\epsilon}}\\
    &\;\times \int dC^L_i \; e^{-i\,(\zeta_i^{}-\zeta_i')\,C^L_i\,\opc{Y}_{L}(t,\tau)^{\epsilon}}\\
    &\;\times \int dC^{SL}_{i\,\zeta} \; e^{-i\,(\zeta_i^{}-\zeta_i')\,C^{SL}_{i\,\zeta}\,\opc{Y}_{SL}(t,\tau)^{\epsilon}} \; p_i(\Delta S_i, C^L_i, C^{SL}_{i\,\zeta}),
    \end{split}
    \end{equation}
    instead of the \eq{Gi_evol_RT_evol_C_eval_aprox_int_eig}, where we have used the same considerations involved in \eq{Gi_evol_RT_evol_C_eval_aprox_int_eig}.
    In \eq{Gi_evol_RT_evol_C_eval_aprox_int_eig_errt2}, we defined

    \begin{subequations}\label{Gi_errt2_poly}
    \begin{equation}\label{Gi_errt2_poly_DeltaS}
    \opc{Y}_{\Delta S}(t,\tau)^{\epsilon} = t-\frac{\epsilon/\kappa}{\kappa+1+\epsilon}\,\tau,
    \end{equation}
    \begin{equation}\label{Gi_errt2_poly_L}
    \begin{split}
    &\opc{Y}_{L}(t,\tau)^{\epsilon} = \frac{1}{2}\bigg\{\left(t + \tau\right)^2 - \left[1+(\kappa+1+\epsilon)^{-1}\right]\,\tau^2\\
    &\qquad-\left[(\kappa+1+\epsilon)^{-1}+(\kappa+1+\epsilon)^{-2}\right]\,\frac{\epsilon}{\kappa}\,\tau^2\bigg\},
    \end{split}
    \end{equation}
    \begin{equation}\label{Gi_errt2_poly_SL}
    \opc{Y}_{SL}(t,\tau)^{\epsilon} = \frac{1}{2}\left(t-\frac{\epsilon/\kappa}{\kappa+1+\epsilon}\,\tau\right)^2.
    \end{equation}
    \end{subequations}
    Under the approximation in the dynamics of $\iiint \prod_{D,n}\left\{\cdot\right\}_{t=0}$, used to obtain the well approximated decoherence function shown in \eq{Gi_evol_RT_evol_C_eval_aprox}, and with $\epsilon/\kappa \ll 1$, we can write \numeq{Gi_evol_RT_evol_C_eval_aprox_int_eig_errt2} as

    \begin{equation}\label{Gi_evol_RT_evol_C_eval_aprox_int_eig_errt2_aprox}
    \begin{split}
    &G_{\{\zeta,\zeta_i'\}}(t, \tau)^{\epsilon} \simeq \int d\Delta S_i \; e^{-i\,(\zeta_i^{}-\zeta_i')\,\Delta S_i\,\left(t-\frac{\epsilon/\kappa}{\kappa+1}\,\tau\right)}\\
    &\;\times \int dC^L_i \; e^{i\,(\zeta_i^{}-\zeta_i')\,C^L_i\,\frac{\tau^2}{2\left(\kappa+1\right)}\,\left[1+\left(\frac{\kappa+2}{\kappa+1}\right)\,\frac{\epsilon}{\kappa}\right]}\\
    &\;\times \int dC^{SL}_{i\,\zeta} \; e^{-i\,(\zeta_i^{}-\zeta_i')\,C^{SL}_{i\,\zeta}\,\frac{1}{2}\left(\frac{\epsilon/\kappa}{\kappa+1}\,\tau\right)^2} \; p_i(\Delta S_i, C^L_i, C^{SL}_{i\,\zeta}).
    \end{split}
    \end{equation}
    The decoherence function \numeq{Gi_evol_RT_evol_C_eval_aprox_int_eig_errt2_aprox}, which is influenced by misadjustments of the reversion time $t_2$, will affect the expectation value of $\widehat{\op{I}}_{\op{y}}$ as can be seen in \eq{Val_Obs_trn_fin_rot}.
    We can see from \eq{Gi_evol_RT_evol_C_eval_aprox_int_eig_errt2_aprox} that the decays produced by the integration of the complex exponential function  under the total reversion time $\tau$ can be compensated excepting for the dynamics under the eigenvalue $C^L_i$. Such decays present the eigen-selection effect and the oscillations in the signals under the reversion time $\tau$, due to the misadjusment of $t_2$, can be cancelled.
    It can also be seen that the decays associated to the eigenvalues $\Delta S_i$ and $C^{SL}_{i\,\zeta}$ can be reverted but the decay due to $C^L_i$ will never be reverted. Such non-reverted decay constitutes an envelope which constrains the signal as a function of the reversion time $\tau$ giving it a bell-like shape, as we will show below.
    By making $\epsilon = 0$ in \eq{Gi_evol_RT_evol_C_eval_aprox_int_eig_errt2_aprox} we recover \eq{Gi_evol_RT_evol_C_eval_aprox}, as expected.
    In \fig{FIG12} (a) we show the effect mentioned above for the reversion experiment in 5CB, where in \fig{FIG12} (b) it can be seen the variations of the amplitudes of the pseudo-FID, obtained extracting the mean values of the amplitudes of the FID around $t=0$, introduced by the error of $\kappa$.
    By comparing \fig{FIG12} and  \fig{FIG6}, it can be noticed that in \fig{FIG6} (a) the amplitudes decay monotonously, indicating that the error $\kappa$ has been satisfactorily mitigated.
    This dynamics induced by  experimental mismatches  cannot produce a definitive signal decay, because of the small number of spin degrees of freedom. The latter is true for any error in the pulse configuration.
    In practice, this effect can be attenuated by varying the time between the two  WHH-4 blocks comprising the MREV8 sequence (see \fig{FIG1} (b)) until the oscillations of the amplitudes as a function of $\tau$ dissapear.
    The results shown in \figs{FIG6} and \numfig{FIG7} were obtained with this procedure.

    The observed signal attenuation might be associated with several causes besides the essentially adiabatic decoherence
    \cite{SegZam2011}, like fluctuations of the spin-spin interactions due to thermal molecular motions \cite{Abragam_cap10} or even experimental non-idealities like inhomogeinity of the static and the rf magnetic fields. In the work of reference \cite{gonza11}, the experimental sources of error in the MS reversion experiment were carefully checked by studying the effect of the sequence on the spin system in the isotropic phase of liquid crystal 5CB.
    Combination of the MS sequence and a $\pi$ pulse to reverse the static field inhomogeinity yielded the same response than the usual Hahn-echo two pulse sequence ($T_2 = 70 ms$). This also showed that the time scale of the decay produced by thermal fluctuations
    is much greater than that of the MS experiment in the nematic phase.

    In the present work, we studied analytically and experimentally the influence of the field inhomogeneity in the FID dynamics and the reversion dynamics performing experiments on isotropic 5CB (see section I in supplementary material\cite{suppmatDQII} for details about such experiment). The experimental results allowed us to conclude that the dynamics produced by the field inhomogeneity has a time scale longer than that of decoherence and it does not present a behaviour with eigen-selectivity.
    Besides, it is worth to mention that MREV8 experiments combined with $\pi$ pulses were carried out by the authors in 5CB in the isotropic phase where it was also obtained an exponential decay with a $T_2$ very similar to the obtained in 5CB in reference \cite{gonza11}.
    On the other hand, the homogeinity of the rf field was optimized by using a low coil filling factor and the signals obtained from samples with different sizes do not present any different behaviour between them.

    These observations give support to the statement that while `bulk' effects like inhomogeneity of the rf pulse and thermal fluctuations of dipole-dipole interaction could eventually perturb the free spin dynamics, their influence is irrelevant in the intermediate time scale of the reversion experiments. Therefore, the observed decay cannot be associated with experimental non-idealities, but it should be assigned to the irreversible spin dynamics induced by quantum decoherence, whose fingerprint is the eigen-selectivity.

    Summaryzing, even when the observed signals are affected by magnetic field inhomogeneities and intermolecular dipole interactions, the contribution of the quantum molecular mechanical dynamics is more important along the different time scales.

\section{Discussion and Conclusions}\label{se:conclu}

    In this work, we studied the spin dynamics which characterizes the irreversible decoherence of a finite quantum interacting spin system coupled with an infinite quantum environment.  We experimentally detected in nematic liquid crystals the salient characteristics of the spin dynamics predicted by a full theory which considers the spins as an open quantum system, namely eigen-selectivity, spectral compression and irreversible decoherence under refocusing of the dipolar spin interactions.\\

    The experiments  were interpreted in the context of the theory presented in the work of reference \cite{SegZam2011}, under the main assumption that irreversible decoherence occurs well before that thermal fluctuations play any significant role, and long after the dephasing by quantum interference and reversible adiabatic decoherence.
    This hypothesis is supported both by our experiments and the work of reference \cite{gonza11}, where it was shown that irreversible decoherence, which cannot be associated with thermalization (neither adiabatic nor nonadiabatic), occurs in an intermediate time scale.
    Besides, the accurate description achieved in reference \cite{SegZam2011} of the time domain FID signal by including quantum interference and reversible adiabatic decoherence, allowed us to show that the time scale characterizing the Liouvillian dynamics and such reversible decoherence (which can be interpreted semiclassically) is much shorter than the time scale of the irreversible quantum decoherence.\\

    The analytical-numerical treatment of \sect{subse:MedRevEigen_DefExp} allowed us to compare the theoretical expressions derived from the proposed theory, with the experiments.
    By introducing decoherence functions with gaussian decay profiles, in the free-evolution and reversion dynamics, the effect of the environment was included. The calculations were performed on a 10-spin model for 5CB \cite{ssnmr09} and on the 8-spin model of PAA$_{d6}$ presented in QD-I \cite{SegZam2011}.
    The results of the calculations are shown in \sect{subse:MedRevEigen_DefExp} and the measurements are shown in \sect{subse:MedRevEigen_Med}.
    By using reversion experiments we confirmed the occurrence of the intermediate time scale and the characteristic behaviour of the irreversible decoherence, and comparison of the analytical FID signals and their spectra under reversion dynamics with the experimental ones confirmed the validity of the theoretical approaches presented in QD-I\cite{SegZam2011} and in this work, beyond the experimental non-idealities.\\

    We presented a detailed analysis of the experimental causes that could affect our measurements, in order to identify their characteristic time scales. In section I of the supplementary material\cite{suppmatDQII}, we analysed the effect of the inhomogeneities of the static magnetic field.
    We conclude that the dynamics produced by this effect has a longer time scale than decoherence, and most importantly, it does not present eigen-selectivity.
    With the aim of checking for a possible frequency dependent behavior induced by pulse misadjustment that might be confused with genuine eigen-selectivity, in \sect{subse:MedRevEigen_ErrMed} we presented a theoretical analysis of the effects of these misadjustments on the FID signals, considering the spins as a closed system.
    The simulations on PAA$_{d6}$ presented in section II of the supplementary material show that misadjustment cannot be the source of the observed eigen-selectivity of a closed spin system.
    Besides, this kind of non-idealities were theoretically analyzed for decoherence produced by the full-quantum dynamics in \sect{subse:MedRevEigen_ErrMed}, as well. The behaviour of the signals extracted by such analysis agrees with the measurement observed.
    Therefore, this detailed analysis leads us to conclude that the observed eigen-selectivity is an evidence of the open quantum system dynamics.\\

    Eigen-selectivity, which introduces a distinction in the response of the diagonal and off-diagonal elements of the density matrix, provides an efficient irreversible mechanism for coherence decay, while preserves the `population' terms which can only change in a much longer time scale.
    This behavior explains the observed buildup of the quasi-equilibrium in liquid crystals.
    The results of this work, together with the conclusions obtained in references \cite{SegZam2011,gonza11}, can contribute to elucidate the underlying quantum mechanisms for decoherence of open quantum systems of interacting spins, for instance the role played by quantum correlations between the observed system and the environment in the damping of the spin coherences \cite{helm09,pernice12}. Certainly, these works showed that in liquid crystals it is essential to assume the quantum character of the spin-environment coupling to explain the observed irreversible decoherence.
    These statements might also apply to interacting spins in ordinary solids, where irreversible adiabatic decoherence could provide an explanation for the quasi-equilibrium states characterized by spin temperatures \cite{skrebnev}.\\

    Summarizing, the occurrence of eigen-selectivity in the spin dynamics under reversion was verified in several nematic liquid crystals, through the direct experimental observation of inhomogeneous decay and spectral compression of the NMR spectrum under refocusing of the dipolar spin interactions, over an intermediate time scale.
    We conclude that the eigen-selection effect is the fingerprint of decoherence associated with a quantum open spin system in liquid crystals.
    Besides, the dynamics of such interacting spins, with few degrees of freedom, can be described in terms of quasi-equilibrium states of each molecule after the irreversible decoherence damps out the coherent part of the spin state.
    Therefore, the observed system reaches these states through a genuine quantum process involving spins and environment, being this a process of different nature than the fluctuations which govern thermalization and relaxation.
    These findings allow to understand the development of the quasi-equilibrium states as being a consequence of the correlated dynamics of the observed system and the quantum environment.
    Accordingly, the quasi-equilibrium representation in liquid crystal needs not being perceived heuristically, instead, it should be considered as a definite stage of the spin system, during its evolution towards equilibrium.


\section{acknowledgement}

    This work was supported by Secretar\'ia de Ciencia y T\'ecnica, Universidad Nacional de C\'ordoba and MINCyT C\'ordoba. The authors would like to thank Dr. C.E. Gonz\'alez for useful discussions. H.H.S. thanks CONICET for financial support.


\appendix

\section{Definition of the order parameter}\label{app:Def_Order_Param}

    In this appendix, we present an expression for the order parameter of a nematic LC under the quantum description of the orientational molecular variables.
    The order parameter $S_{zz}$ is defined as

        \begin{equation}\label{order_param_def_BIS}
        \begin{split}
        S_{zz} \equiv \langle \op{S}_{\op{zzi}}^{(f)} \rangle_f &= tr_f\left\{\op{S}_{\op{zzi}}^{(f)}\,\rho_{L(eq)}^{(f)}\right\}\\
        &= \sum_f S_i(f) \bra{f}\rho_{L(eq)}^{(f)}\ket{f},
        \end{split}
        \end{equation}
    where $\left\{\ket{f}\right\}$ is an eigenbasis of the molecular orientational operator $\op{S}_{\op{zzi}}^{(f)}$ with
    $\op{S}_{\op{zzi}}^{(f)}\ket{f} = S_i(f)\ket{f}$.
    If the environment states form a continuous and dense space we can replace in \eq{order_param_def_BIS} the sum in the trace by an integral, this is

        \begin{equation}\label{order_param_def_int}
        \begin{split}
        S_{zz} = \int df\,S_i(f)\,\bra{f}\rho_{L(eq)}^{(f)}\ket{f} = \int dS_i\,S_i\,p^{\{S\}}_i\left(S_i\right).
        \end{split}
        \end{equation}
    In \eq{order_param_def_int}, we changed the state integration variables $df$ by the eigenvalue integration variables $dS_i$ and
    we defined the function
        \begin{equation}\label{Func_Deco_dist_S}
        \begin{split}
        p^{\{S\}}_i\left(S_i\right) = \int dE_f \,\rho_{L(eq)}(E_f)\Bigg[\int df^{E} \int df \;\abs{k_{f,f^{E}}}^2\Bigg]^{E_f}_{S_i},
        \end{split}
        \end{equation}
    where $k_{f,f^{E}} \equiv \braket{f}{f^{E}}$ and
    \[\rho_{L(eq)}\left(E_f\right) \equiv e^{-\beta_T\,E_f}/\opc{N}_f.\]
    Besides, $\left\{\ket{f^E}\right\}$ is defined as the eigenbasis of the environment or lattice Hamiltonian $\opc{H}_L$ with the eigenvalues $E_f\left(f^E\right)$, that is $\opc{H}^{(f)}_{L}\,\ket{f^E} = E_f\left(f^E\right)\,\ket{f^E}$, and we used the closure relation $\int df^E \ket{f^E}\bra{f^E} = \op{1}^{(f)}$.
    In \eq{Func_Deco_dist_S}, the symbol $[\cdot]^{E_f}_{S_i}$ indicates that the integrals over $df^{E}$ and $df$ are calculated integrating over the states with the eigenvalues ranging from $E_f$ to $E_f + dE_f$ and from $S_i$ to $S_i + dS_i$, respectively.
    We can see that
    \[\int dS_i\,p^{\{S\}}_i\left(S_i\right) = 1,\]
    thus $p^{\{S\}}_i\left(S_i\right)$ can be interpreted as a probability distribution function of the eigenvalues $S_i$.

    If we consider an homogeneous environment for each molecule, i.e. absence of border effects, the order parameter \numeq{order_param_def_int} has the same value for different molecules, therefore
        \begin{equation}\label{order_param_homogenv_BIS}
        \begin{split}
        S_{zz} = \int dS_1\,S_1\,p^{\{S\}}_1\left(S_1\right).
        \end{split}
        \end{equation}

\section{Decoherence function for a continuous and dense lattice Hilbert space}\label{app:Deco_Fun_CDSpace}

    This appendix is dedicated to extract an expression for the decoherence function \numeq{Gi_evol_RT_evol_C_eval} under the condition that the environment states form a continuous and dense space. Accordingly, we can replace in \eq{Gi_evol_RT_evol_C_eval} the sum in the trace by an integral over the lattice space, this is

        \begin{equation}\label{Gi_evol_RT_evol_C_eval_aprox_int}
        \begin{split}
        &G_{\{\zeta,\zeta_i'\}}(t, \tau) = \int df \int dg \int dh \; k_{f,g}\,k_{g,h}\;\rho_{L(eq)}^{h,f}\\
        &\times e^{-i\,(\zeta_i^{}-\zeta_i')\Delta S_i(f)\,t} \, e^{-i\,(\zeta_i^{}-\zeta_i')C^{SL}_{i\,\zeta}(h)\,t^2/2}\\
        &\times e^{-i\,(\zeta_i^{}-\zeta_i')C^L_i(g)\,\left\{\left(t + \tau\right)^2 - \left[1+(\kappa+1)^{-1}\right] \, \tau^2 \right\}/2},
        \end{split}
        \end{equation}
    where $\tau = (\kappa+1)\,t_1$ is the total time under reversion, $k_{f,g} \equiv \braket{f}{g}$, $k_{g,h} \equiv \braket{g}{h}$ and $\rho_{L(eq)}^{h,f} \equiv \bra{h}\rho_{L(eq)}^{(f)}\ket{f}$.
    We have used in \numeq{Gi_evol_RT_evol_C_eval_aprox_int} the eigenbasis $\left\{\ket{f}\right\}$, $\left\{\ket{g}\right\}$ and $\left\{\ket{h}\right\}$ of the Hermitian operators $\opc{H}^{(f)}_{SLi}$, $i\op{C}^{(f)}_{i,L}$ and $i\op{C}^{(f)}_{i,SL}(\zeta)$, respectively,
    with $\opc{H}^{(f)}_{SLi}\ket{f} = \Delta S_i(f)\ket{f}$, $\Delta S_i(f) = S_i(f) - S_{zz}$, $i\op{C}^{(f)}_{i,L}\ket{g} = C^L_i(g)\ket{g}$ and $i\op{C}^{(f)}_{i,SL}(\zeta)\ket{h} = C^{SL}_{i\,\zeta}(h)\ket{h}$, where $\Delta S_i(f)$, $C^L_i(g)$ and $C^{SL}_{i\,\zeta}(h)$ are real numbers. Also, we used the closure relations
    $\int dg \ket{g}\bra{g} = \op{1}^{(f)}$ and $\int dh \ket{h}\bra{h} = \op{1}^{(f)}$.
    Now, defining the eigenbasis of the environment or lattice Hamiltonian $\opc{H}_L$ as $\left\{\ket{f^E}\right\}$, with the eigenvalues $E_f\left(f^E\right)$, where $\opc{H}^{(f)}_{L}\,\ket{f^E} = E_f\left(f^E\right)\,\ket{f^E}$ and using the closure relation
    $\int df^E \ket{f^E}\bra{f^E} = \op{1}^{(f)}$, we can write:

        \begin{equation}\label{rho_L_elem_int}
        \rho_{L(eq)}^{h,f} = \int df^E \;k_{h,f^E}\;k_{f^E,f}\; \rho_{L(eq)}\left(f^E\right),
        \end{equation}
    where $k_{h,f^{E}} \equiv \braket{h}{f^{E}}$, $k_{f^{E},f} \equiv \braket{f^{E}}{f}$, and
        \[\rho_{L(eq)}\left(f^E\right) \equiv \bra{f^E}\rho_{L(eq)}^{(f)}\ket{f^E} \equiv  e^{-\beta_T\,E_f\left(f^E\right)}/\opc{N}_f.\]

    The final decoherence expression is obtained changing in \eq{Gi_evol_RT_evol_C_eval_aprox_int} the state integration variables $df$, $dg$ and $dh$, by the eigenvalue integration variables $d\Delta S_i$, $dC^L_i$ and $dC^{SL}_{i\,\zeta}$, thus we can finally write

        \begin{equation}\label{Gi_evol_RT_evol_C_eval_aprox_int_eig_BIS}
        \begin{split}
        &G_{\{\zeta,\zeta_i'\}}(t, \tau) = \int d\Delta S_i \; e^{-i\,(\zeta_i^{}-\zeta_i')\Delta S_i\,t}\\
        &\quad\times \int dC^L_i \; e^{-i\,(\zeta_i^{}-\zeta_i')C^L_i\,\left\{\left(t + \tau\right)^2 - \left[1+(\kappa+1)^{-1}\right] \, \tau^2 \right\}/2}\\
        &\quad\times \int dC^{SL}_{i\,\zeta} \; e^{-i\,(\zeta_i^{}-\zeta_i')C^{SL}_{i\,\zeta}\,t^2/2} \; p_i(\Delta S_i, C^L_i, C^{SL}_{i\,\zeta}),
        \end{split}
        \end{equation}
    where it is defined the function
        \begin{equation}\label{Func_Deco_dist_S_CL_CSL}
        \begin{split}
        &p_i(\Delta S_i, C^L_i, C^{SL}_{i\,\zeta}) = \int dE_f \,
        \rho_{L(eq)}(E_f)\,\Bigg[\int df^{E}\\
        &\;\times \int df \int dg \int dh \;k_{f,g}\;k_{g,h}\;k_{h,f^{E}}\;k_{f^{E},f}\Bigg]^{E_f, \Delta S_i}_{C^L_i, C^{SL}_{i\,\zeta}},
        \end{split}
        \end{equation}
    and $[\cdot]^{E_f, \Delta S_i}_{C^L_i, C^{SL}_{i\,\zeta}}$ indicates that the integrals over $df^{E}$, $df$, $dg$ and $dh$ are calculated integrating over the states with the eigenvalues ranging from $E_f$ to $E_f + dE_f$, from $\Delta S_i$ to $\Delta S_i + d\Delta S_i$,
    from $C^L_i$ to $C^L_i + dC^L_i$ and from $C^{SL}_{i\,\zeta}$ to $C^{SL}_{i\,\zeta} + dC^{SL}_{i\,\zeta}$, respectively.
    Under the condition $\iiint \prod_{D,n}\left\{\cdot\right\}_{t=0}$ (see \eq{Gi_evol_RT_evol_C_eig_complete} for a definition of this condition), the decoherence function \numeq{Gi_evol_RT_evol_C_eval_aprox_int_eig_BIS} has the form

        \begin{equation}\label{Gi_evol_RT_evol_C_eval_aprox_BIS}
        \begin{split}
        &G_{\{\zeta,\zeta_i'\}}(t, \tau) \simeq \int d\Delta S_i \; e^{-i\,(\zeta_i^{}-\zeta_i')\Delta S_i\,t}\\
        &\quad\times \int dC^L_i \; e^{i\,(\zeta_i^{}-\zeta_i')C^L_i\,\tau^2/[2(\kappa+1)]} \; p_i(\Delta S_i, C^L_i),
        \end{split}
        \end{equation}
    where it is defined the function
        \begin{equation}\label{Func_Deco_dist_S_CL}
        \begin{split}
        &p_i(\Delta S_i, C^L_i) \equiv \int dC^{SL}_{i\,\zeta} \; p_i(\Delta S_i, C^L_i, C^{SL}_{i\,\zeta})\\
        &\;\; = \int dE_f \, \rho_{L(eq)}(E_f)\\
        &\quad\times \Bigg[\int df^{E} \int df \int dg \;k_{f,g}\;k_{g,f^{E}}\;k_{f^{E},f}\Bigg]^{E_f, \Delta S_i}_{C^L_i},
        \end{split}
        \end{equation}
    and $[\cdot]^{E_f, \Delta S_i}_{C^L_i}$ indicates that the integrals over $df^{E}$, $df$ and $dg$ are calculated integrating over the states with the eigenvalues ranging from $E_f$ to $E_f + dE_f$, from $\Delta S_i$ to $\Delta S_i + d\Delta S_i$ and from $C^L_i$ to $C^L_i + dC^L_i$, respectively.

\section{Fourier transform of the self-spin coherences and the AQD contributions to the dynamics}\label{app:FT_FIDrev}

    In the following, we will calculate the Fourier transform of the self-spin coherences and the AQD contributions to the FID signal under reversion used in \eq{Val_Obs_MREV_FID_FT}.
    First, we use that the Fourier transform of a complex exponential function is a Dirac delta function,
        \begin{equation}\label{FT_expoc}
        \opc{F}_t\left\{e^{-i\,(\zeta_1^{}-\zeta_1') S_{zz} t}\right\}(\omega) = 2\pi\,\delta\left[\omega-(\zeta_1'-\zeta_1^{})S_{zz}\right].
        \end{equation}
    Next, we will calculate the Fourier transform of the AQD function, that is
        \begin{equation}\label{FT_AQD_int}
        \begin{split}
        &\opc{F}_t\left\{G_{\{\zeta_1^{},\zeta_1'\}}(t)\right\}(\omega) = \int^{\infty}_{-\infty} dt\; e^{-i\,\omega\,t}\\
        &\qquad\times \int^{\infty}_{-\infty} d\Delta S_1 \; e^{-i\,(\zeta_1^{}-\zeta_1')\,\Delta S_1\,t} \; p^{\{\Delta S\}}_1\left(\Delta S_1\right)\\
        &= 2\pi \int^{\infty}_{-\infty} d\Delta S_1 \; \delta\left[\omega-(\zeta_1'-\zeta_1^{})\Delta S_1\right] \; p^{\{\Delta S\}}_1\left(\Delta S_1\right),
        \end{split}
        \end{equation}
    where we used that
        \[\delta\left[\omega-(\zeta_1'-\zeta_1^{})\Delta S_1\right] = \frac{1}{2\pi}\int^{\infty}_{-\infty} dt\; e^{-i\,\left[\omega-(\zeta_1'-\zeta_1^{})\Delta S_1\right]\,t}.\]
    Using in \eq{FT_AQD_int} the delta function property
        \[\delta\left[\omega-(\zeta_1'-\zeta_1^{})\Delta S_1\right] = \frac{1}{\abs{\zeta_1' - \zeta_1^{}}}\,\delta\left[\Delta S_1-\omega/(\zeta_1'-\zeta_1^{})\right],\]
    we have the following result for the Fourier transform of the AQD function
        \begin{equation}\label{FT_AQD}
        \opc{F}_t\left\{G_{\{\zeta_1^{},\zeta_1'\}}(t)\right\}(\omega) = \frac{2\pi}{\abs{\zeta_1' - \zeta_1^{}}}\,p^{\{\Delta S\}}_1\left(\frac{\omega}{\zeta_1' - \zeta_1^{}}\right).
        \end{equation}
    The last step for obtaining the Fourier transform of the self-spin coherences and the AQD contribution is to calculate the Fourier transform of the product of a complex exponential function with the AQD function, which can be calculated using the convolution\cite{OppenheimBook} of the Fourier transforms \numeq{FT_expoc} and \numeq{FT_AQD}, as follow
        \begin{equation}\label{FT_AQD_int_result}
        \begin{split}
        &\opc{F}_t\left\{e^{-i\,(\zeta_1^{}-\zeta_1')\,S_{zz}\,t}\,G_{\zeta_1^{},\zeta_1'}(t)\right\}(\omega)\\
        &=\frac{1}{2\pi}\,\opc{F}_t\left\{e^{-i\,(\zeta_1^{}-\zeta_1')\,S_{zz}\,t}\right\}(\omega)\ast\opc{F}_t\left\{G_{\{\zeta_1^{},\zeta_1'\}}(t)\right\}(\omega)\\
        &=\frac{2\pi}{\abs{\zeta_1' - \zeta_1^{}}} \int^{\infty}_{-\infty} d\omega'\; \delta\left[\omega-\omega'-(\zeta_1'-\zeta_1^{})S_{zz}\,\right]\\ &\qquad\qquad\qquad\qquad\times p^{\{\Delta S\}}_1\left(\frac{\omega'}{\zeta_1' - \zeta_1^{}}\right)\\
        &=\frac{2\pi}{\abs{\zeta_1' - \zeta_1^{}}}\,p^{\{\Delta S\}}_1\left(\frac{\omega-(\zeta_1'-\zeta_1^{})S_{zz}}{\zeta_1' - \zeta_1^{}}\right).
        \end{split}
        \end{equation}
    where $A(\omega) \ast B(\omega)$ is the convolution between the functions $A$ and $B$, and we used in \numeq{FT_AQD_int_result} that
        \[\delta\left[\omega-\omega'-(\zeta_1'-\zeta_1^{})S_{zz}\,\right] = \delta\left\{\omega'-\left[\omega-(\zeta_1'-\zeta_1^{})S_{zz}\,\right]\right\}.\]

\section{Relationship between the decoherence function and the distribution probability function}\label{app:Rel_Deco_Dist_Funs}

    In order to gain insight into the relationship between the decoherence function and the distribution probability function $p_i$, we use that, as shown in \eq{Gi_evol_RT_evol_C_eig_complete}, the decoherence function is the result of a superposition of complex exponential functions weighted by a distribution of their frequencies.
    Such kind of superposition can be generically expressed as follows:

        \begin{equation}\label{G_genera_int}
        \opc{P}_i(\Delta \zeta_i, t,\tau) = \int dA_i \; e^{-i\,\Delta \zeta_i\,\opc{Y}_{A_i}(t,\tau)\,A_i} \; p_i(A_i,\Delta \zeta_i,t,\tau),
        \end{equation}
    with  $\Delta \zeta_i \equiv \zeta_i^{}-\zeta_i'$ and $\opc{Y}_{A_i}(t,\tau)$ is some polynomial in $t$ and $\tau$, the distribution of frequencies of the complex exponential is determined by $p_i(A_i,\Delta \zeta_i,t,\tau)$.
    The expression  \numeq{G_genera_int} is a general form for the different integrals in \eq{Gi_evol_RT_evol_C_eig_complete}. Therefore, we can use \numeq{G_genera_int} to extract conclusions about \numeq{Gi_evol_RT_evol_C_eig_complete}.

    We consider that the spin system is coupled with an environment whose states belong to a continuous and dense Hilbert space, where a distribution of the eigenvalues of each complex exponential in \eq{Gi_evol_RT_evol_C_eig_complete} and \eq{G_genera_int} can be defined. Besides, we suppose that such distribution have a general bell-shape form around the mean value of the eigenvalues, converging to zero fast enough so that
    integrations in \eq{Gi_evol_RT_evol_C_eig_complete} and \eq{G_genera_int} can be extended to $\pm\infty$.

    Accordingly, we can observe that \eq{G_genera_int} is the Fourier transform on the variable $A_i$ of the function $p_i(A_i,\Delta \zeta_i,t,\tau)$ valued in $\Delta \zeta_i\,\opc{Y}_{A_i}(t,\tau)$, that is

        \begin{equation}\label{G_genera_int_FT}
        \begin{split}
        &\opc{P}_i(\Delta \zeta_i, t,\tau) = \left[\opc{F}_{A_i}\left\{p_i(A_i,\Delta \zeta_i,t,\tau)\right\}(\alpha)\right]_{\alpha = \Delta \zeta_i\,\opc{Y}_{A_i}(t,\tau)}\\
        &\;= \left[\int^{+\infty}_{-\infty} dA_i \; e^{-i\,\alpha\,A_i} \; p_i(A_i,\Delta \zeta_i,t,\tau)\right]_{\alpha = \Delta \zeta_i\,\opc{Y}_{A_i}(t,\tau)}.
        \end{split}
        \end{equation}
    Therefore, \eq{Gi_evol_RT_evol_C_eig_complete} is the Fourier transform over all the variables of $p_i$, which is valuated in some polynomial in $t$ and/or $\tau$ multiplied by $\Delta \zeta_i$ for each variable.\\

\section*{Caption in figures}

\insertfig{h}{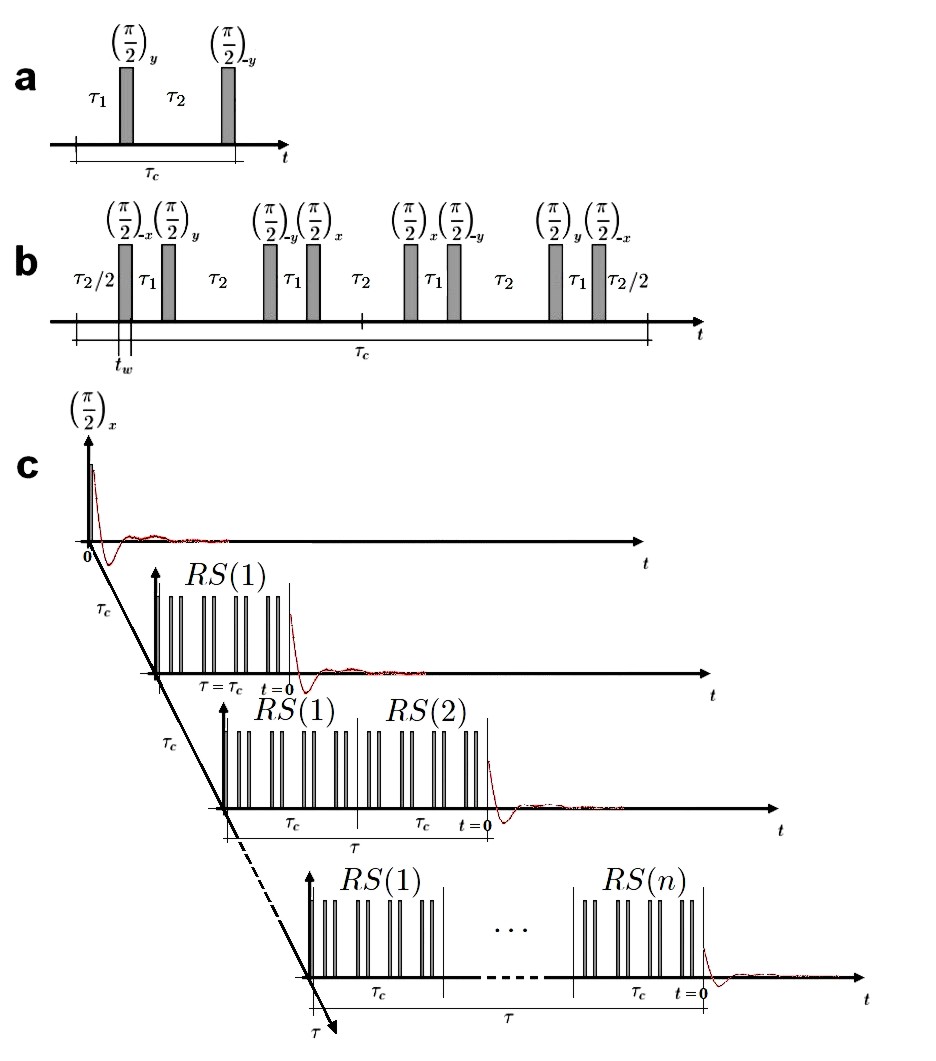}{8.85}{9.98}{Experimental pulse sequences. a: Single reversion. b: MREV8. c: Compound reversion sequence, where each reversion block $RS(i)$ constitutes a single reversion sequence like the MREV8.}

\insertfig{c}{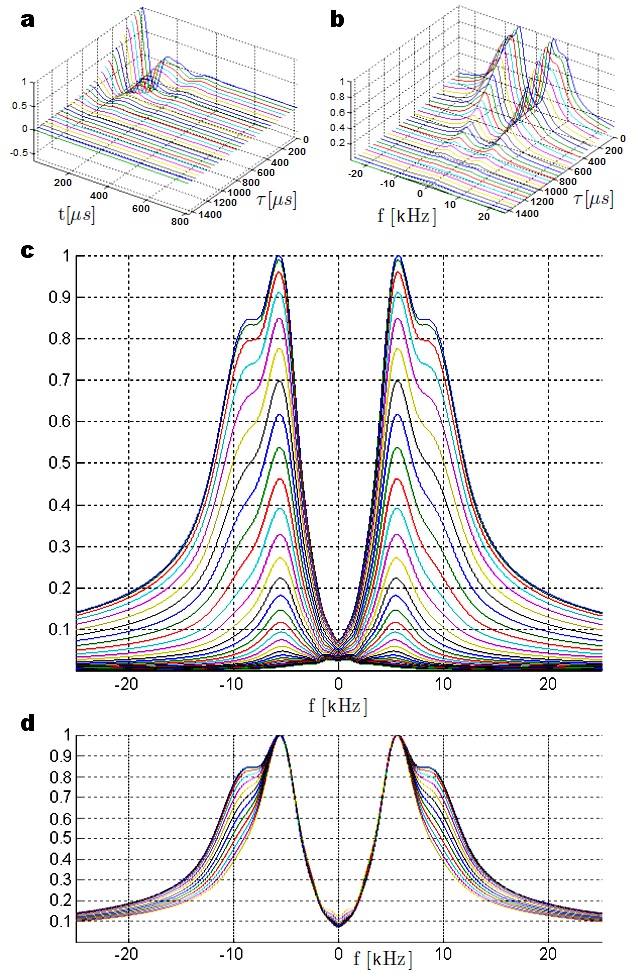}{8.89}{13.73}{Calculation of the FID signals under spin reversion dynamics, using MREV8 pulse blocks, for the nematic 5CB molecule in the resonance condition. Decoherence is represented with gaussian profiles in $t$ and $\tau$.
Parameters: $S_{zz} = 0.5403$, $\sigma_t = 0.07$ (standard deviation in t), $\sigma_{\tau} = 0.04$ (standard deviation in $\tau$).
a: FID signals, depending on time $t$, as a function of the reversion time $\tau$.
b: Amplitude spectra for the results in (a).
c: Frontal detail of the spectra shown in (b).
d: Spectra of (c) normalized to compare the evolution of their frequency components (for clarity only the spectra for the shorter $\tau$ values are shown).
It is observed in (d) the different decay rates of a set of spectral lines close to
5.65kHz and 8.50kHz as well as the spectral compression which are evidences of the eigen-selectivity in the decoherence process.}

\insertfig{h}{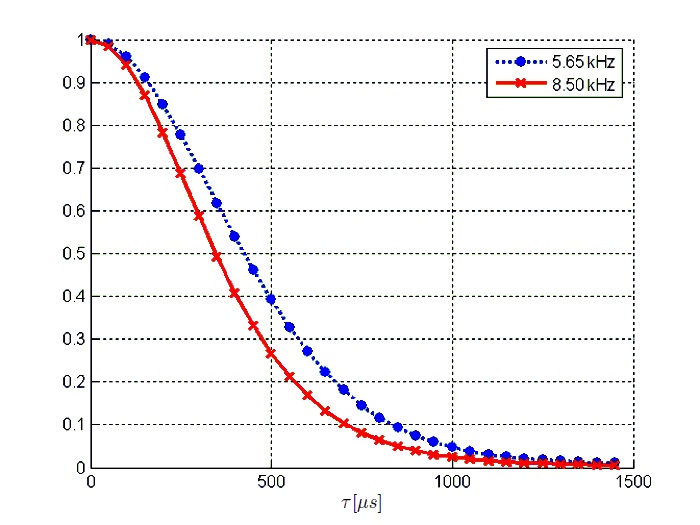}{8.89}{6.68}{Calculated evolution of the normalized amplitude as a function of the reversion time $\tau$ for the spectral lines at 5.65kHz and 8.50kHz of the spectra shown in \fig{FIG2} (b) and (c) for the nematic 5CB.}

\insertfig{h}{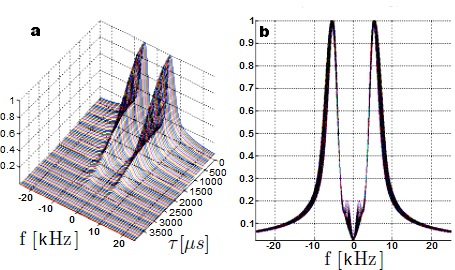}{8.89}{5.28}{Calculation of the FID signals under spin reversion dynamics, using MREV8 pulse blocks, for the nematic PAA$_{d6}$ molecule in the resonance condition. Decoherence is represented with gaussian profiles in $t$ and $\tau$.
Parameters: $S_{zz} = 0.53$, $\sigma_t = 0.06$ (standard deviation in t), $\sigma_{\tau} = 0.015$ (standard deviation in $\tau$).
a: Amplitude spectra for the calculated FID's.
b: Spectra of (a) normalized to compare the evolution of their frequency components (for clarity only the spectra for the shorter $\tau$ values are shown).
It is observed in (b) the spectral compression which is evidence of the eigen-selectivity in the decoherence process.}

\insertfig{h}{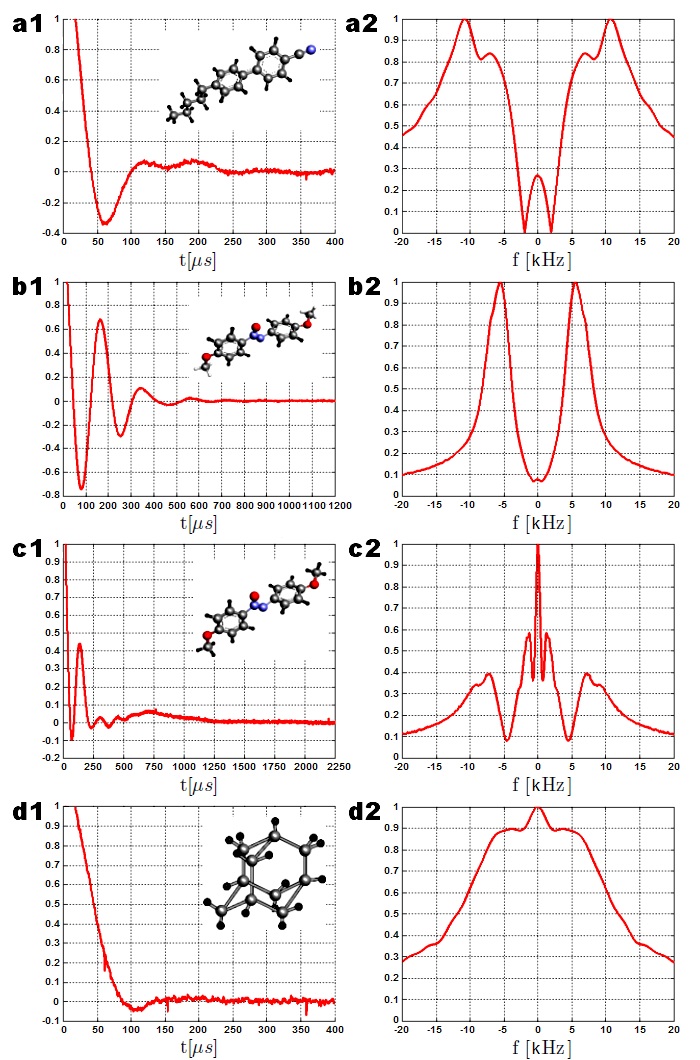}{8.89}{13.46}{Molecular sketch (inner detail in the figures), FID evolution (index 1 in the figures) and the FID spectrum (index 2 in the figures) measured for nematic 5CB (a1,a2) at room temperature (namely T = 27$^o$C), PAA$_{d6}$ (b1,b2) at T = 115$^o$C, PAA (c1,c2) at T = 110$^o$C, and solid adamantane (d1,d2).}

\insertfig{h}{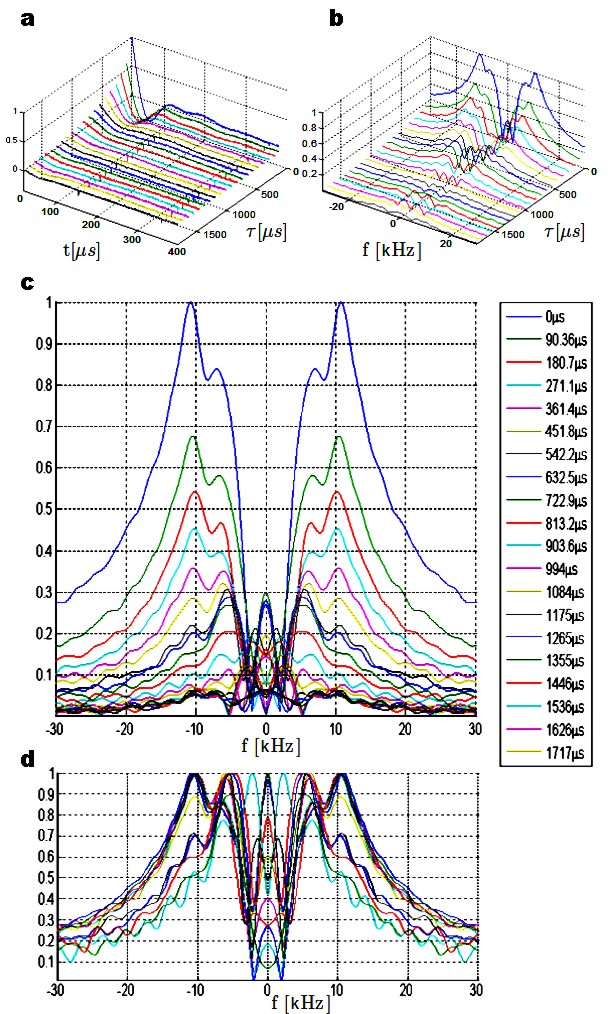}{8.89}{14.72}{Experimental results of the FID signals under spin reversion dynamics, using MREV8 pulse blocks, for the nematic 5CB at room temperature (namely T = 27$^o$C) in the resonance condition.
a: Measured signals for the FID experiments, depending on the time $t$, as a function of the reversion time $\tau$.
b: Amplitude spectra for the measurements shown in (a).
c: Frontal detail of the spectra shown in (b).
d: Spectra of (c) normalized to compare the evolution of their frequency components (for clarity only the spectra for the shorter $\tau$ values are shown). In the center-right box is detailed the time $\tau$ for each spectrum.}

\insertfig{h}{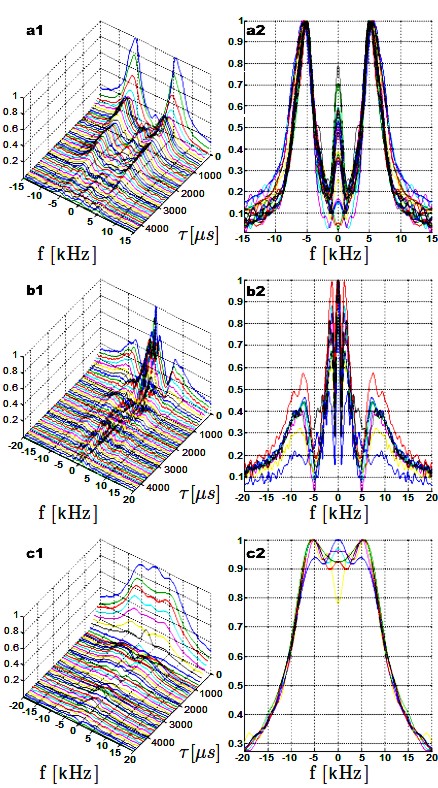}{8.89}{16}{Experimental results of the FID signals under spin reversion dynamics, using MREV8 pulse blocks, in the resonance condition.
The figures with index 1 show the amplitude spectra for the FID measurements and the ones with index 2 show the measured FID spectra, which are normalized to compare the evolution of their frequency components (for clarity only the spectra for the shorter $\tau$ values are shown).
a1,a2: PAA$_{d6}$ at T=115$^o$C. b1,b2: PAA at T=110$^o$C. c1,c2: Solid adamantane.}

\insertfig{h}{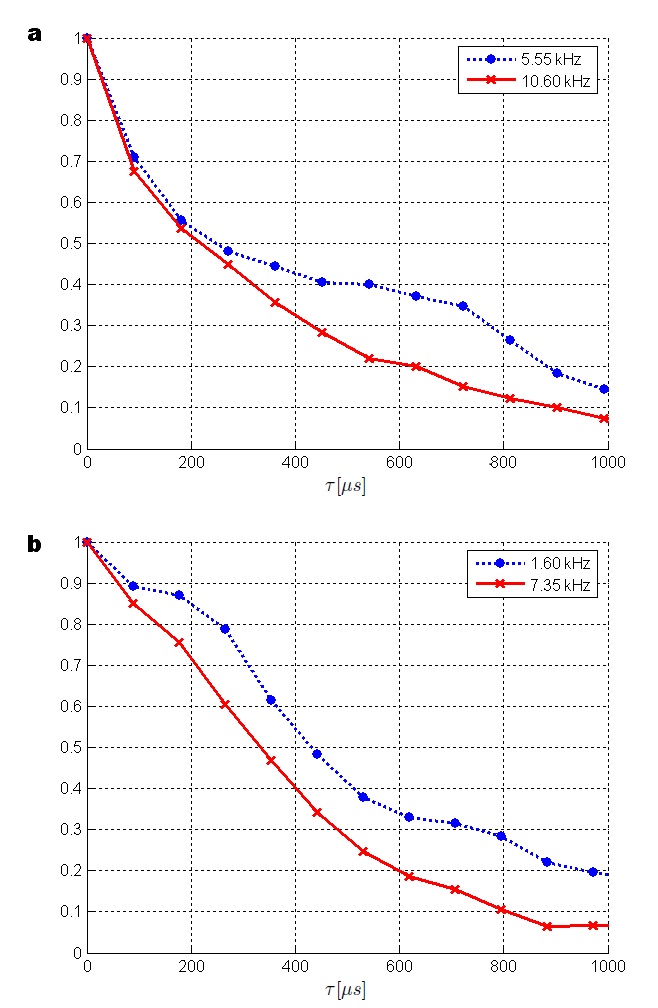}{8.89}{13.61}{Measured evolution of the normalized amplitude as a function of the reversion time $\tau$ for different spectral lines of the FID spectra.
a: Spectral lines at 5.55kHz and 10.60kHz for the spectra shown in \figs{FIG6} (b) and (c) for nematic 5CB at room temperature (namely T = 27$^o$C).
b: Spectral lines at 1.60kHz and 7.35kHz for the spectra shown in \fig{FIG7} (b1) for the PAA at T = 110$^o$C.}

\insertfig{c}{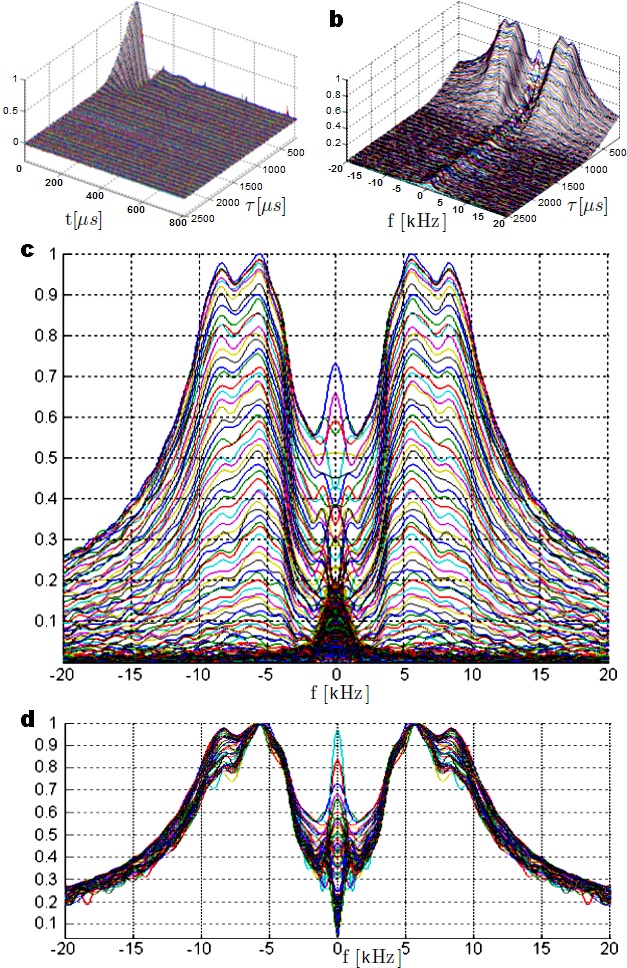}{8.89}{13.73}{Experimental results of the FID signals under spin reversion dynamics, using a single MREV8 pulse sequence with a continuous variation of its setting times, for nematic 5CB at room temperature (namely T = 27$^o$C) in the resonance condition.
a: Measured signals for the FID experiments, depending on the time $t$, as a function of the reversion time $\tau$.
b: Amplitude spectra for the measurements shown in (a).
c: Frontal detail of the spectra shown in (b).
d: Spectra of (c) normalized to compare the evolution of their frequency components (for clarity only the spectra for the shorter $\tau$ values are shown).}

\insertfig{c}{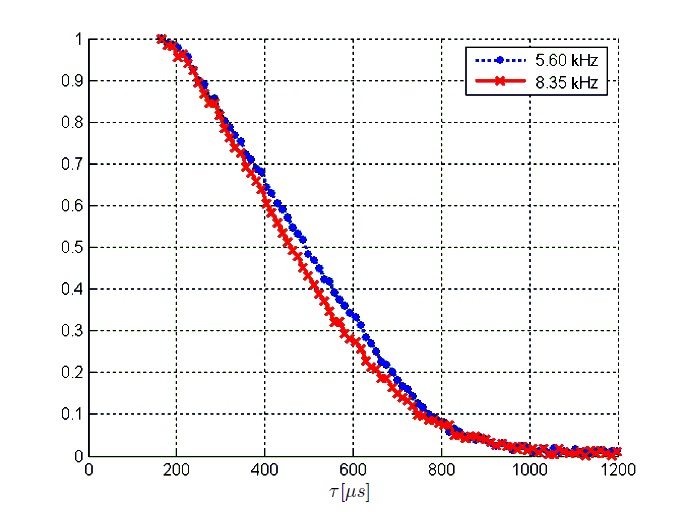}{8.99}{6.67}{Measured evolution of the normalized amplitude as a function of the reversion time $\tau$ for the spectral lines at 5.60kHz and 8.35kHz for the spectra shown in \figs{FIG9} (b) and (c) for the nematic 5CB at room temperature (namely T = 27$^o$C).
It can be seen the eigen-selectivity where the higher the frequency of the line is, the faster its decay will be.
The continuous variation of the setting time allows to see clearly the bell-like form for the decay of the spectral lines.}

\insertfig{h}{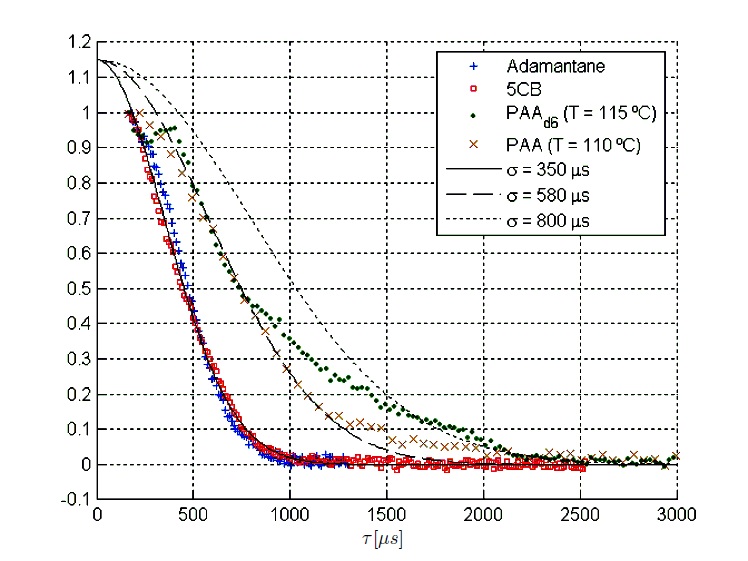}{9.5}{7.13}{Experimental results of the normalized evolution of the maximum of the FID signals under spin reversion dynamics, using a single MREV8 pulse sequence with a continuous variation of its setting times, for adamantane and 5CB at room temperature (namely T = 27$^o$C), PAA$_{d6}$ at T = 115$^o$C and PAA at T = 110$^o$C.
Three gaussian decays with different standard deviation, $\sigma$, are plotted for comparison with the measurements.}

\insertfig{h}{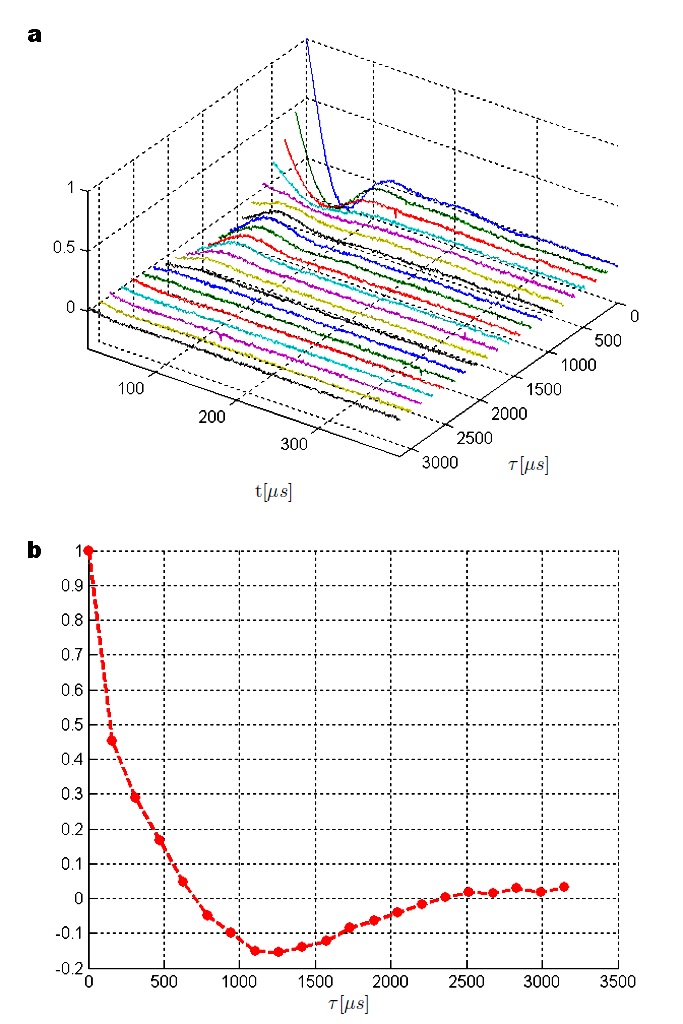}{8.89}{13.34}{Experimental results with a misadjustment in the reversion parameter $\kappa$ of the FID signals under spin reversion dynamics, using MREV8 pulse blocks, in the resonance condition for nematic 5CB at room temperature (namely T = 27$^o$C).
a: Measured signals for the FID experiments, depending on the time $t$, as a function of the reversion time $\tau$.
b: Pseudo-FID obtained for the evolutions in $\tau$ of the FID-signal values in (a) at $t=18\mu s$.
Such pseudo-FID presents a signal form similar to the FID signal, but with lower characteristic frequencies.}

\end{document}


\title{Supplementary Material\\
    Quantum irreversible decoherence behaviour in open quantum systems with few degrees of freedom. Application to $^1$H NMR reversion experiments in nematic liquid crystals.}
    \author{H. H. Segnorile and  R. C. Zamar}
    \affiliation{Instituto de F\'isica Enrique Gaviola - CONICET,
    Facultad de Matem\'atica, Astronom\'{\i}a y F\'{\i}sica,
    Universidad Nacional de C\'ordoba \\ M.Allende y H. de la Torre - Ciudad Universitaria,
    X5016LAE - C\'ordoba, Argentina}
    \date{\today}

\maketitle

The material below is intended as supplementary material to the article \textbf{Quantum irreversible decoherence behaviour in open quantum systems with few degrees of freedom. Application to $^1$H NMR reversion experiments in nematic liquid crystals}. Such manuscript will be referred to as QD-II in the following text.\\

In \sect{se:Field_effect} we analyse analytically and experimentally the time scale of the effects produced by the inhomogeneity of the strong magnetic field on the signals obtained from spin observables in liquid crystals (LC).
\sect{se:Misadj_rev} is dedicated to estimate the effects of misadjustments of the characteristic times of the reversion pulse sequence, as well as the influence of the non-secular parts of the dipolar Hamiltonian in such experiments.
Hereinafter, the Hamiltonians will be expressed in units of $\hbar$.

\section{Effects of the field inhomogeneity}\label{se:Field_effect}

    This section is concerned with the analysis of the field inhomogeneity effects on the experimental measurements presented in QD-II.
    The field inhomogeneity causes the decay of the NMR signals due to a distribution of the $\op{\hat{z}}$ component of the static magnetic field. Therefore, it is important to estimate the characteristic behaviour and time scale of that decay in order to compare with the decay produced by decoherence.\\

    To observe the effects of the field inhomogeneity in the nematic phase, we first study the FID signals obtained by applying the Zeeman reversion sequence shown in \fig{Paper_Fig_SecPulsos_sup} (a).
    In a nematic LC, the static magnetic field does not vary significantly inside each molecule but there is a distribution of the field strength along the whole sample. Such distribution produces a dispersion of the Larmor frequency in the sample.
    Therefore, the evolution operator due to the field inhomogeneity for a single molecule in the `on resonance' condition will be
    \begin{equation}\label{Zeeman_inh_Op}
    \op{\widehat{U}}^{Z}(t) = e^{-i\Delta\op{I}_{\op{z}}t},
    \end{equation}
    where $\Delta = \omega-\omega_0$. The frequencies $\omega = \gamma B$ and $\omega_0 = \gamma B_0$, with $\gamma$ the proton gyromagnetic ratio,  are associated with the strength $B$ of the magnetic field inside the molecule and the mean value $B_0$ of the strength of the magnetic field of the sample, thus $\omega_0$ is the frequency of the rotating-frame and it defines the resonance condition.\\
    The complete dynamics in this condition for the intermediate time scale (see QD-II for details about the existence of different time scales in the dynamics) will be produced by the secular full-quantum dipolar Hamiltonian and the environment or lattice Hamiltonian, studied in QD-II, together with the Zeeman-like Hamiltonian in \eq{Zeeman_inh_Op}.
    Since the Hamiltonian in the evolution operator \numeq{Zeeman_inh_Op} commutes with the other ones, the dynamics produced by such evolution operator can be studied independently of the other evolution operators and we can readily include the effects of the inhomogeneity on the complete dynamics.\\
    Keeping the last statement in mind, we can extract the evolution operator due to the field inhomogeneity of the dynamics under the Zeeman reversion experiment shown in \fig{Paper_Fig_SecPulsos_sup} (a) for a single molecule and then to conclude how it affects the complete dynamics.
    Such operator is written as
    \begin{equation}\label{Zeeman_inh_rev_Op}
    \op{\widehat{U}}^{Z}_{RT}(\tau) = \op{R}^{\dag\pi}_{\op{x}}\,e^{-i\Delta\op{I}_{\op{z}}\tau/2}\,\op{R}^{\pi}_{\op{x}}\,e^{-i\Delta\op{I}_{\op{z}}\tau/2} = \op{1},
    \end{equation}
    where the operator $\op{R}^{\pi}_{\op{x}} \equiv e^{i\,\op{I}_{\op{x}}\pi}$ is defined for a $\pi$ pulse in the direction $\op{\hat{x}}$, and we also used $\op{R}^{\dag\pi}_{\op{x}}\op{I}_{\op{z}}\op{R}^{\pi}_{\op{x}} = -\op{I}_{\op{z}}$.
    It is important to note that the secular dipolar Hamiltonian is invariant under $\pi$ pulses, thus the total Hamiltonian in QD-II is not altered by the sequence in \fig{Paper_Fig_SecPulsos_sup} (a).
    Therefore, the expression in \eq{Zeeman_inh_rev_Op} tells us that the field inhomogeneity effect in the whole sample can be removed and the complete dynamics will produce the corresponding signals for the observables, for instance the FID, in the same way as if the inhomogeneity would not exist for an ideal experiment.\\
    The results of such experiments in the nematic LC's 5CB and PAA$_{d6}$ are shown in \fig{Paper_Fig_FID_EcoHahn_5CB_PAAd6}.
    By comparing with the measured FID's of Figure 5 (a1) and (b1) in QD-II, it can be seen that the reversion sequence does not produce changes of the time scale, which indicates that the field inhomogeneity has not important effects on the FID time scale.
    It is worth to note that the changes observed in the profiles of \fig{Paper_Fig_FID_EcoHahn_5CB_PAAd6} are due to experimental issues, like the finite width of the $\pi$ pulses, but they do not change the time scale of the output signals nor the conclusions extracted about that.\\

    Then, we studied the effect of the field inhomogeneity on the dynamics of the pseudo-FID's obtained under the reversion pulse sequence shown in \fig{Paper_Fig_SecPulsos_sup} (c) using as $RS(i)$ blocks an MREV8-like sequence as that shown in \fig{Paper_Fig_SecPulsos_sup} (b).
    The aim of such reversion experiment, in our work, is to remove the dynamics produced by the dipolar Hamiltonian.
    Nevertheless, the environmental decoherence (which is studied in QD-II) and the non-ideal effects or misadjustments in the reversion dynamics (which are studied in \sect{se:Misadj_rev} in this supplementary material) prevent the complete removal of the dipolar dynamics, thus introducing a final decay of the signals of the measured observables.
    The field inhomogeneity will introduce in this time scale an extra dynamics produced by the Zeeman-like Hamiltonian \numeq{Zeeman_inh_Op}, and this dynamics will be observed as an extra decay in the measured signals (as we will see), which will overlap with the other mentioned effects.

    To extract the time scale of the decay due to the field inhomogeneity under the mentioned reversion sequence, we performed the experiment described in \fig{Paper_Fig_SecPulsos_sup} (c) in a 5CB sample in the isotropic phase. In such phase, the dynamics is governed by the Zeeman-like Hamiltonian of the inhomogeneity, which is the same Hamiltonian as in the nematic phase, but here the dipolar Hamiltonian is averaged out.
    In the longest time scale, the thermalization, which is studied by relaxation theories, will produce a final decay of the signals, but its time scale is much longer than that of the inhomogeneity.
    Therefore, the fast decays observed in the pseudo-FID's under the reversion dynamics in the isotropic phase can be associated mainly to the field inhomogeneity.

    In the same way that in the former experiment, where it is obtained the evolution operator \numeq{Zeeman_inh_rev_Op}, we first calculate the evolution operator due to the field inhomogeneity for the sequence shown in \fig{Paper_Fig_SecPulsos_sup} (b), with $\tau_2 = 2\tau_1$,  for a single molecule.
    Such evolution operator is written as

    \begin{equation}\label{MREV8_Zeeman_rev_Op}
    \begin{split}
    &\op{\widehat{U}}^{Z}_{RS}(\tau_1) = e^{-i\Delta\op{I}_{\op{z}}\tau_1}\,e^{i\Delta\op{I}_{\op{y}}\tau_1}\,e^{i\Delta\op{I}_{\op{x}}2\tau_1}\,
    e^{i\Delta\op{I}_{\op{y}}\tau_1}\\
    &\quad\times e^{-i\Delta\op{I}_{\op{z}}2\tau_1}\,e^{-i\Delta\op{I}_{\op{y}}\tau_1}\,
    e^{-i\Delta\op{I}_{\op{x}}2\tau_1}\,e^{-i\Delta\op{I}_{\op{y}}\tau_1}\,e^{-i\Delta\op{I}_{\op{z}}\tau_1}.
    \end{split}
    \end{equation}
    It can be observed in \eq{MREV8_Zeeman_rev_Op} that there are exponential operators with the same exponent and with opposite sign, thus it will produce a cancellation effect of their dynamics.
    For small enough values of the time $\tau_1$, the operator \numeq{MREV8_Zeeman_rev_Op} can be approximated as

    \begin{equation}\label{MREV8_Zeeman_rev_Op_approx}
    \op{\widehat{U}}^{Z}_{RS}(\tau) \simeq e^{-i\Delta\op{I}_{\op{z}}\tau/3},
    \end{equation}
    where $\tau = 12\tau_1$ for a single MREV8 block.
    A reversion experiment with n blocks of the MREV8 pulse sequence will have for the field inhomogeneity a similar evolution operator as in \eq{MREV8_Zeeman_rev_Op_approx}, in the approximation of small times, with $\tau = 12 n \tau_1$.

    Comparing \eq{MREV8_Zeeman_rev_Op_approx} with \eq{Zeeman_inh_Op}, we can see that the dynamics under reversion using MREV8 blocks is a third slower than the free-evolution dynamics, thus the reversion experiment mitigates the decay effect over the signals due to the field inhomogeneity.\\

    Next, using the evolution operators for a free-evolution dynamics, \eq{Zeeman_inh_Op}, and for a reversion dynamics, \eq{MREV8_Zeeman_rev_Op_approx}, we obtain the signals of the whole sample of an isotropic LC affected by a field inhomogeneity as

    \begin{equation}\label{Val_Obs_MREV_FID_iso_inh_sum}
    \begin{split}
    \left<\widehat{\op{I}}_{\op{y}}(t, \tau)\right>^{iso} &=  \frac{\beta_T\,\omega_0}{\opc{N}_{S_1}} \sum_i\sum_{\zeta_i^{}s_i^{},\zeta_i's_i'}
    \abs{\bra{\zeta_i^{}s_i^{}}\op{I}_{\op{y}}^{(s_i)}\ket{\zeta_i's_i'}}^2\\
    &\qquad\qquad\times e^{\pm(\zeta_i^{}s_i^{},\zeta_i's_i')\,i\Delta_i (t+\tau/3)},
    \end{split}
    \end{equation}
    where the index $i$ labels the $i$-th molecule (the sum runs over the molecules of the whole sample), $\{\ket{\zeta_i s_i}\}$ is a base in the Hilbert space of the spins of the $i$-th molecule which are eigenstates of the spin part of the interaction dipolar Hamiltonian.
    Since the dipolar  and the Zeeman Hamiltonians commute, such base can be selected as a common eigenbase of the Zeeman Hamiltonian as well.
    In \eq{Val_Obs_MREV_FID_iso_inh_sum}, the symbol $\pm(\zeta_i^{}s_i^{},\zeta_i's_i')$ represents the possible values $+1$ or $-1$ depending on the difference in the Zeeman eigenvalues of the states $\ket{\zeta_i^{}s_i^{}}$ and $\ket{\zeta_i's_i'}$, such difference can only have the values $\pm1$ due that the non-null matrix elements of the operator $\op{I}_{\op{y}}^{(s_i)}$ connect eigenstates with such difference of Zeeman eigenvalues.
    Besides, we defined $\opc{N}_{S_1} \equiv tr_{s_1}\left\{\op{1}^{(s_1)}\right\}$ and $\beta_T \equiv \frac{1}{k_B T}$, with $k_B$ the Boltzmann constant and $T$ the absolute temperature.\\
    The matrix elements $\bra{\zeta_i^{}s_i^{}}\op{I}_{\op{y}}^{(s_i)}\ket{\zeta_i's_i'}$ are the same for different molecules.
    We suppose a dense and continuous distribution of the Larmor frequency shifts $\Delta$ for the molecules (which is realistic for a typical field inhomogeneity).
    Due to this, we can replace the sum over the molecules in the sample as an integration over the values of $\Delta$. Therefore, \eq{Val_Obs_MREV_FID_iso_inh_sum} can be written as

    \begin{equation}\label{Val_Obs_MREV_FID_iso_inh}
    \begin{split}
    \left<\widehat{\op{I}}_{\op{y}}(t, \tau)\right>^{iso} &=  \frac{\beta_T\,\omega_0\,N}{\opc{N}_{S_1}} \sum_{\zeta_1^{}s_1^{},\zeta_1's_1'}
    \abs{\bra{\zeta_1^{}s_1^{}}\op{I}_{\op{y}}^{(s_1)}\ket{\zeta_1's_1'}}^2\\
    &\qquad\qquad\times G_{Z}(t+\tau/3).
    \end{split}
    \end{equation}
    In \eq{Val_Obs_MREV_FID_iso_inh} it is defined

    \begin{equation}\label{Deco_Zeeman}
    G_{Z}(t) = \int d\Delta\, e^{i\Delta t}\, p_Z(\Delta),
    \end{equation}
    where $p_Z(\Delta)$ is the distribution function of $\Delta$ and
    $p_Z(\Delta)d\Delta = n(\Delta)/N$ with $n(\Delta)$ as the number of molecules with the value of $\Delta$ and $N$ as the number of molecules of the whole sample. Given that $p_Z(\Delta)$ is supposed a symmetric bell-form distribution with a null mean value (which is realistic as well), the integration \numeq{Deco_Zeeman}, whose limits of integration can be extended to $\pm\infty$, produces the same result if the complex exponential has positive or negative sign, thus it is independent of the values of $\pm(\zeta_1^{}s_1^{},\zeta_1's_1')$ and we only keep the positive sign.
    To extract an expression for \eq{Deco_Zeeman}, we will calculate it using a Lorentzian distribution of $\Delta$, only for numerical purpose, this is

    \begin{equation}\label{prob_Lorenz_inh}
    p_Z(\Delta) = \frac{1}{\pi}\;\frac{1/t_0}{1/t_0^2+\Delta^2}.
    \end{equation}
    Using \eq{prob_Lorenz_inh} in \eq{Deco_Zeeman} we obtain

    \begin{equation}\label{Deco_Zeeman_Lorentzian}
    G_{Z}(t+\tau/3) = e^{-t/t_0}\,e^{-\tau/\tau_0}.
    \end{equation}
    with $t_0$ as the characteristic decay time and $\tau_0 = 3t_0$.\\

    In \fig{Paper_Fig_REVT8_5CB_iso} we showed the measured results for the maximum values of the FID's, i.e. the pseudo-FID obtained by the values for an averaged amplitude close to $t=0$, which are extracted by the reversion experiment in \fig{Paper_Fig_SecPulsos_sup} (c) using MREV8-like blocks of pulses as $RS(i)$ in the isotropic 5CB (at $T=37^{\circ} C$). Besides, in \fig{Paper_Fig_REVT8_5CB_iso} the obtained data are fitted with an exponential decay function with characteristic time $\tau_0 = 4.5ms$.
    Comparing with the results of the same experiment in nematic LC's, whose measurements are shown in Figure 11 in QD-II, we can see that for a reversion time of $\tau \simeq 2.5 ms$ the signals in the nematic phase are vanishing but the decay due to inhomogeneity only can produce a reduction around of the 40\% in such signals.
    Therefore, in the reversion experiment in nematic LC's the field inhomogeneity has minor effects on the signal decay and it can be neglected.
    It is worth to note that it is possible to configure a new MREV8-like sequence introducing four $\pi$ pulses between the $\pi/2$ pulses in \fig{Paper_Fig_SecPulsos_sup} (b) with the aim of removing more efficiently the dynamics due to the field inhomogeneity. Nevertheless, such sequence would double of total time under pulses which adds non-ideal effects over the measurements like those who will be commented in in \sect{se:Misadj_rev} of this supplementary material. Therefore, to avoid extra non-ideal effects and given that the decay produced by the field inhomogeneity has a longer time scale than the observed decays, we used the standard MREV8 sequence.\\

    Finally, using \eq{Val_Obs_MREV_FID_iso_inh} corresponding to the isotropic phase and the result for the FID signals under refocusing experiments on nematic LC's, which was obtained in Section II.B in QD-II, it is ready to obtain the corresponding signals for an open quantum spin system in the nematic LC's under such reversion experiments, and affected by the field inhomogeneity, as

    \begin{equation}\label{Val_Obs_MREV_FID_inh}
    \begin{split}
    &\left<\widehat{\op{I}}_{\op{y}}(t, \tau)\right> =  \frac{\beta_T\,\omega_0\,N}{\opc{N}_{S_1}} \sum_{\zeta_1^{}s_1^{},\zeta_1's_1'}
    \abs{\bra{\zeta_1^{}s_1^{}}\op{I}_{\op{y}}^{(s_1)}\ket{\zeta_1's_1'}}^2\\
    &\qquad\times e^{-i\,(\zeta_1^{}-\zeta_1')\, S_{zz}\,t}\,
    G_{\zeta_1^{},\zeta_1'}(t) \,G_{\zeta_1^{},\zeta_1'}^{\,(rt)}(\tau) \,G_{Z}(t+\tau/3).
    \end{split}
    \end{equation}
    In \eq{Val_Obs_MREV_FID_inh} $G_{\zeta_1^{},\zeta_1'}$ and $G_{\zeta_1^{},\zeta_1'}^{\,(rt)}$ are the decoherence functions for the free-evolution dynamics and the reversion dynamics, respectively.
    It is important to remark here that these decoherence functions depend on the eigenvalues $\zeta_1^{}$ and $\zeta_1'$ so imposing eigen-selectivity (see QD-II for details about eigen-selectivity). On the contrary, $G_{Z}$ does not depend on such eigenvalues thus it does not involve eigen-selectivity effects.\\

    We conclude in this section that the dynamics produced by the field inhomogeneity has a time scale longer than the decoherence one and it does not involve eigen-selectivity.

\section{Misadjustments in the dynamics under reversion}\label{se:Misadj_rev}

    In this section, we study with numerical simulations the effects of parameter misadjustments or errors in reversion experiments under the dynamics of a closed spin system.
    The aim of this study is to check for possible eigen-selectivity behaviours due to such misadjustment affecting only the spin dynamics.
    To do that, we use the eight-spin model of a PAA$_{d6}$ molecule, which was used in QD-I\cite{SegZam2011}.\\
    We simulated the dynamics under the reversion experiment of  \fig{Paper_Fig_SecPulsos_sup} (c), where each block of reversion $RS(i)$ is a sequence like the MREV8 shown in \fig{Paper_Fig_SecPulsos_sup} (b).\\
    To calculate the effect of this sequence on the spin dynamics, first we  define the dipolar spin Hamiltonian of a nematic LC molecule as\cite{SegZam2011}

    \begin{equation}\label{dipolar_sol_SI}
    \opc{\widehat{H}}_{S1} \equiv \pi \sqrt{6}\, \sum_{j \neq k} D_{j k} \, \op{T}^{jk}_{2,0},
    \end{equation}
    with the following 2nd order tensor
    \begin{equation}\label{T20}
    \begin{split}
    \op{T}^{jk}_{2,0} &= \frac{1}{\sqrt{6}}\,\left(3\,\op{I_{zj}I_{zk}}-\vop{I_j}\pe\vop{I_k}\right)\\
    &= \frac{1}{\sqrt{6}}\,\left[2\,\op{I_{zj}I_{zk}}-\frac{1}{2}\,\left(\op{I_{+j}I_{-k}}+\op{I_{-j}I_{+k}}\right)\right],
    \end{split}
    \end{equation}
    and the dipolar couplings defined as (in Hz units)

    \begin{equation}\label{acop_dip}
    D_{j k}\equiv \frac{\mu_0\gamma^{2}\hbar}{8\pi^2} \frac{\left(1-3\cos^{2}\beta_{j k}\right)}{2 r_{jk}^{3}}\,S_{zz},
    \end{equation}
    under the international system of units (SI), where the indices $j,k$ run over all the proton sites within the molecule.
    In \eq{acop_dip}, $\gamma$ is the proton gyromagnetic ratio, $\mu_0$ is the vacuum permeability, $r_{jk}$ is the internuclear distance between spins $j$ and $k$, $\beta_{jk}$ stand for the polar angle of the vector $\vect{r_{jk}}$ with respect to the system fixed to the molecule, $S_{zz}$ is the mean molecular order, $\vop{I_j} \equiv \op{I_{xj}}\,\vers{x} + \op{I_{yj}}\,\vers{y} + \op{I_{zj}}\,\vers{z}$ is the vector of the i-th spin angular momentum and $\op{I_{\alpha i}}$ is the i-th spin angular momentum operator, with $\alpha\equiv\{x, y, z, +, -\}$.
    Since all the spin operators in this section act on the spin Hilbert space of one molecule, for instance with the molecular index 1, for simplicity we will not explicit in such operators the upper index $(s_1)$ pointing the corresponding molecular spin Hilbert space.\\
    It is useful for describing the spin dynamics under reversion to calculate a set of operators which result from applying pulses with an angle $\theta$ and opposite phases on the dipolar Hamiltonian of one molecule, $\opc{\widehat{H}}_{S1}$.
    Such expressions are the following

    \begin{subequations}\label{HS1_pulses}
    \begin{equation}\label{HS1_pulses_px}
    \begin{split}
    &\opc{\widehat{H}}^{x}_{S1}(\theta) \equiv  \op{R}^{\theta}_{\op{x}}\,\opc{\widehat{H}}_{S1}\,\op{R}^{\dag\theta}_{\op{x}} =
    \frac{3\cos^2{\theta}-1}{2}\,\opc{\widehat{H}}_{S1}\\
    &\quad -i\,\sqrt{\frac{3}{2}}\,\sin{\theta}\cos{\theta}\,\opc{\widehat{H}}^{(1)}_{S1}-
    \sqrt{\frac{3}{2}}\frac{\sin^2{\theta}}{2}\,\opc{\widehat{H}}^{(2)}_{S1},
    \end{split}
    \end{equation}
    \begin{equation}\label{HS1_pulses_mx}
    \begin{split}
    &\opc{\widehat{H}}^{\bar{x}}_{S1}(\theta) \equiv \op{R}^{\dag\theta}_{\op{x}}\,\opc{\widehat{H}}_{S1}\,\op{R}^{\theta}_{\op{x}} =
    \frac{3\cos^2{\theta}-1}{2}\,\opc{\widehat{H}}_{S1}\\
    &\quad +i\,\sqrt{\frac{3}{2}}\,\sin{\theta}\cos{\theta}\,\opc{\widehat{H}}^{(1)}_{S1}-
    \sqrt{\frac{3}{2}}\frac{\sin^2{\theta}}{2}\,\opc{\widehat{H}}^{(2)}_{S1},
    \end{split}
    \end{equation}
    \begin{equation}\label{HS1_pulses_py}
    \begin{split}
    &\opc{\widehat{H}}^{y}_{S1}(\theta) \equiv \op{R}^{\theta}_{\op{y}}\,\opc{\widehat{H}}_{S1}\,\op{R}^{\dag\theta}_{\op{y}} =
    \frac{3\cos^2{\theta}-1}{2}\,\opc{\widehat{H}}_{S1}\\
    &\quad -\sqrt{\frac{3}{2}}\,\sin{\theta}\cos{\theta}\,\opc{\widehat{H}}^{(\bar{1})}_{S1}+
    \sqrt{\frac{3}{2}}\frac{\sin^2{\theta}}{2}\,\opc{\widehat{H}}^{(2)}_{S1},
    \end{split}
    \end{equation}
    \begin{equation}\label{HS1_pulses_my}
    \begin{split}
    &\opc{\widehat{H}}^{\bar{y}}_{S1}(\theta) \equiv \op{R}^{\dag\theta}_{\op{y}}\,\opc{\widehat{H}}_{S1}\,\op{R}^{\theta}_{\op{y}} =
    \frac{3\cos^2{\theta}-1}{2}\,\opc{\widehat{H}}_{S1}\\
    &\quad +\,\sqrt{\frac{3}{2}}\,\sin{\theta}\cos{\theta}\,\opc{\widehat{H}}^{(\bar{1})}_{S1}+
    \sqrt{\frac{3}{2}}\frac{\sin^2{\theta}}{2}\,\opc{\widehat{H}}^{(2)}_{S1},
    \end{split}
    \end{equation}
    \end{subequations}
    where the pulses of an angle $\theta$ and direction $\alpha$ are defined as $\op{R}^{\theta}_{\alpha} \equiv e^{i\,\op{I}_{\op{\alpha}}\theta}$, with $\op{I}_{\op{\alpha}} \equiv \sum_i \op{I_{\alpha i}}$.
    In \eq{HS1_pulses} we defined the Hamiltonians

    \begin{subequations}\label{HS1_tensor}
    \begin{equation}\label{HS1_tensor_1}
    \opc{\widehat{H}}^{(1)}_{S1} \equiv \pi \sqrt{6}\, \sum_{j \neq k} D_{j k} \left(\op{T}^{jk}_{2,+1}+\op{T}^{jk}_{2,-1}\right),
    \end{equation}
    \begin{equation}\label{HS1_tensor_m1}
    \opc{\widehat{H}}^{(\bar{1})}_{S1} \equiv \pi \sqrt{6}\, \sum_{j \neq k} D_{j k} \left(\op{T}^{jk}_{2,+1}-\op{T}^{jk}_{2,-1}\right),
    \end{equation}
    \begin{equation}\label{HS1_tensor_2}
    \opc{\widehat{H}}^{(2)}_{S1} \equiv \pi \sqrt{6}\, \sum_{j \neq k} D_{j k} \left(\op{T}^{jk}_{2,+2}+\op{T}^{jk}_{2,-2}\right),
    \end{equation}
    \end{subequations}
    with the following 2nd order tensors

    \begin{subequations}\label{T212}
    \begin{equation}\label{T2p1}
    \op{T}^{jk}_{2,+1} = -\frac{1}{2}\,\left(\op{I_{zj}I_{+k}}+\op{I_{+j}I_{zk}}\right),
    \end{equation}
    \begin{equation}\label{T2m1}
    \op{T}^{jk}_{2,-1} = \frac{1}{2}\,\left(\op{I_{zj}I_{-k}}+\op{I_{-j}I_{zk}}\right),
    \end{equation}
    \begin{equation}\label{T2p2}
    \op{T}^{jk}_{2,+2} = \frac{1}{2}\,\op{I_{+j}I_{+k}},
    \end{equation}
    \begin{equation}\label{T2m2}
    \op{T}^{jk}_{2,-2} = \frac{1}{2}\,\op{I_{-j}I_{-k}}.
    \end{equation}
    \end{subequations}
    The Hamiltonian $\opc{\widehat{H}}_{S1}$ is called the secular part of the dipolar Hamiltonian and the Hamiltonians $\opc{\widehat{H}}^{(1)}_{S1}$, $\opc{\widehat{H}}^{(\bar{1})}_{S1}$ and $\opc{\widehat{H}}^{(2)}_{S1}$ constitute the non-secular parts of the dipolar Hamiltonian.\\

    In order to take into account experimental misadjustemnts, consider the case that the rf pulses in the MREV8 pulse sequence of \fig{Paper_Fig_SecPulsos_sup} (b) are of an arbitrary angle $\theta$ instead of $\pi/2$ pulses.
    In such a case, and supposing the same angle $\theta$ for all the pulses, the evolution operator of the sequence in \fig{Paper_Fig_SecPulsos_sup} (b) will be

    \begin{equation}\label{Evol_Op_MREV8_theta}
    \begin{split}
    &\op{\widehat{U}}^{\theta}_{RS}(\tau_2, \tau_1) \equiv\\
    &\op{R}^{\dag\theta}_{\op{x}}\, \op{\widehat{U}^{x,\theta}_{rt}}(\tau_2/2) \op{\widehat{U}}(\tau_1) \op{\widehat{U}^{y,\theta}_{rt}}(\tau_2) \op{\widehat{U}}(\tau_1) \op{\widehat{U}^{x,\theta}_{rt}}(\tau_2/2)\, \op{R}^{\theta}_{\op{x}}\\
    &\times \op{R}^{\theta}_{\op{x}}\, \op{\widehat{U}^{\bar{x},\theta}_{rt}}(\tau_2/2) \op{\widehat{U}}(\tau_1) \op{\widehat{U}^{\bar{y},\theta}_{rt}}(\tau_2) \op{\widehat{U}}(\tau_1) \op{\widehat{U}^{\bar{x},\theta}_{rt}}(\tau_2/2)\, \op{R}^{\dag\theta}_{\op{x}},
    \end{split}
    \end{equation}
    where
    \begin{equation}\label{Evol_Op_tau1}
    \op{\widehat{U}}(\tau_1) = e^{-i\,\opc{\widehat{H}}_{S1}\,\tau_1},
    \end{equation}
    \begin{equation}\label{Evol_Op_tau2}
    \op{\widehat{U}^{\alpha,\theta}_{rt}}(\tau_2) = e^{-i\,\opc{\widehat{H}}^{\alpha}_{S1}(\theta)\,\tau_2},
    \end{equation}
    with $\alpha\equiv\{\op{x, y, \bar{x}, \bar{y}}\}$ and we applied the identity operator 
    $\op{1} \equiv \op{R}^{\theta}_{\op{x}}\op{R}^{\dag\theta}_{\op{x}} \equiv \op{R}^{\dag\theta}_{\op{x}}\op{R}^{\theta}_{\op{x}}$ at the ends and the middle of the sequence.\\
    When the times $\tau_1$ and $\tau_2$ are small enough, it is possible to approximate the evolution operator \numeq{Evol_Op_MREV8_theta} by puting together the exponent of all the operators \numeq{Evol_Op_tau1} and \numeq{Evol_Op_tau2}.
    Besides, if the angle $\theta$ is close to $\pi/2$, the contribution to the dynamics of the non-secular Hamiltonians vanishes due to cancellation of the terms with $\opc{\widehat{H}}^{(1)}_{S1}$, $\opc{\widehat{H}}^{(\bar{1})}_{S1}$ and $\opc{\widehat{H}}^{(2)}_{S1}$,
    and the resulting evolution operator will be

    \begin{equation}\label{Evol_Op_MREV8_theta_aprox}
    \op{\widehat{U}}^{\theta}_{RS}(t_2, t_1) \simeq \op{R}^{\dag\theta}_{\op{x}}\,\op{\widehat{U}_{rt}}(t_2)\op{\widehat{U}}(t_1)\,\op{R}^{\theta}_{\op{x}},
    \end{equation}
    where we defined
    \begin{equation}\label{Evol_Op_Free}
    \op{\widehat{U}}(t_1) \equiv e^{-i\,\opc{\widehat{H}}_{S1}\,t_1},
    \end{equation}
    \begin{equation}\label{Evol_Op_RT}
    \op{\widehat{U}_{rt}}(t_2) \equiv e^{i\,\opc{\widehat{H}}_{S1}\,t_2/\kappa},
    \end{equation}
    with $t_1 = 4\tau_1$, $t_2 = 4\tau_2$ and $\kappa \equiv 2/(1-3\cos^2{\theta})$.
    It is worth to note that \eq{Evol_Op_MREV8_theta_aprox} has the form of the general evolution operator for reversion experiments analysed in QD-II, except for the pulse operators at the ends in \eq{Evol_Op_MREV8_theta_aprox}.
    Nevertheless, the presence of such pulses does not alter the expectation values of the observables as calculated in QD-II; thus they can be neglected for simplicity in the study of decoherence effects in reversion experiments using the MREV8 sequence.
    If $t_2 = \kappa\,t_1$,  we can see that $\op{\widehat{U}}^{\theta}_{RS}(t_2, t_1) \simeq \op{1}$.
    Therefore, for that time setting, the spin dynamics during $t_1$ will be perfectly compensated (or will be very close to that) by the dynamics during the evolution time $t_2$,  and the initial state is recovered after the total time evolution.\\
    The pulse sequence of \fig{Paper_Fig_SecPulsos_sup} (c) consists of chains of blocks of MREV8 units.
    In this case, under the same approximation of the time parameters that yields the operator \numeq{Evol_Op_MREV8_theta_aprox}, we will obtain the same evolution operator as in  \eq{Evol_Op_MREV8_theta_aprox} for the experiment with $n$ blocks $RS$ of reversion, but with $t_1 = 4n\tau_1$ and $t_2 = 4n\tau_2$.\\
    
    In the case of an exact (and ideal) experimental setting of the $\pi/2$ pulses, the expressions in \eq{HS1_pulses}  simplify to
    \begin{subequations}\label{HS1_pulses_pi2}
    \begin{equation}\label{HS1_pulses_pi2_x}
    \begin{split}
    \opc{\widehat{H}}^{x}_{S1} &\equiv \opc{\widehat{H}}^{x}_{S1}(\pi/2) = \opc{\widehat{H}}^{\bar{x}}_{S1}(\pi/2)\\
    &\equiv -\frac{1}{\kappa}\,\opc{\widehat{H}}_{S1}-\frac{1}{2}\sqrt{\frac{3}{2}}\,\opc{\widehat{H}}^{(2)}_{S1},
    \end{split}
    \end{equation}
    \begin{equation}\label{HS1_pulses_pi2_y}
    \begin{split}
    \opc{\widehat{H}}^{y}_{S1} &\equiv \opc{\widehat{H}}^{y}_{S1}(\pi/2) = \opc{\widehat{H}}^{\bar{y}}_{S1}(\pi/2)\\
    &\equiv -\frac{1}{\kappa}\,\opc{\widehat{H}}_{S1}+\frac{1}{2}\sqrt{\frac{3}{2}}\,\opc{\widehat{H}}^{(2)}_{S1},
    \end{split}
    \end{equation}
    \end{subequations}
    with $\kappa = 2$.\\

    To observe the effects over the dynamics under reversion due to the non-secular Hamiltonians and/or experimental misadjustments, we performed numerical simulations of the experiment of \fig{Paper_Fig_SecPulsos_sup} (c), using the Hamiltonians \numeq{HS1_pulses_pi2} for a proton system model of a PAA$_{d6}$ molecule\cite{SegZam2011}, and including a misadjustment parameter of the reversion sequence, using
    $\tau_2 = (\kappa+\epsilon)\,\tau_1$ (in this case $\kappa = 2$).
    
    In order to observe how simulations using the last assumptions can take into account errors in the pulse sequence as well as non-secular contributions in the dynamics, we can see that in case of small error in the setting of the $\pi/2$ pulses, the value of $\kappa$ (i.e $2/(1-3\cos^2{\theta})$) in \eq{HS1_pulses} will be slightly different of $2$.
    Besides, the factors `$\sin{\theta}\cos{\theta}$' will be near to zero and thus there will be a small contribution to the reversion dynamics coming from the non-secular Hamiltonians $\opc{\widehat{H}}^{(1)}_{S1}$ and $\opc{\widehat{H}}^{(\bar{1})}_{S1}$.
    On the other hand, the factor `$\sqrt{\frac{3}{2}}\frac{\sin^2{\theta}}{2}$' will be close to $\frac{1}{2}\sqrt{\frac{3}{2}}$.
    Therefore, a small misadjustment in the $\pi/2$ pulses (the same for all the pulses in the sequence) can be taken into account by putting $\kappa$ different of 2 in the Hamiltonians \numeq{HS1_pulses_pi2}. Notice that, in the limit of small enough $\tau_1$ (as well as $\tau_2$) the contribution to the dynamics coming from the non-secular parts will be negligible and the resultant effect will be an evolution operator under reversion equal to \eq{Evol_Op_RT} with $\kappa \neq 2$.\\
    In such situation, if we set $\tau_2 = 2\tau_1$, the error in the pulse angle will be equivalent to a misadjustment of $\tau_2$ for a correct setting of the reversion sequence, due to $\kappa \neq 2$; in this way we should set $\tau_2 = (\kappa+\epsilon)\,\tau_1$ with $\kappa+\epsilon = 2$.
    Accordingly, we introduce the same kind of misadjustment using \eq{HS1_pulses_pi2} with $\kappa = 2$ and setting $\tau_2 = (\kappa+\epsilon)\,\tau_1$. In other words, using a sequence with ideal pulses and time misadjustments will take into account both effects, the time error setting and the slightly non-ideal pulses.\\
    
    Since there are only two Hamiltonians in \numeq{HS1_pulses_pi2} with different contributions of the non-secular Hamiltonian $\opc{\widehat{H}}^{(2)}_{S1}$, the sequence of operators given by \eq{Evol_Op_MREV8_theta} will actually be conformed by two equal sets of reversion evolution operators, i.e. there will be a duplication of such operators. Therefore, for the sake of simplicity we used in the simulation of the reversion sequence, for each block $RS$, the following evolution operator

    \begin{equation}\label{Evol_Op_RS1_sim}
    \op{\widehat{U}}^{\kappa,\epsilon}_{RS}(\tau_1) \equiv \op{\widehat{U}^{y}_{rt}}[(\kappa+\epsilon)\,\tau_1]\op{\widehat{U}}(\tau_1)\op{\widehat{U}^{x}_{rt}}[(\kappa+\epsilon)\,\tau_1]\op{\widehat{U}}(\tau_1),
    \end{equation}
    where
    \begin{equation}\label{Evol_Op_tau1_sim}
    \op{\widehat{U}}(\tau_1) = e^{-i\,\opc{\widehat{H}}_{S1}\,\tau_1},
    \end{equation}
    \begin{equation}\label{Evol_Op_tau2x_sim}
    \op{\widehat{U}^{x}_{rt}}[(\kappa+\epsilon)\,\tau_1] = e^{-i\,\opc{\widehat{H}}^{x}_{S1}\,(\kappa+\epsilon)\,\tau_1},
    \end{equation}
    \begin{equation}\label{Evol_Op_tau2y_sim}
    \op{\widehat{U}^{y}_{rt}}[(\kappa+\epsilon)\,\tau_1] = e^{-i\,\opc{\widehat{H}}^{y}_{S1}\,(\kappa+\epsilon)\,\tau_1},
    \end{equation}
    with $\kappa = 2$.\\
    The complete reversion evolution operator is $n$ times the operator of \eq{Evol_Op_RS1_sim}, namely
    \begin{equation}\label{Evol_Op_RSn_sim}
    \op{\widehat{U}}^{\kappa,\epsilon}_{RS(n)}(\tau_1) \equiv \op{\widehat{U}}^{\kappa,\epsilon}_{RS}(\tau_1)\times\cdots\times\op{\widehat{U}}^{\kappa,\epsilon}_{RS}(\tau_1)
    \equiv \prod^{(n)} \op{\widehat{U}}^{\kappa,\epsilon}_{RS}(\tau_1),
    \end{equation}
    which takes into account arbitrary values of $\tau_1$ and allows for the non-secular effects in the dynamics that can be expected at long times.
    The evolution time of each reversion block $RS$ is
    \begin{equation}\label{def_tau_c}
    \tau_c = (\kappa+1+\epsilon)\,2\tau_1,
    \end{equation}
    and the total evolution time has the discrete values of $\tau = n\tau_c$ ($n=0,1,2,\dots$). Thus, we can alternatively define the evolution operator \numeq{Evol_Op_RSn_sim} as
    \begin{equation}\label{Evol_Op_RSn_sim_tau}
    \op{\widehat{U}}^{\kappa,\epsilon,\tau_1}_{RS}(\tau) \equiv \op{\widehat{U}}^{\kappa,\epsilon}_{RS(n)}(\tau_1),
    \end{equation}
    where $n = \tau / \left[(\kappa+1+\epsilon)\,2\tau_1\right]$.\\
    
    Finally, the signals which we obtained from the simulations of the reversion experiments were the normalized expectation values of $\op{I}_{\op{y}}$ and they were calculated as

    \begin{equation}\label{Val_Obs_RSn_FID_errHd}
    \begin{split}
    &\left<\widehat{\op{I}}_{\op{y}}(t, \tau)^{\kappa,\epsilon}_{\tau_1}\right> = \frac{1}{tr_{s_1}\left\{\op{I}^2_{\op{y}}\right\}}\, tr_{s_1}\bigg\{\op{I}_{\op{y}}\,e^{-i\,\opc{\widehat{H}}_{S1}\,t}\\
    &\qquad\qquad\qquad\times \op{\widehat{U}}^{\kappa,\epsilon,\tau_1}_{RS}(\tau)\,
    \op{I}_{\op{y}}\,\op{\widehat{U}}^{\dag\,\kappa,\epsilon,\tau_1}_{RS}(\tau)\,e^{i\,\opc{\widehat{H}}_{S1}\,t}\bigg\}.
    \end{split}
    \end{equation}
    We performed numerical simulations of \eq{Val_Obs_RSn_FID_errHd} for different values of $\tau_1$ and $\epsilon$ with $\kappa = 2$. The results are shown in \figs{Paper_Fig_RTFID_FIDtau_PAAd6}-\numfig{Paper_Fig_RTFID_FFTcutN_PAAd6}.
    \fig{Paper_Fig_RTFID_FIDtau_PAAd6} shows the behavior of \eq{Val_Obs_RSn_FID_errHd} with the free-evolution time $t$ and the reversion time $\tau$. \fig{Paper_Fig_RTFID_FIDcut0_PAAd6} shows the dependence with $\tau$ of \eq{Val_Obs_RSn_FID_errHd} for $t=0$,
    which is useful to appreciate the reversion dynamics. \fig{Paper_Fig_RTFID_FFTtau_PAAd6} shows the $\tau$ dependent
    Fourier transforms of \eq{Val_Obs_RSn_FID_errHd} in the time $t$; in this picture we can test the eigen-selection effects on the spectral lines of the signals. For such test, several cuts of the spectra for different normalized frequency lines as function of $\tau$ are shown in \fig{Paper_Fig_RTFID_FFTcut_PAAd6}.

    \fig{Paper_Fig_RTFID_FFTNtau_PAAd6} contanins the spectra of \fig{Paper_Fig_RTFID_FFTtau_PAAd6} with the spectrum normalized for each value of $\tau$. In \fig{Paper_Fig_RTFID_FFTcutN_PAAd6} several cuts of the spectra of \fig{Paper_Fig_RTFID_FFTNtau_PAAd6} for different normalized frequency lines are shown. With these normalized spectra it is possible to compare the decay rate in the time $\tau$ for different spectral line.\\

    The evolution operator \numeq{Evol_Op_RSn_sim_tau} can be approximated for small enough values of $\tau_1$ times as:

    \begin{equation}\label{Evol_Op_RSn_sim_tau_approx}
    \begin{split}
    &\op{\widehat{U}}^{\kappa,\epsilon,\tau_1}_{RS}(\tau) \simeq e^{-i\,\left[2\,\opc{\widehat{H}}_{S1}+\left(\kappa+\epsilon\right)\left(\opc{\widehat{H}}^{x}_{S1}+\opc{\widehat{H}}^{y}_{S1}\right)\right]\,n\tau_1}\\
    &\qquad= e^{-i\,\left[\opc{\widehat{H}}_{S1}-\left(1+\epsilon/\kappa\right)\opc{\widehat{H}}_{S1}\right]\,2n\tau_1}
    = e^{i\,\opc{\widehat{H}}_{S1}\,\frac{\epsilon/\kappa}{\kappa+1+\epsilon}\,\tau},
    \end{split}
    \end{equation}
    where we used $\opc{\widehat{H}}^{x}_{S1}+\opc{\widehat{H}}^{y}_{S1} = -2\,\opc{\widehat{H}}_{S1}/\kappa$, with $\kappa = 2$, and
    $2n\tau_1 = \tau / \left(\kappa+1+\epsilon\right)$. Therefore, under this time approximation, using \eq{Evol_Op_RSn_sim_tau_approx} in \eq{Val_Obs_RSn_FID_errHd}, we obtain

    \begin{equation}\label{Val_Obs_RSn_FID_errHd_approx}
    \begin{split}
    &\left<\widehat{\op{I}}_{\op{y}}(t, \tau)^{\kappa,\epsilon}_{\tau_1}\right> \simeq \frac{1}{tr_{s_1}\left\{\op{I}^2_{\op{y}}\right\}}\, tr_{s_1}\bigg\{\op{I}_{\op{y}}\,e^{-i\,\opc{\widehat{H}}_{S1}\,t}\\
    &\qquad\times e^{i\,\opc{\widehat{H}}_{S1}\,\frac{\epsilon/\kappa}{\kappa+1+\epsilon}\,\tau}\,
    \op{I}_{\op{y}}\,e^{-i\,\opc{\widehat{H}}_{S1}\,\frac{\epsilon/\kappa}{\kappa+1+\epsilon}\,\tau}\,
    e^{i\,\opc{\widehat{H}}_{S1}\,t}\bigg\}.
    \end{split}
    \end{equation}
    We can note that \eq{Val_Obs_RSn_FID_errHd_approx} is the normalized expression of $\left<\widehat{\op{I}}_{\op{y}}(t, \tau)^{\epsilon}\right>$ shown in Section II.D in QD-II.
    The results  showing the effects of the dynamics under the approximation that yields \numeq{Val_Obs_RSn_FID_errHd_approx} are those corresponding to the settings $\{\tau_1 = 5\mu s, \epsilon/\kappa = 0.1\}$ and $\{\tau_1 = 5\mu s, \epsilon/\kappa = 0.3\}$ in the \figs{Paper_Fig_RTFID_FIDtau_PAAd6}-\numfig{Paper_Fig_RTFID_FFTcutN_PAAd6}, where the effects of the non-secular Hamiltonians in the dynamics have been mitigated by adopting a small value of $\tau_1$.
    We can see, using the trace property $tr\{AB\} = tr\{BA\}$, that for $t = 0$ the signal obtained in \eq{Val_Obs_RSn_FID_errHd_approx} as a function of the reversion time $\tau$ is the same as a single FID signal, but with a factor of time scaling of $\frac{\epsilon/\kappa}{\kappa+1+\epsilon}$.
    Such kind of signals are shown in \fig{Paper_Fig_RTFID_FIDcut0_PAAd6} for the mentioned setting, where it is observed that the behavior of the signals presents oscillations, which are typical of the FID's signals, where the smaller the ratio $\epsilon/\kappa$ is, the lower the frequency of the oscillation will be.

    For general cases of values of $\tau_1$ and $\epsilon/\kappa$, we can write the evolution operator \numeq{Evol_Op_RS1_sim} as
    \begin{equation}\label{Evol_Op_RSn_gen}
    \op{\widehat{U}}^{\kappa,\epsilon}_{RS}(\tau_1) \equiv e^{-i\,\opc{\widehat{H}}^{\ddag\,\kappa,\epsilon}_{S1}(\tau_1)\,\tau_c},
    \end{equation}
    with $\tau_c$ given in \numeq{def_tau_c}.
    We define the eigenbasis $\left\{\ket{\alpha_1}\right\}$ common to the evolution operator $\op{\widehat{U}}^{\kappa,\epsilon}_{RS}(\tau_1)$ and the Hamiltonian $\opc{\widehat{H}}^{\ddag\,\kappa,\epsilon}_{S1}(\tau_1)$, and we define the eigenvalues
    $A^{\kappa,\epsilon}_1(\tau_1) = \bra{\alpha_1}\op{\widehat{U}}^{\kappa,\epsilon}_{RS}(\tau_1)\ket{\alpha_1}$
    and $\alpha^{\kappa,\epsilon}_1(\tau_1) = \bra{\alpha_1}\opc{\widehat{H}}^{\ddag\,\kappa,\epsilon}_{S1}(\tau_1)\ket{\alpha_1}$.
    Therefore the Hamiltonian $\opc{\widehat{H}}^{\ddag\,\kappa,\epsilon}_{S1}(\tau_1)$ can be obtained writing its matrix representation in such eigenbase, where $\alpha^{\kappa,\epsilon}_1(\tau_1) = i\,\ln{A^{\kappa,\epsilon}_1(\tau_1)}/\tau_c$.
    Using \eq{Evol_Op_RSn_gen} in \eq{Evol_Op_RSn_sim}, the evolution operator \numeq{Evol_Op_RSn_sim_tau} for the $n$ blocks of reversion will be
    \begin{equation}\label{Evol_Op_RSn_gen_tau}
    \op{\widehat{U}}^{\kappa,\epsilon,\tau_1}_{RS}(\tau) \equiv e^{-i\,\opc{\widehat{H}}^{\ddag\,\kappa,\epsilon}_{S1}(\tau_1)\,\tau},
    \end{equation}
    with $\tau = n\tau_c$, and the expectation value \numeq{Val_Obs_RSn_FID_errHd} for general cases can be written as
    \begin{equation}\label{Val_Obs_RSn_FID_errHd_gen}
    \begin{split}
    &\left<\widehat{\op{I}}_{\op{y}}(t, \tau)^{\kappa,\epsilon}_{\tau_1}\right> = \frac{1}{tr_{s_1}\left\{\op{I}^2_{\op{y}}\right\}}\, tr_{s_1}\bigg\{\op{I}_{\op{y}}\,e^{-i\,\opc{\widehat{H}}_{S1}\,t}\\
    &\qquad\qquad\times e^{-i\,\opc{\widehat{H}}^{\ddag\,\kappa,\epsilon}_{S1}(\tau_1)\,\tau}\,
    \op{I}_{\op{y}}\,e^{i\,\opc{\widehat{H}}^{\ddag\,\kappa,\epsilon}_{S1}(\tau_1)\,\tau}\,e^{i\,\opc{\widehat{H}}_{S1}\,t}\bigg\}.
    \end{split}
    \end{equation}
    The signals in \eq{Val_Obs_RSn_FID_errHd_gen} are the normalized form of the expectation values $\left<\widehat{\op{I}}_{\op{y}}(t, \tau)^\ddag\right>$  which were obtained in Section II.D in QD-II, while in \eq{Val_Obs_RSn_FID_errHd_gen} the dependence of $\opc{\widehat{H}}^\ddag_{S1}$ with the parameters $\kappa$, $\epsilon$ and $\tau_1$ has been explicitly written .\\

    It can be seen by comparing \eq{Evol_Op_RSn_sim_tau_approx} and \eq{Evol_Op_RSn_gen_tau} that under the approximation of small values of $\tau_1$

    \begin{equation}\label{HamS1_gen_approx}
    \opc{\widehat{H}}^{\ddag\,\kappa,\epsilon}_{S1}(\tau_1) \simeq -\opc{\widehat{H}}_{S1}\,\frac{\epsilon/\kappa}{\kappa+1+\epsilon},
    \end{equation}
    and the expression of \eq{Val_Obs_RSn_FID_errHd_gen} is approximated by \eq{Val_Obs_RSn_FID_errHd_approx}.\\

    In other calculated cases, if we set $\epsilon = 0$ for different values of $\tau_1$, the signals \numeq{Val_Obs_RSn_FID_errHd_gen} so obtained will show the effects on the reversion dynamics due exclusively to the non-secular Hamiltonians.
    The results which expose these cases correspond with the settings $\{\tau_1 = 10\mu s, \epsilon/\kappa = 0\}$ and
    $\{\tau_1 = 50\mu s, \epsilon/\kappa = 0\}$ in the \figs{Paper_Fig_RTFID_FIDtau_PAAd6}-\numfig{Paper_Fig_RTFID_FFTcutN_PAAd6}.
    In the extreme misadjustment setting with $\tau_1 = 50\mu s$ we can observe from \fig{Paper_Fig_RTFID_FIDcut0_PAAd6} a fast bell-shape decay of the signal, but such experimental setting can be improved drastically in the practice.
    On the other hand, such decay does not present an eigen-selection effect as can be observed from the comparison between the evolution in $\tau$ of different spectral lines shown in \fig{Paper_Fig_RTFID_FFTtau_PAAd6} and \fig{Paper_Fig_RTFID_FFTcut_PAAd6} or the normalized spectra in \fig{Paper_Fig_RTFID_FFTNtau_PAAd6}.
    Besides, in \fig{Paper_Fig_RTFID_FFTcutN_PAAd6} there is not a monotonous and eigen-selective decay for each spectral line with different frequency.\\

    Other results shown are for the settings $\{\tau_1 = 20\mu s, \epsilon/\kappa = 0.1\}$, which constitutes an intermediate case between the ones with small values of $\tau_1$ and $\epsilon \neq 0$ (expressed by \eq{Val_Obs_RSn_FID_errHd_approx}) and with large values of $\tau_1$ and $\epsilon = 0$, and $\{\tau_1 = 5\mu s, \epsilon/\kappa = 0\}$. The latter case would be close to the the best possible practical condition and it is used as contrast with the other setting cases.\\

    The conclusion, extracted from the results of the simulations, is that for the closed spin system there are no eigen-selection effects present in the signals produced by experimental misadjustments or contribution to the dynamics coming from the non-secular Hamiltonians, as can be observed from the spectra evolution and the spectral lines evolution in the results shown in \figs{Paper_Fig_RTFID_FFTtau_PAAd6} - \numfig{Paper_Fig_RTFID_FFTcutN_PAAd6}.
    This conclusion reproduces the one extracted from the theoretical analysis in Section II.D in QD-II.


\section*{Figures}

\insertfig{h}{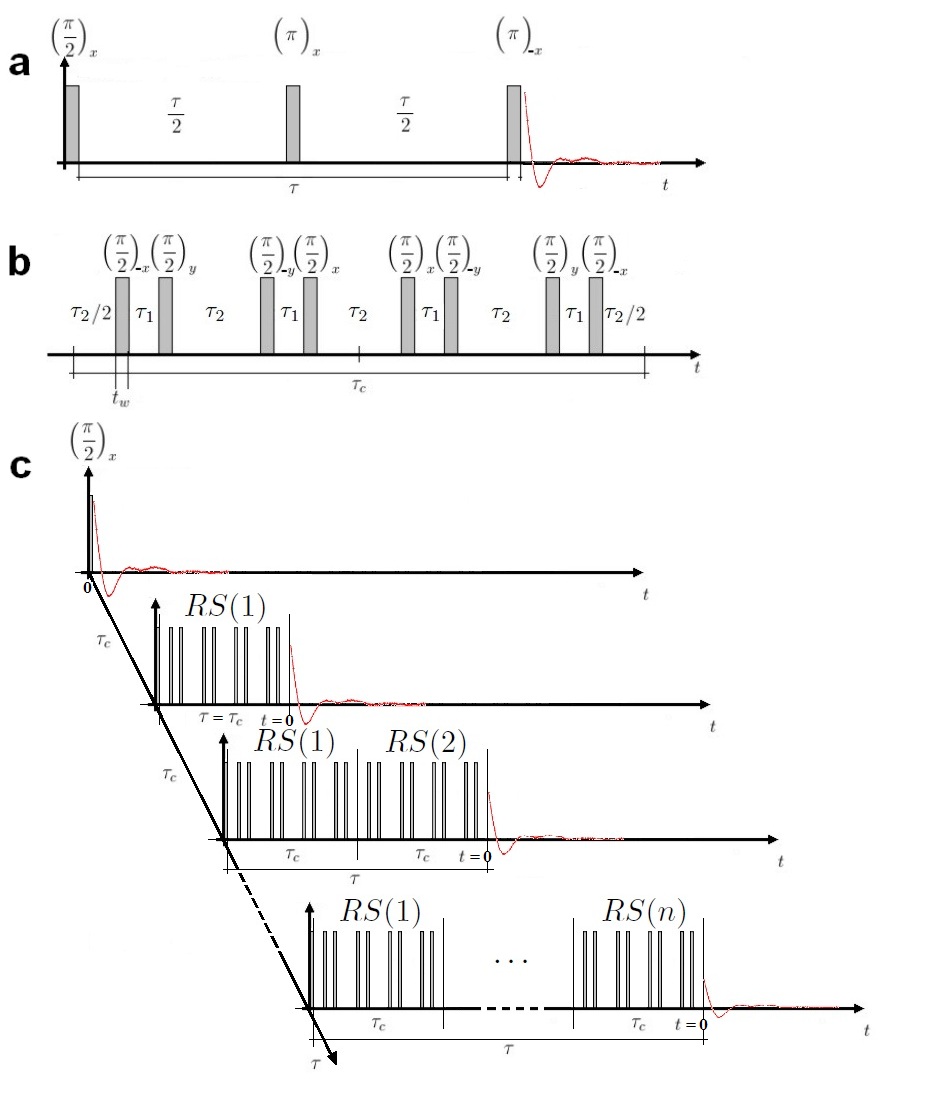}{8.81}{10.33}{Experimental pulse sequences. a: Zeeman reversion. b: MREV8. c: Compound reversion sequence, where each reversion block $RS(i)$ constitutes a single reversion sequence like the MREV8.}

\insertfig{h}{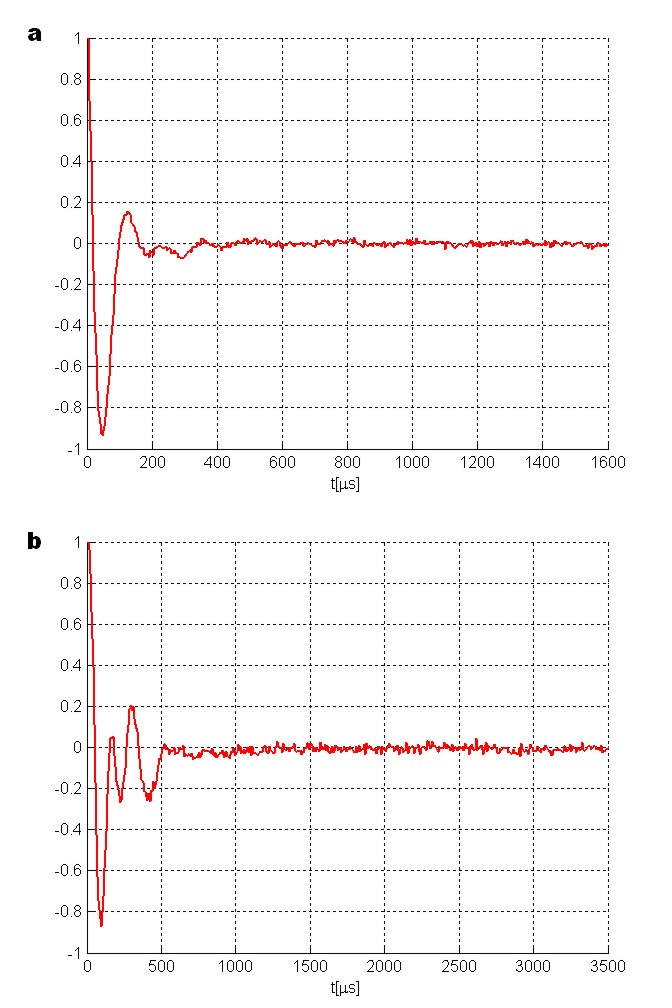}{8.89}{13.34}{Zeeman reversion experiment in nematic phase under resonance condition. a: 5CB ($T=27^{\circ} C$). b: PAA$_{d6}$ ($T=115^{\circ} C$).}

\insertfig{h}{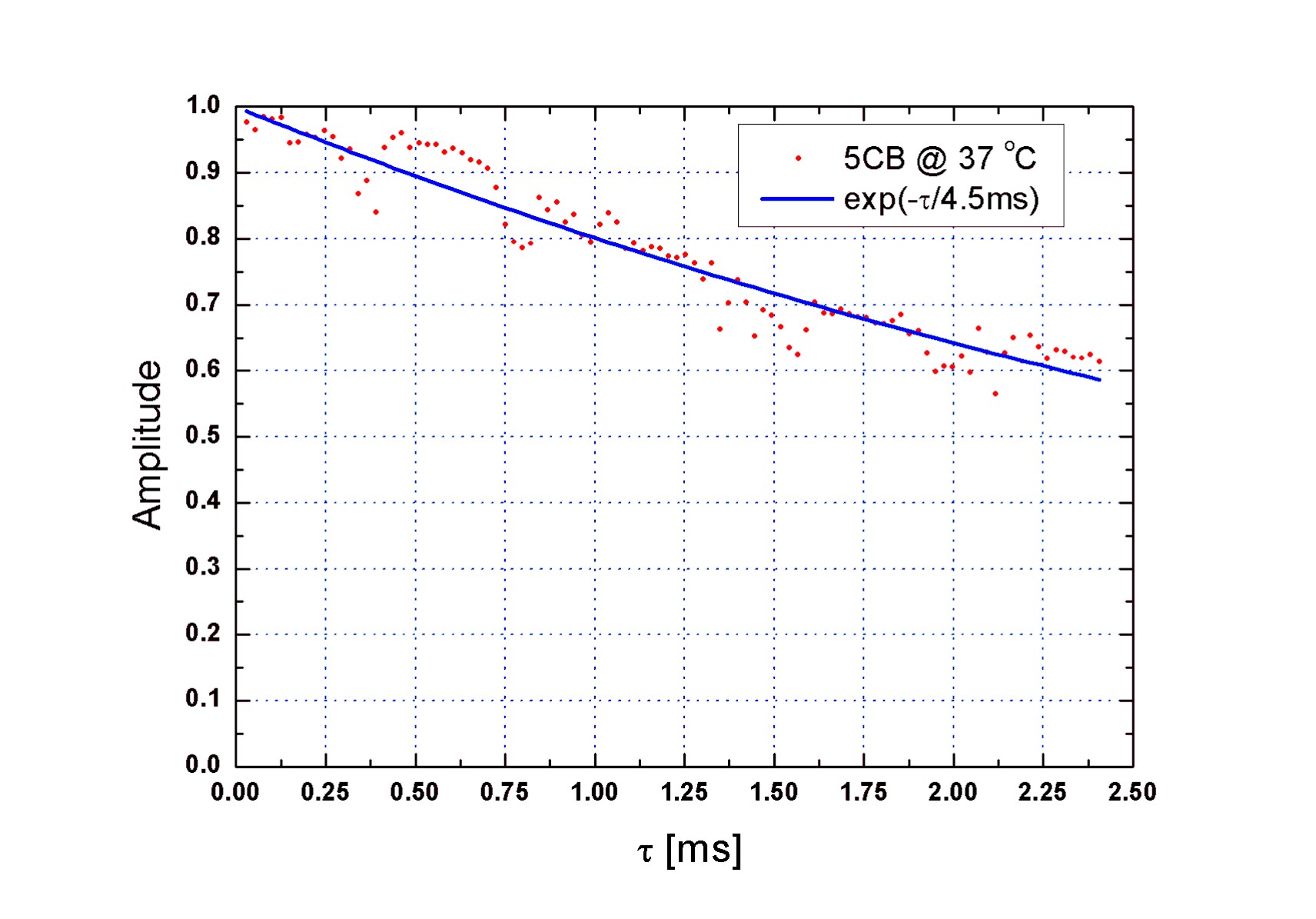}{8.89}{6.21}{Measured values of the maximum amplitude of the FID under a single MREV8 reversion, with continuous variation of the setting times, for the 5CB at $T=37^{\circ} C$ in isotropic phase under resonance condition (dots). The solid line is a fitting of the measured values with an exponential decay function with a characteristic decay time of 4.5ms.}

\insertfig{h}{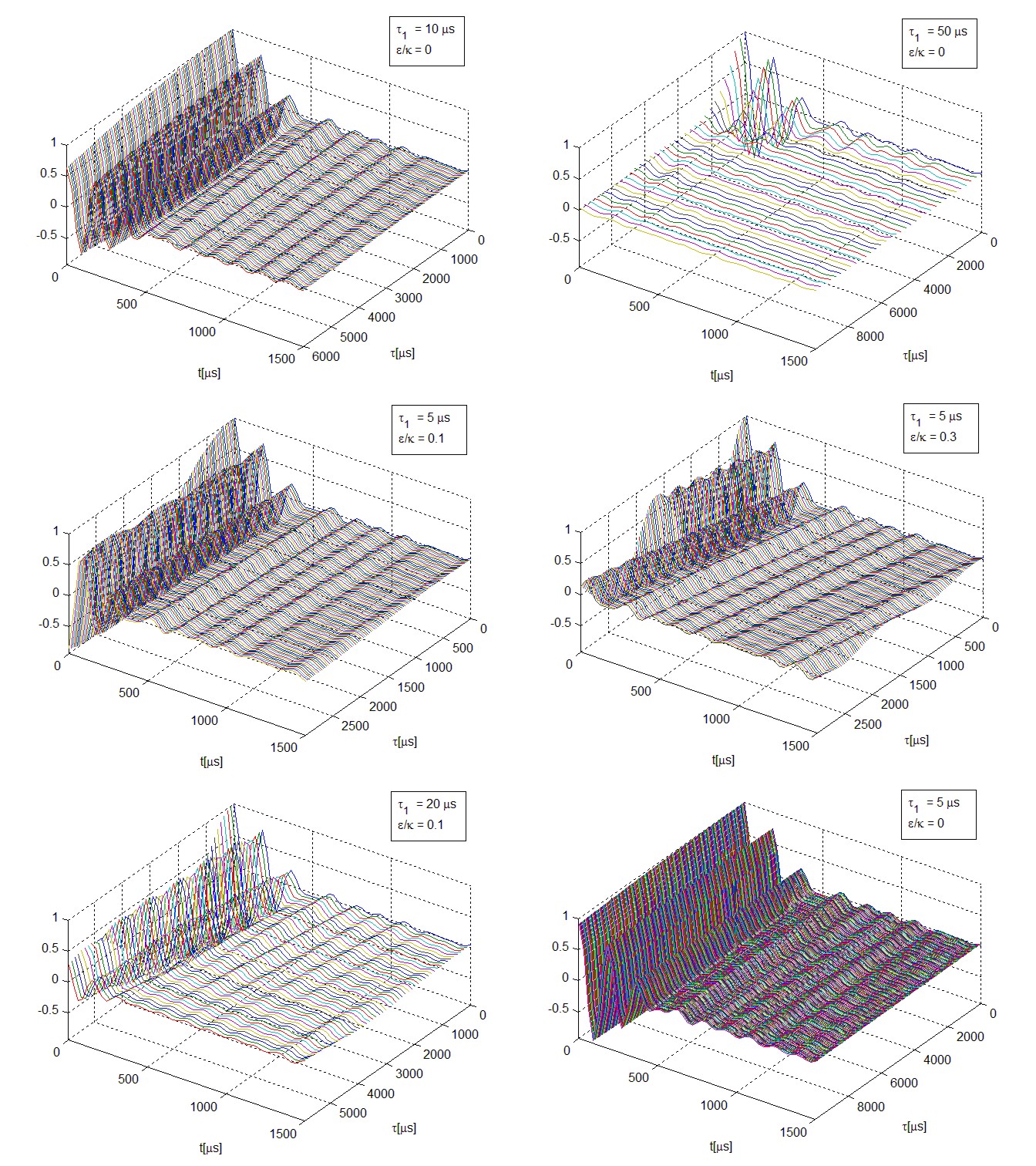}{17.78}{20}{Simulated results for the FID's as a function of the free-evolution time $t$ and the reversion time $\tau$, produced by reversion experiments with blocks of MREV8 sequence in a nematic PAA$_{d6}$ molecule under resonance condition. The results corresponding to different setting of the characteristic reversion time $\tau_1$ and the error time factor $\epsilon/\kappa$ are shown.}

\insertfig{h}{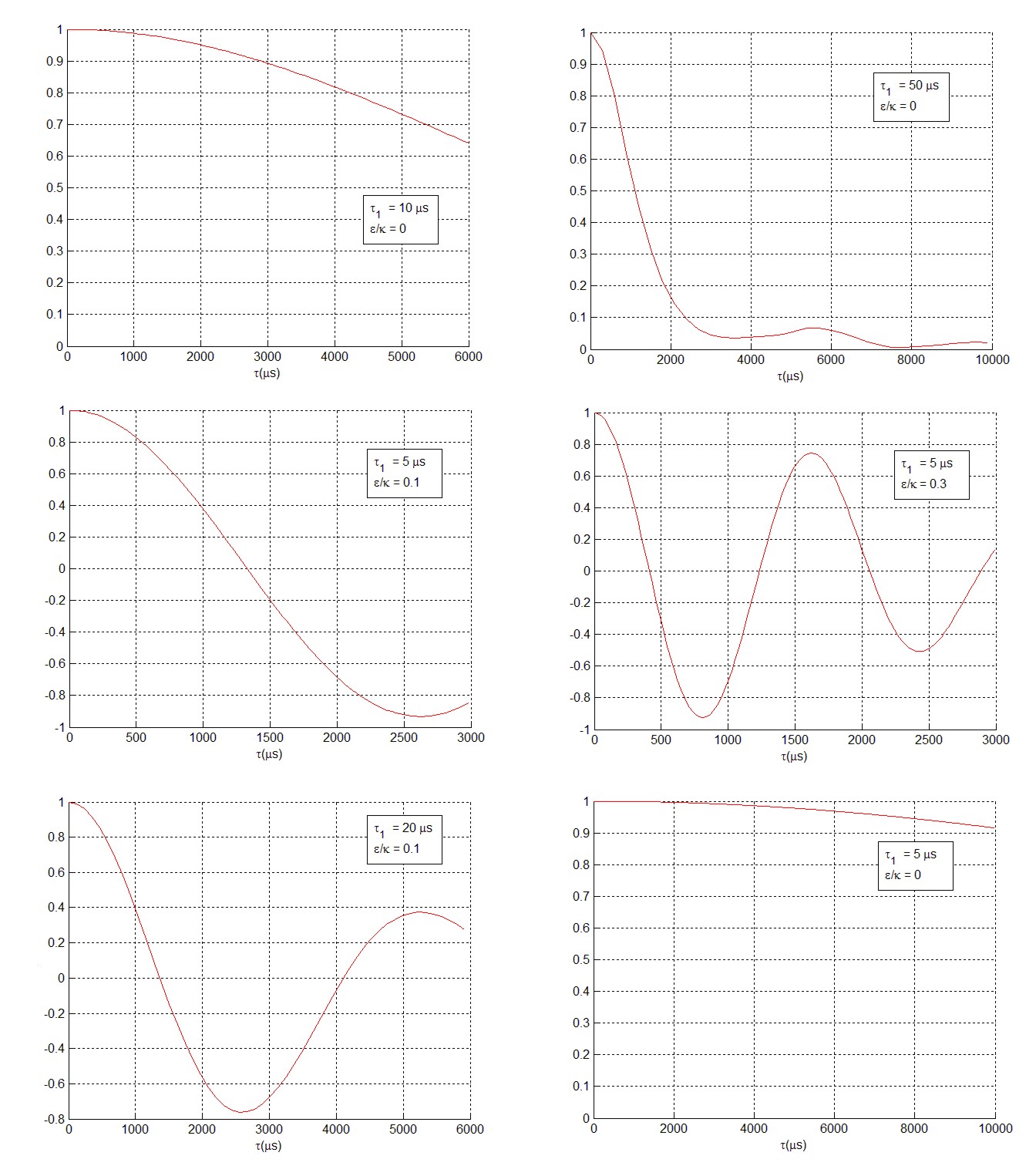}{17.78}{20}{Simulated results for the amplitude in $t=0$ of the FID's shown in \fig{Paper_Fig_RTFID_FIDtau_PAAd6} as a function of the reversion time $\tau$, produced by reversion experiments with blocks of MREV8 sequence in a nematic PAA$_{d6}$ molecule under resonance condition. The results corresponding to different setting of the characteristic reversion time $\tau_1$ and the error time factor $\epsilon/\kappa$ are shown.}

\insertfig{h}{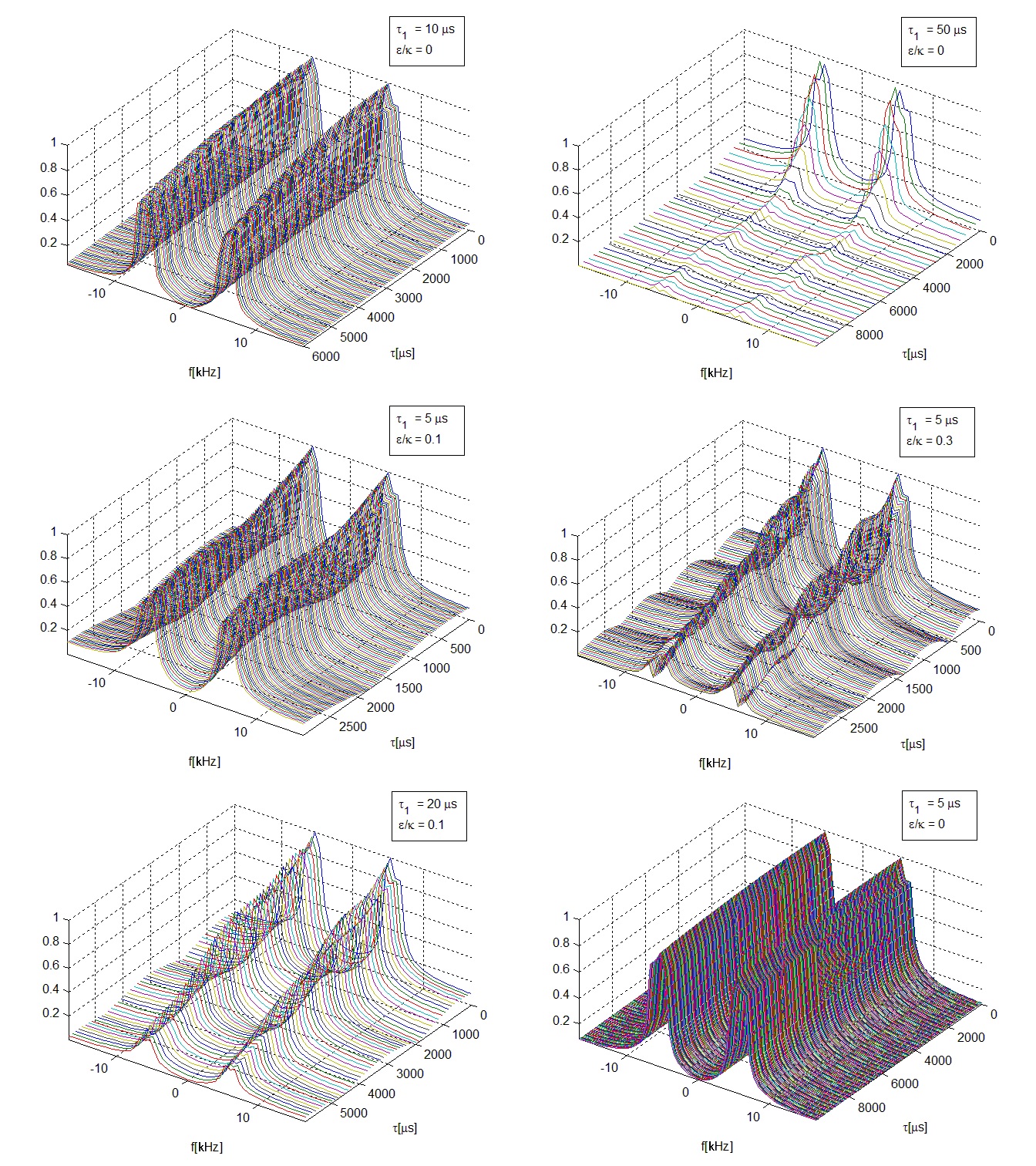}{17.78}{20}{Simulated results for the Fourier transform in the free-evolution time $t$ of the FID's as a function  of the reversion time $\tau$, produced by reversion experiments with blocks of MREV8 sequence in a nematic PAA$_{d6}$ molecule under resonance condition. The results corresponding to different setting of the characteristic reversion time $\tau_1$ and the error time factor $\epsilon/\kappa$ are shown.}

\insertfig{h}{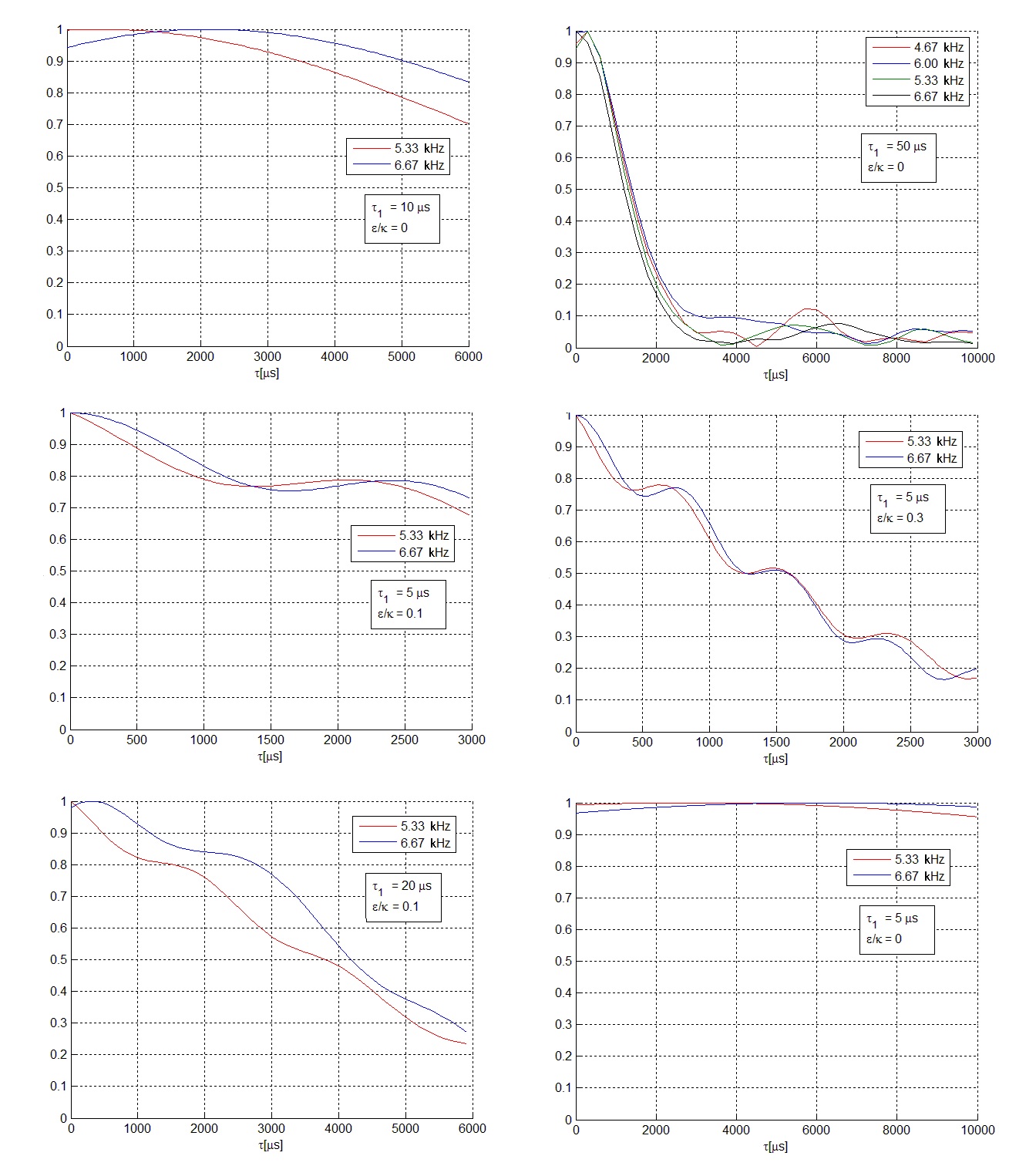}{17.78}{20}{Simulated results for the evolution of different normalized spectral lines of the spectra shown in \fig{Paper_Fig_RTFID_FFTtau_PAAd6} as a function of the reversion time $\tau$, in a nematic PAA$_{d6}$ molecule under resonance condition. The results corresponding to different setting of the characteristic reversion time $\tau_1$ and the error time factor $\epsilon/\kappa$ are shown.}

\insertfig{h}{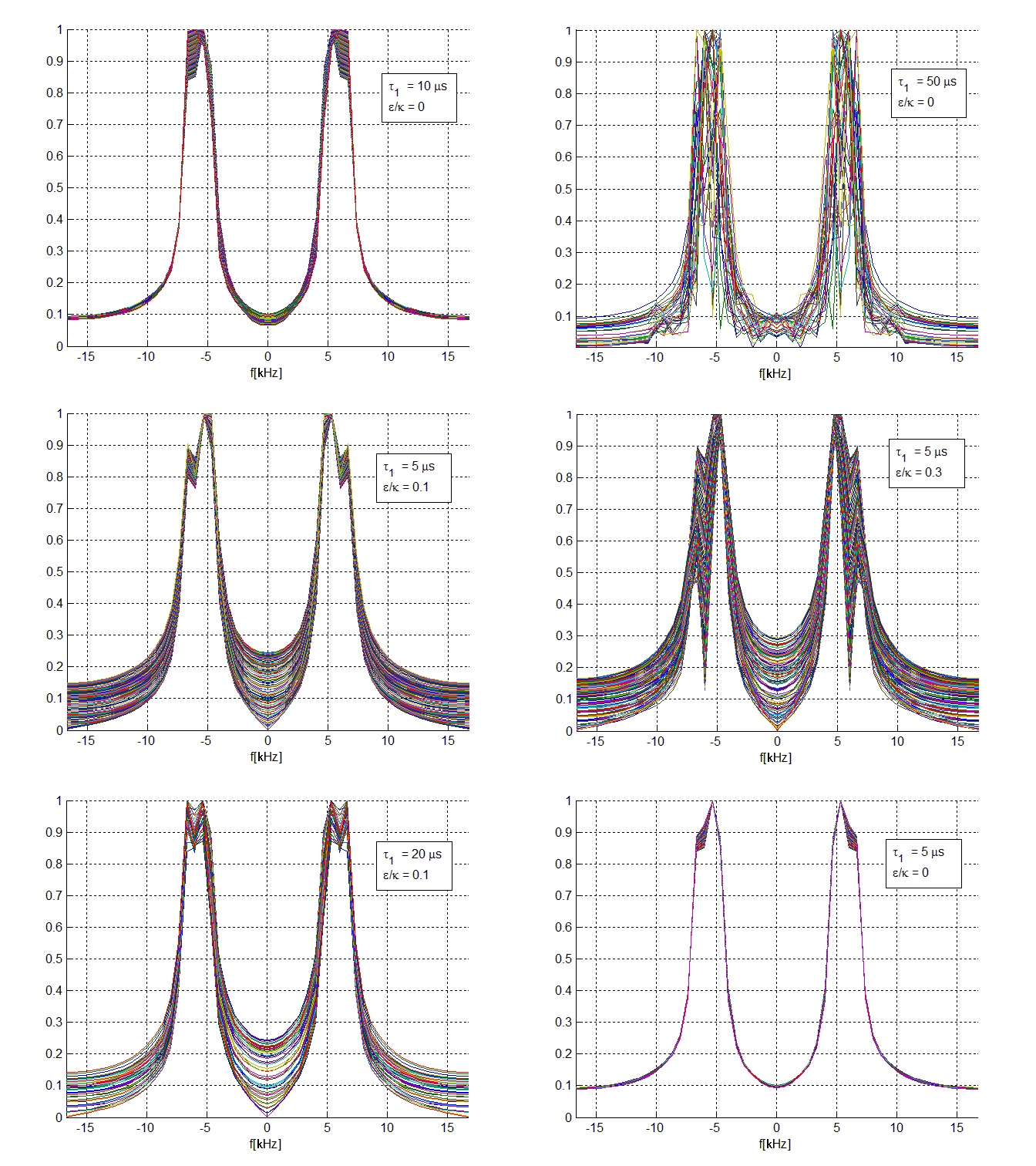}{17.78}{20}{Simulated results for the normalized Fourier transform in the free-evolution time $t$ of the FID's as a function of the reversion time $\tau$, produced by reversion experiments with blocks of MREV8 sequence in a nematic PAA$_{d6}$ molecule under resonance condition. The spectra were normalized for each $\tau$ value. The results corresponding to different setting of the characteristic reversion time $\tau_1$ and the error time factor $\epsilon/\kappa$ are shown.}

\insertfig{h}{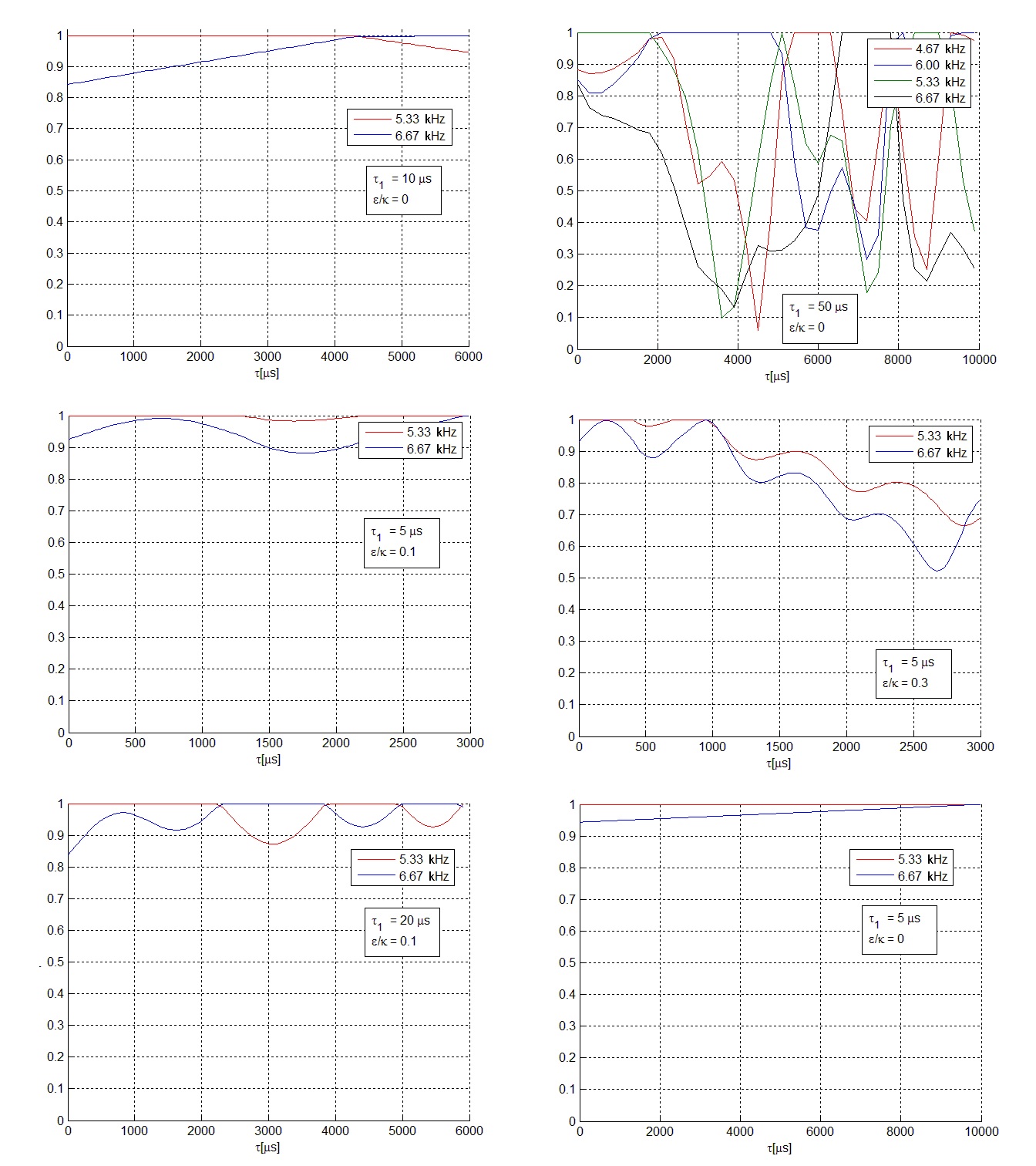}{17.78}{20}{Simulated results for the evolution of different normalized spectral lines of the spectra shown in \fig{Paper_Fig_RTFID_FFTNtau_PAAd6} as a function of the reversion time $\tau$, in a nematic PAA$_{d6}$ molecule under resonance condition, where the spectra have been normalized for each $\tau$ value. The results corresponding to different setting of the characteristic reversion time $\tau_1$ and the error time factor $\epsilon/\kappa$ are shown.}